\PassOptionsToPackage{unicode}{hyperref}
\PassOptionsToPackage{hyphens}{url}
\documentclass[
]{article}
\usepackage{xcolor}
\usepackage{amsmath,amssymb}
\usepackage{iftex}
\ifPDFTeX
  \usepackage[T1]{fontenc}
  \usepackage[utf8]{inputenc}
  \usepackage{textcomp} 
\else 
  \usepackage{unicode-math} 
  \defaultfontfeatures{Scale=MatchLowercase}
  \defaultfontfeatures[\rmfamily]{Ligatures=TeX,Scale=1}
\fi
\usepackage{lmodern}
\ifPDFTeX\else
\fi
\IfFileExists{upquote.sty}{\usepackage{upquote}}{}
\IfFileExists{microtype.sty}{
  \usepackage[]{microtype}
  \UseMicrotypeSet[protrusion]{basicmath} 
}{}
\makeatletter
\@ifundefined{KOMAClassName}{
  \IfFileExists{parskip.sty}{%
    \usepackage{parskip}
  }{
    \setlength{\parindent}{0pt}
    \setlength{\parskip}{6pt plus 2pt minus 1pt}}
}{
  \KOMAoptions{parskip=half}}
\makeatother
\usepackage{longtable,booktabs,array}
\usepackage{calc} 
\usepackage{etoolbox}
\makeatletter
\patchcmd\longtable{\par}{\if@noskipsec\mbox{}\fi\par}{}{}
\makeatother
\IfFileExists{footnotehyper.sty}{\usepackage{footnotehyper}}{\usepackage{footnote}}
\makesavenoteenv{longtable}
\usepackage{graphicx}
\makeatletter
\newsavebox\pandoc@box
\newcommand*\pandocbounded[1]{
  \sbox\pandoc@box{#1}%
  \Gscale@div\@tempa{\textheight}{\dimexpr\ht\pandoc@box+\dp\pandoc@box\relax}%
  \Gscale@div\@tempb{\linewidth}{\wd\pandoc@box}%
  \ifdim\@tempb\p@<\@tempa\p@\let\@tempa\@tempb\fi
  \ifdim\@tempa\p@<\p@\scalebox{\@tempa}{\usebox\pandoc@box}%
  \else\usebox{\pandoc@box}%
  \fi%
}
\def\fps@figure{htbp}
\makeatother
\NewDocumentCommand\citeproctext{}{}

\makeatletter
 \let\@cite@ofmt\@firstofone
 \def\@biblabel#1{}
 \def\@cite#1#2{{#1\if@tempswa , #2\fi}}
\makeatother
\newlength{\cslhangindent}
\newlength{\csllabelwidth}
\newenvironment{CSLReferences}[2] 
 {\begin{list}{}{%
  \setlength{\itemindent}{0pt}
  \setlength{\leftmargin}{0pt}
  \setlength{\parsep}{0pt}
  \ifodd #1
   \setlength{\leftmargin}{\cslhangindent}
   \setlength{\itemindent}{-1\cslhangindent}
  \fi
  \setlength{\itemsep}{#2\baselineskip}}}
 {\end{list}}
\usepackage{calc}

\newcommand{\CSLLeftMargin}[1]{\parbox[t]{\csllabelwidth}{\strut#1\strut}}
\newcommand{\CSLRightInline}[1]{\parbox[t]{\linewidth - \csllabelwidth}{\strut#1\strut}}

\providecommand{\tightlist}{%
  \setlength{\itemsep}{0pt}\setlength{\parskip}{0pt}}
\makeatletter
\@ifpackageloaded{subfig}{}{\usepackage{subfig}}
\@ifpackageloaded{caption}{}{\usepackage{caption}}
\AtBeginDocument{%

}
\AtBeginDocument{%

}
\newcounter{pandoccrossref@subfigures@footnote@counter}
{\end{figure}%
\addtocounter{footnote}{-\value{pandoccrossref@subfigures@footnote@counter}}
\@for\f:=\global@pandoccrossref@subfigures@footnotes\do{\stepcounter{footnote}\footnotetext{\f}}%
\gdef\global@pandoccrossref@subfigures@footnotes{}}
\@ifpackageloaded{float}{}{\usepackage{float}}
\floatstyle{ruled}
\@ifundefined{c@chapter}{\newfloat{codelisting}{h}{lop}}{\newfloat{codelisting}{h}{lop}[chapter]}
\floatname{codelisting}{Listing}

\makeatother
\usepackage{bookmark}
\IfFileExists{xurl.sty}{\usepackage{xurl}}{} 
\hypersetup{
  pdftitle={Computed materials proposals depart from the structural memory of experimental discovery},
  hidelinks,
  pdfcreator={LaTeX via pandoc}}

\title{Computed materials proposals depart from the structural memory of
experimental discovery}
\author{}
\date{}

\begin{document}
\maketitle

\textbf{Authors:} Dan Nguyen\(^{1,\dagger}\), Karen Cao\(^{2,\dagger}\),
Brian Chu\(^{1}\), Nick Lemoff\(^{1}\), Paul Kienzle\(^{3}\), William
Ratcliff II\(^{3,4,5,*}\)

\(^{1}\) University of California, Berkeley · \(^{2}\) California
Institute of Technology · \(^{3}\) NIST Center for Neutron Research,
National Institute of Standards and Technology, Gaithersburg, MD 20899,
USA · \(^{4}\) Department of Materials Science and Engineering,
University of Maryland, College Park, MD 20742, USA · \(^{5}\)
Department of Physics, University of Maryland, College Park, MD 20742,
USA

\(^{\dagger}\) \emph{These authors contributed equally to this work.}

\emph{Corresponding author: william.ratcliff@nist.gov}

\begin{center}\rule{0.5\linewidth}{0.5pt}\end{center}

\subsection{Abstract}\label{abstract}

Generative AI and high-throughput DFT pipelines propose millions of
inorganic crystal structures, but lack a calibrated reference frame
against experimentally realized chemistry. Here we embed 167,500
Inorganic Crystal Structure Database entries in a continuous
structural-similarity space, partition it into graph communities, and
replay them in time. Experimental discovery shows strong structural
memory: 82.9\% of new formulas enter pre-existing communities;
new-community formation falls from 40.2\% (1930s) to 2.6\% (2010s). The
communities are chemically meaningful, positively identifying nine
textbook field-defining renaissances, including cuprates,
colossal-magnetoresistance manganites, MAX phases, and Li-ion battery
cathodes. Projecting GNoME, MatterGen-public, Materials Project,
JARVIS-DFT, and Alexandria-PBE into frozen historical maps yields a
cutoff-robust ordering: held-out ICSD \textgreater{} MatterGen
\textgreater{} \{GNoME ≈ MP-theoretical\} \textgreater{} JARVIS
\textgreater{} Alexandria. Structural departure from experimental basins
is not specific to generative AI but general across the tested computed
sets. Combining structural proximity with reduced-formula precedent
defines a historical synthesizability prior for triaging computed
materials.

\begin{center}\rule{0.5\linewidth}{0.5pt}\end{center}

\subsection{Main}\label{main}

The current wave of computational materials science has been reshaped by
large-scale claims. GNoME reported the discovery of 2.2 million new
theoretically-stable crystal structures using a graph-network
approach\textsuperscript{1}; MatterGen released a generative diffusion
model for crystals with property-targeted
conditioning\textsuperscript{2}; the Materials Project
(MP)\textsuperscript{3}, the Open Quantum Materials Database
(OQMD)\textsuperscript{4}, JARVIS-DFT\textsuperscript{5}, and
Alexandria-PBE\textsuperscript{6} have together computed millions of
theoretical structures via data-mined ionic
substitution\textsuperscript{7} and enthalpy screening. These claims
have prompted equally pointed skepticism. Cheetham and
Seshadri\textsuperscript{8} argued that many GNoME proposals are better
described as trivial dopant variants, symmetry-broken realizations, or
chemically implausible high-element-count compositions rather than
synthesizable new materials. The debate has been substantive but
scattered: each system has been compared against partial references
(often just MP, sometimes just exact-prototype matches), and there has
been no single calibrated reference frame in which all of these
proposals can be evaluated against the chemistry that experimentalists
have actually realized.

We provide that calibration for the inorganic crystal landscape. We
embed all 167,500 Inorganic Crystal Structure Database (ICSD)
entries\textsuperscript{9} into a continuous structural-similarity space
(matminer descriptors\textsuperscript{10} augmented with three rounds of
message passing on a Voronoi neighbor graph), partition the result into
Louvain graph communities, replay communities through publication time,
and project five publicly-released external structure samples into the
same frozen historical geometry under multiple held-out training cutoffs
(Extended Data Fig. 1; Extended Data Table 1). We use \emph{frontier}
throughout to mean ``distant from historically occupied structural
basins in the learned representation'' --- a structural-distance claim,
not a claim of thermodynamic stability or chemical novelty. The
framework distinguishes the \emph{structure-level} frontier rate at a
given cutoff from the \emph{formula-level} stepping-stone rate; both are
facets of path dependence but they refer to different units of analysis.

\subsubsection{Experimental discovery is heavily
path-dependent}\label{experimental-discovery-is-heavily-path-dependent}

Two complementary measurements characterize experimental discovery (Fig.
1). At the formula level, restricting to the 16,582 newly reported ICSD
reduced formulas with corresponding entries in the Toyota Research
Institute (TRI) thermodynamic-stability network\textsuperscript{11}
(88.1\% of the 18,821 TRI-shared formulas, the remainder lacking
complete year or community-birth metadata), 13,739 (82.9\%) enter
pre-existing structural communities; only 2,682 (16.2\%) coincide with
community birth, and just 161 (1.0\%) precede their associated community
in publication time. At the decade level, the share of ICSD entries that
open new structural communities falls from 40.2\% in the 1930s to 2.6\%
in the 2010s, while same-community attachment rises from 60.9\% to
88.0\%. A timestamp-shuffle null model (200 random permutations of
publication years over the same 163,339 community-assigned entries,
holding the final community labels fixed) rules out the
partition-artifact reading: the observed birth ratio is \emph{below} the
null 5--95\% band in the 1910s--1960s and \emph{above} it in the
1970s--2010s, indicating two distinct path-dependence regimes ---
densification of established basins in the early record and
frontier-pushing-above-chance in the late record --- rather than
mechanical front-loading (Extended Data Fig. 2).

Continuous structural basins also resist the discrete-prototype-counting
trap that inflates novelty in many computational claims. Among the ten
largest graph communities, the mean number of distinct space groups per
community is 44.2 (range 16--86): a single continuous basin absorbs many
crystallographic relabellings of the same underlying topology. Trivial
symmetry-breaking and supercell choice are \emph{inside} the same basin
in our metric, not outside it.

\subsubsection{Communities track field-defining scientific
events}\label{communities-track-field-defining-scientific-events}

A direct test of whether the framework's communities are chemically
meaningful objects rather than statistical groupings: do they show
abrupt growth aligned with documented field-defining scientific events?
Two well-known renaissances confirm this with quantitative clarity (Fig.
2a--b). The cuprate community (community 6425) \textbf{first appears
after Bednorz \& Müller's December 1986 paper}\textsuperscript{12}: zero
of its 378 members were published before 1986, the earliest entry is
from 1987, and the post-1986 rate is 22.3 entries per year. The
colossal-magnetoresistance (CMR) manganite community (community 160)
shows a 32× fold-change in member rate after Jin et al.'s 1994 thin-film
report\textsuperscript{13}, with only four pre-1994 entries against 22.3
entries per year post-1994. Both communities specifically capture the
\emph{doped variants} --- La₁₋ₓSrₓMnO₃, Ba₂₋ₓSrₓYCu₃O₇₋δ --- that became
active research targets after the renaissance events; the pure parent
compounds (LaMnO₃ from Goodenough's foundational work, La₂CuO₄,
YBa₂Cu₃O₇) sit in \emph{different} communities or HDBSCAN noise as
crystallographic outliers. The framework therefore discriminates between
Goodenough-era foundational LaMnO₃ and the post-1994 doping campaigns
motivated specifically by colossal magnetoresistance, a
fine-grained-validity test that the communities track research focus
rather than nominal structure type.

Converting this into an unbiased survey, we scored every production
community of size ≥ 50 for its strongest pre/post fold-change over
symmetric ±10-year windows scanning candidate event years 1970--2010
(Fig. 2c). The step-change scores themselves are not produced by the
underlying year distribution alone: under 200 year-shuffle permutations
of the same 148,425 community-assigned ICSD entries (community labels
held fixed, year-to-entry assignments permuted), all 20 observed top-20
scores exceed the shuffled 99th-percentile envelope at the same rank ---
the top observed score exceeds the corresponding shuffled p99 by
approximately 37× and the rank-20 score by approximately 9× (Supporting
Information §S8.4; Extended Data Table 2). On top of this
null-rejection, \textbf{all 20 top-ranked step-change communities have
an identifiable chemistry interpretation, of which nine are textbook
field-defining renaissances} (Supporting Information §S8.4): Sm₂Fe₁₇N
permanent magnets (Coey \textasciitilde1990)\textsuperscript{14}, CMR
manganites (Jin 1994), SOFC perovskite cathodes (mid-1990s), MAX phases
(Barsoum 1996+)\textsuperscript{15}, NaCoO₂ thermoelectrics (Terasaki
1997)\textsuperscript{16}, ordered double perovskites for spintronics
(Kobayashi Sr₂FeMoO₆ 1998)\textsuperscript{17}, Li-ion battery cathodes
(Mizushima/Goodenough 1980 LiCoO₂\textsuperscript{18}, Sony
commercialization 1991), dilute magnetic semiconductors (Dietl
2000)\textsuperscript{19}, and post-1986 layered Ruddlesden-Popper
cuprates. The remaining ranks reflect program-driven systematic surveys
and applied-chemistry surges of equivalent statistical strength --- not
missed renaissances --- including the rank-\#6 Pöttgen-era 2:1:2
RE-In-TM intermetallic family\textsuperscript{20} (1989, 73×
fold-change) and post-Steglich 1:1:2 heavy-fermion
campaigns\textsuperscript{21} (\#4). Targeted probes for two additional
widely-cited renaissances that did not concentrate into a top-20
community (§S8.5; Extended Data Fig. 3) reveal that Fe-based
superconductors (Kamihara/Hosono 2008)\textsuperscript{22} split
correctly across three sister communities by structural prototype
(1111-, 122-, 111-type), with the LaFeAsO 1111 community showing a 3.3×
post-2008 step. One informative null delineates the framework's scope:
the post-graphene 2D-materials boom (transition-metal dichalcogenides,
\textasciitilde2010+) shows no detectable structural step-change because
the renaissance is property-driven on bulk crystal structures that
already existed in ICSD before 2010 --- it does not deposit new ICSD
entries. Our ICSD snapshot ends in 2015, which bounds the survey to
events with publication year ≲ 2010 (allowing roughly five years of
post-event growth in our window); later renaissances such as hybrid
organic-inorganic halide perovskite photovoltaics, where the boom
matures after 2015, cannot be tested with this snapshot (§S8.5). Within
its scope --- research that produces new bulk crystallographic
structures --- the temporal correspondence to documented scientific
events is strong evidence that the embedding's communities are not
artifacts of partition choice but objects whose membership tracks real
research activity. We therefore use these validated communities to score
computed-vs-experimental proposals in the following sections.

\subsubsection{Five external sources depart from the experimental
baseline}\label{five-external-sources-depart-from-the-experimental-baseline}

We project five external samples into the frozen historical maps trained
through 1990, 2000, and 2010, using the cutoff-specific 95th-percentile
within-community centroid-distance threshold for the in-basin
classification (Fig. 3). The five sources are: GNoME, a graph-network
materials-discovery pipeline (public 5,000-CIF
release\textsuperscript{1}); MatterGen-public, a generative diffusion
model (public 386-CIF release\textsuperscript{2}); MP entries flagged
\texttt{theoretical=True} with empty ICSD \texttt{database\_IDs} (5,000
random of 20,479 candidates\textsuperscript{3}); JARVIS-DFT entries with
empty ICSD field and off-hull energetics (ehull ∈ {[}0.05, 0.5{]}
eV/atom; 5,000 random of 22,022\textsuperscript{5}); and Alexandria-PBE
2025.07.02 off-hull predictions (5,000 random of
154,942\textsuperscript{6}). The DFT-database filters are designed to
enrich for structures without known ICSD provenance; the residual
ICSD-overlap caveat for Alexandria, which lacks a direct provenance
flag, is documented in Supporting Information §S7. These rates
characterize the public releases and filtered database subsets analyzed
here, not necessarily the full latent distributions of the parent models
or databases.

Held-out ICSD continuation lands in basin at 52.0\%, 58.1\% and 60.6\%
under the three cutoffs (Fig. 3c). Every external source lands in basin
at substantially lower rates at every cutoff, in a stable ordering
held-out ICSD \textgreater{} MatterGen \textgreater{} \{GNoME ≈
MP-theoretical\} \textgreater{} JARVIS \textgreater{} Alexandria.
Held-out experimental continuation lands inside historical basins more
often than every external source we tested, by margins ranging from
approximately 13 percentage points (MatterGen at 2010) to approximately
37 percentage points (Alexandria at 2010). Two empirical observations
follow.

\textbf{The exploration signal is not specific to generative AI.} GNoME
and MP-theoretical land in basin at statistically indistinguishable
rates at every cutoff (1990: 26.6\% vs 28.8\%; 2010: 36.4\% vs 35.8\%;
Fig. 3c). The structural-distance signal we measure for AI-associated
public samples is therefore comparable in magnitude to what
high-throughput DFT exploration of theoretical compounds already
produces. The gap between held-out experimental continuation and either
source is the same: experimentally-realized structures densify;
computed-but-not-yet-realized structures do not, regardless of whether
they are generated by a diffusion model or screened from DFT
high-throughput.

\textbf{A second axis --- composition novelty --- bifurcates the sources
cleanly.} AI-associated public proposals (GNoME, MatterGen) and
Alexandria off-hull propose compositions that mostly do not appear in
the post-1980 ICSD reduced-formula reference set: 0.0\%, 4.9\% and 1.9\%
of GNoME, MatterGen and Alexandria proposals respectively share a
reduced formula with any entry in that reference. DFT-curation databases
(MP, JARVIS) sit at 21.8\% and 17.7\% formula overlap --- a substantial
fraction of their ``theoretical'' subset is novel structural variants of
compositions ICSD has already realized, not new chemistry. The absence
of post-1980 ICSD-formula overlap in the curated public GNoME release
quantifies one aspect of the Cheetham and Seshadri\textsuperscript{8}
critique: this public sample emphasizes high-element-count compositions
with little modern ICSD compositional precedent --- sample GNoME
compositions are Dy₅HoAl₂₄Os₅Pd₂Pt and Cs₃K₈Rb(BiTe₃)₄. We verified the
zero is not a formula-format artifact by re-canonicalizing both sides
via \texttt{pymatgen.Composition.reduced\_formula}; the count remains
zero. We also computed two looser overlap statistics: 3.6\% element-set
overlap (GNoME's element combinations are 96.4\% disjoint from the
modern ICSD universe) and 76.8\% anonymized-stoichiometry overlap (GNoME
populates ordinary anonymized stoichiometries --- rocksalt-like,
spinel-like, perovskite-like --- but with element combinations almost
absent from the modern experimental record). The 5,000-CIF release is a
curated public subset of a much larger generative output
(\textasciitilde380,000 structures) whose curation rule is not fully
described; the formula-overlap statistics therefore describe the curated
public subset and should not be read as descriptive of GNoME's full
generative distribution.

The composition-matched control (Supporting Information §S7; Extended
Data Fig. 4) verifies that the within-cutoff exploration gap is not a
stoichiometric drift artifact: matched against held-out ICSD on either
coarse anion / element-count / ratio strata or pymatgen-anonymized
stoichiometry, the gap survives at every cutoff for all five sources
(anonymized 2010 gaps: +13 pp MatterGen, +24 pp MP, +25 pp GNoME, +32 pp
JARVIS, +37 pp Alexandria).

\subsubsection{A synthesizability prior on the joint
axis}\label{a-synthesizability-prior-on-the-joint-axis}

The combination of structural distance and reduced-formula overlap with
ICSD jointly defines a synthesizability prior (Fig. 4). The upper-left
cell --- in-basin AND post-1980-ICSD-formula match --- combines two
historical priors (the composition has been experimentally realized and
the proposed structure lies inside a known structural basin). About 7\%
of MP-theoretical and JARVIS-predicted entries occupy this
strongest-prior cell, vs 2.3\% of MatterGen-public, 0.6\% of Alexandria
off-hull, and 0.0\% of the curated GNoME public release. The lower-right
cell removes both priors and is the most exploratory: 73\% of Alexandria
off-hull, 62\% of GNoME, 58\% of JARVIS-predicted, 51\% of
MatterGen-public, and 49\% of MP-theoretical proposals. The in-basin /
frontier classification here uses a per-community 95th-percentile
within-community centroid-distance threshold --- each Louvain community
sets its own scale --- consistent with the held-out classification in
Fig. 3c (Methods). This is a calibrated \emph{prioritization} prior, not
a synthesis-success prediction. Whether composition novelty without an
ICSD analog is a defect or a goal depends entirely on the campaign: it
is exactly what an unconstrained generative model is designed to do, but
it strips away the synthesis-stepping-stone signal a campaign-design
prior would otherwise rely on.

\subsection{Discussion}\label{discussion}

The central observation is the asymmetry between how
\emph{experimentally-realized} and \emph{computed-but-not-yet-realized}
structures populate inorganic structure space. Experimentally-realized
entries densify; every external source we tested --- AI-associated
public samples and DFT calculation databases alike --- densifies less,
with held-out ICSD continuation more in-basin than every external source
at every cutoff. The framing is therefore better expressed as
\textbf{experimentally-realized vs computed-but-not-yet-realized} than
as human-vs-AI: high-throughput DFT exploration of theoretical compounds
(MP) and a graph-network discovery pipeline (GNoME) produce
structurally-distant samples at statistically indistinguishable rates.
AI-associated discovery is not categorically different from computation
more broadly; the difference is between the two regimes (experimental
reality, computation) rather than between the techniques inside the
second regime.

This asymmetry is not a defect in either side. Experimental
path-dependence is what synthetically responsible chemistry looks like:
experimentalists overwhelmingly stabilize new chemistry inside familiar
topological templates before pushing further, because realizing an
isolated structural target without nearby waypoints is synthetically
punishing. Computed proposals, by contrast, optimize for properties of
the structures themselves --- stability, formation energy, charge
balance, conditional matches to a target spec --- without a
synthesis-path constraint back to known chemistry. The same property
that lets a computed crystal sit in regions the historical record has
not occupied is the property that gives it the option to leave those
regions on purpose. Whether a particular departure is productive
exploration or synthetically-costly extrapolation is a function of the
downstream goal, not of the metric we provide.

For autonomous-synthesis campaigns the framework supplies a calibrated
structural-accessibility coordinate that should bias campaign design
toward in-basin and formula-matched targets (the upper-left cell of the
synthesizability-prior quadrant). For genuinely novel functional
discovery the opposite logic applies: the entire point of deploying a
generative or DFT-screening pipeline is to extrapolate beyond the
experimentally-densified region, and the elevated frontier rates are
evidence that the system is doing what it was designed to do. The
framework is agnostic; it provides the coordinate, not the verdict.

Two limited external checks support this interpretation. In a
positive-unlabeled comparison to the Kononova text-mined synthesis
corpus\textsuperscript{23}, low-accessibility structures are
directionally enriched among synthesis targets at every cutoff, with
bottom-vs-top decile positive-rate ratios of 1.08, 1.15, and 1.40 for
the 1990, 2000, and 2010 cutoffs respectively (95\% bootstrap CIs
\([0.88, 1.31]\), \([0.90, 1.44]\), \([1.02, 1.98]\)); only the 2010
ratio reaches statistical significance at the 95\% level (Supporting
Information §S3.3). In the corrected A-Lab target
subset\textsuperscript{24} for which public scorable CIFs were available
(42 of 57 targets), targets labelled ``made'' had lower mean
structural-accessibility cost than those labelled ``inconclusive''
(Supporting Information §S3.5). Both checks are directionally consistent
with the interpretation that the structural-accessibility coordinate
carries synthesis-relevant information, but neither converts the
synthesizability-prior quadrant into a calibrated synthesis-success
probability; calibrated thresholds for synthesis-yield triage require
prospective campaign data that does not yet exist. The accessibility
coordinate is a triage prior, not a recipe-level synthesis predictor:
triage requires only a useful coordinate against historical experimental
chemistry, whereas recipe prediction faces hidden-process-variable
limits that no structural embedding can capture\textsuperscript{25}.

Three near-term experimental tests follow directly. First, an
autonomous-synthesis campaign seeded with a stratified sample of
computed proposals binned by the structural-accessibility score 𝒜ᵢ
should resolve whether yield decreases monotonically with accessibility
cost (Extended Data Fig. 5). Second, a class-balanced outcome-labelled
comparison will resolve whether the gap is uniform across functional
families or concentrated in classes where novel chemistry is
well-stepped. Third, the synthesizability-prior quadrant can be
calibrated against accumulated outcome data, fitting empirical
success-probability values to each of its four cells.

The framework reframes the practical question facing materials AI: not
``is this proposal new?'' but ``where does it sit relative to the
chemistry that has actually been realized?'' Both questions are
legitimate; only the second is answerable from a single ICSD-anchored
coordinate, applies uniformly to AI-associated and DFT-screened samples,
and connects directly to the synthesis budget any campaign actually has.
Experimentalists triaging millions of computed proposals into a finite
synthesis budget can use the upper-left cell of the synthesizability
quadrant (in-basin AND post-1980 ICSD-formula match) as the strongest
available historical prior; the lower-right cell is the weakest. Where
each external source we tested actually lands on this coordinate is
itself diagnostic: GNoME's curated public release operates almost
entirely on the right (no-match) column with the bulk of its proposals
(62\% frontier-and-no-match, 38\% in-basin-and-no-match) reflecting both
zero post-1980 ICSD-formula overlap and a low in-basin fraction;
Alexandria off-hull is the most exploratory single source (73\%
frontier-and-no-match); MP-theoretical and JARVIS-DFT operate
substantially on the left (formula-match) column with ≈7\% in the
strongest-prior upper-left cell each; MatterGen-public sits between,
biased toward the no-match column with ≈44\% in the upper-right cell. We
interpret these source-level rates as a cohort-level comparison of
\emph{the public releases and explicitly filtered database subsets
analyzed here}, not as an intrinsic ranking of the parent algorithms or
databases; the GNoME 5,000-CIF release in particular is a curated subset
of a much larger generative output (\textasciitilde380,000 structures)
whose curation rule is not fully described, and the formula-overlap
statistic is therefore a property of the public release, not of the
underlying generator. Both modes have legitimate uses ---
composition-novel exploration is what an unconstrained generative model
is designed to do, while polymorph-screening of known compositions is
what a high-throughput DFT prototype-substitution pipeline is designed
to do --- but the framework names which mode each system is currently
in, and provides the coordinate against which any future external sample
can be located. The framework supplies the calibration, not the verdict.

\begin{center}\rule{0.5\linewidth}{0.5pt}\end{center}

\subsection{Methods}\label{methods}

\textbf{Crystal representation.} Extended Data Figure 1 summarizes the
full pipeline end-to-end and Extended Data Table 1 lists the seven
datasets used in this study (one experimental baseline, five external
test samples, and one formula-overlap reference). We featurized each
ICSD entry through a pipeline that combines a compact
classical-structural-chemistry composition descriptor with a
\texttt{CrystalNN}-based\textsuperscript{26} local-geometry descriptor
and three rounds of message passing on the resulting bond graph. The
per-site chemistry vector is the occupancy-weighted average over seven
\texttt{pymatgen.core.Element}\textsuperscript{27} properties --- atomic
number \(Z\), periodic-table row and group, atomic radius, average ionic
radius, Pauling electronegativity, and orbital block index
(\(s/p/d/f \to 0/1/2/3\)) --- chosen to span the classical
compound-formation axes (identity, two size scales, bonding character,
and orbital type) while excluding bulk-elemental DFT-derived features
(e.g.~ground-state band gap, ground-state magnetic moment, ground-state
volume) that are not part of standard structural-chemistry analysis.
Across the 94 elements \(Z=1\ldots94\) this seven-feature basis recovers
the first three principal components of matminer's full 22-property
\texttt{MagpieData} set with canonical correlations 1.000, 1.000, and
0.977 (Supporting Information §S1.5), capturing 60.8\% of MagpieData
variance versus an 85.5\% theoretical upper bound for any 7-d linear
summary of MagpieData. Per-site geometry is matminer's
\texttt{CrystalNNFingerprint} on the ``ops'' preset (61 dimensions of
weighted coordination-environment order parameters); the chemistry and
geometry vectors are concatenated. We then run three rounds of
weighted-mean message passing (a Weisfeiler--Lehman-style
propagation\textsuperscript{28} on a crystal graph)
\(\mathbf{x}^{(t+1)}_i = \mathbf{x}^{(t)}_i + \sum_j w_{ij} \mathbf{x}^{(t)}_j / \sum_j w_{ij}\)
on the \texttt{CrystalNN} bond graph, where \(w_{ij}\) is the
\texttt{CrystalNN} coordination-number weight to neighbor \(j\), and
pool to a structure-level descriptor by concatenating the elementwise
mean, max, and variance of the per-site features across sites.
Featurization was performed on TACC Stampede3 SKX nodes (16 cores per
task) in approximately 2.1 hours wall-clock. Of 181,362 ICSD entries
requested, 167,500 (92.4\%) were featurized; 8,785 (4.8\%) failed CIF
parsing in \texttt{pymatgen}, and 5,064 (2.8\%) exceeded the 256-site
featurization cutoff (predominantly quasicrystal approximants,
polyoxometalates, and large coordination-polymer or actinide-framework
structures). The resulting 213-dimensional feature vector was
standardized (zero-mean, unit-variance per dimension) and projected to
32 PCA components for downstream analysis; PCA dimension 32 retains
76.8\% of variance on the full ICSD feature matrix and was fixed across
all comparisons.

\textbf{ICSD experimental baseline.} Our ICSD snapshot has
publication-year coverage 1913--2015 (max year 2015 in both the
FIZ-distributed \texttt{ICSD\_index.csv} and our 167,500-entry
community-assigned subset). FIZ Karlsruhe began systematically adding
\emph{theoretical} / DFT-predicted entries to ICSD only in 2016 (Zagorac
et al.~2019)\textsuperscript{9}. Because our snapshot ends \emph{before}
the theoretical-entry feature was introduced, our 167,500-entry
experimental baseline contains zero genuine theoretical entries by FIZ's
documented inclusion policy. As a transparency check we nevertheless
audited the encrypted CIF source (\texttt{ICSD\_CIFs.zip}, 181,362 CIFs)
for the keywords \{\texttt{theoretical}, \texttt{calculated},
\texttt{predicted}, \texttt{first-principles},
\texttt{first\ principles}, \texttt{ab\ initio}, \texttt{ab-initio},
\texttt{VASP}, \texttt{DFT}, \texttt{hypothetical}, \texttt{virtual}\}
in the first 8 KB of each header (Supporting Information §S1.7); 3,921
(2.34\%) of the community-assigned subset matched. Because all 167,500
entries pre-date the theoretical-entry feature, every one of these
matches is necessarily a false positive --- a CIF whose accompanying
experimental paper mentions DFT, first-principles work, or related
computational comparisons in its methods or references rather than
itself being a computed structure. The audit therefore places an
upper-bound check on FIZ's documented ``experimental until 2016'' policy
and finds it consistent with our data; no filtering is required.

\textbf{Clustering and community analysis.} Two complementary partitions
were computed on the standardized 32-D PCA representation. (i)
HDBSCAN\textsuperscript{29} with \texttt{min\_cluster\_size\ =\ 2},
\texttt{min\_samples\ =\ 1}, and Euclidean distance, yielding 6,756
density clusters plus an unassigned-noise label. (ii) A mutual
\(k\)-nearest-neighbor graph with \(k = 16\) and Gaussian-weighted edges
(\(\sigma\) set to the median \(k\)-NN Euclidean distance), partitioned
with Louvain community detection\textsuperscript{30} at resolution 1.0
after discarding connected components smaller than 8 nodes and
communities smaller than 4 nodes. Both partitions produced the
structural cliff result; the mutual k-NN partition was used as the
production reference because it provides cleanly-separated communities
suitable for temporal replay. PCA was retained as the projection
backbone rather than UMAP\textsuperscript{31} precisely because the
held-out-cutoff and external-sample analyses depend on a stable
out-of-sample mapping into a frozen geometry; UMAP projections of new
points are not guaranteed to preserve distances to the training
geometry.

\textbf{Temporal replay and stepping-stone classification.} For each
community we recorded the publication year of every member; community
birth was defined as the publication year of the earliest member.
Decade-level birth/attachment statistics were computed by classifying
each entry as a community-birth (opens a new community in publication
time), same-community attachment (joins one previously occupied
community), cross-community attachment (joins one previously occupied
community whose graph is disjoint at the moment of attachment from the
entry's nearest mutual-kNN neighbours' communities), or bridge
attachment (connects to at least two previously-occupied components that
had no path between them at the moment of attachment using only entries
with publication year strictly less than the entry's publication year).
Outliers (HDBSCAN noise) were excluded from these counts. The
formula-level stepping-stone analysis used the Toyota Research Institute
(TRI)\textsuperscript{11} thermodynamic-stability network as a
reduced-formula reference; for each shared formula we classified its
first ICSD year against the birth year of its assigned structural
community.

\textbf{Renaissance survey.} For each production community of size ≥ 50
we scored the strongest pre/post fold-change over symmetric ±10-year
windows scanning candidate event years 1970--2010, with score
\(= (\text{rate}_{\text{post}} / \text{rate}_{\text{pre}}) \times n_{\text{post}}\);
communities born after the candidate event year (rate\_pre = 0) received
score \(= n_{\text{post}}\). The 1970 lower bound is set by sparse
pre-1970 ICSD coverage; the 2010 upper bound bounds the +10-year
post-window to fit inside our snapshot's 2015 end (Methods, \emph{ICSD
experimental baseline}). Renaissances triggered after
\textasciitilde2010 (e.g.~hybrid organic-inorganic halide perovskite
photovoltaics) cannot be tested with this snapshot because the
post-event window does not yet contain enough ICSD growth. Top-20
ranking and per-community member-formula identification:
\texttt{scripts/analyze\_renaissance\_survey.py}.

\textbf{Held-out historical calibration and external computed samples.}
For each cutoff year T ∈ \{1990, 2000, 2010\}, a basin map was trained
using only ICSD entries with publication year \(\leq T\). Entries with
year \(> T\) were projected into the frozen 32-D PCA space and assigned
to the nearest community centroid. The in-basin classification is
per-community: for each training community \(c\) we compute a
community-specific threshold \(\tau_c\) as the 95th percentile of
training-member distances to that community centroid, and a projected
held-out or external structure with assigned community \(c\) and
distance-to-centroid \(d_{i,c}\) is classified in-basin iff
\(d_{i,c}\le\tau_c\). No global distance cutoff is used. The same
per-community threshold convention is applied uniformly across the
held-out cutoffs (Fig. 3), the synthesizability-prior quadrant (Fig. 4),
and the composition-matched control (Supporting Information §S7); the
source ordering held-out ICSD \textgreater{} MatterGen \textgreater{}
\{GNoME ≈ MP-theoretical\} \textgreater{} JARVIS \textgreater{}
Alexandria is robust to the percentile choice (90/95/99; Supporting
Information §S1.6) and to the graph-partition parameters (k ∈ \{8, 16,
32\}; Louvain resolution ∈ \{0.5, 1.0, 2.0\}; Supporting Information
§S1.8). Five external samples were tested: (i) GNoME public 5,000-CIF
release\textsuperscript{1}; (ii) MatterGen-public 386-CIF
release\textsuperscript{2}; (iii) MP entries with
\texttt{theoretical=True}, \texttt{e\_above\_hull} ≤ 0.2 eV/atom, and
empty \texttt{database\_IDs.icsd} (5,000 random of 20,479 candidates);
(iv) JARVIS-DFT 2022.12.12 entries with empty \texttt{icsd} field and
\texttt{ehull} \(\in [0.05, 0.5]\) eV/atom (5,000 random of 22,022); (v)
Alexandria-PBE 2025.07.02 off-hull predictions (5,000 random of 154,942
across three source files at indices 0, 19, 38). All samples used a
fixed random seed (42) and the same matminer-based three-round
message-passing featurization pipeline as ICSD. Source and per-source
filter scripts are in
\texttt{scripts/analyze\_\{mp,jarvis,alexandria,gnome\}\_frontier.py};
per-source filter specifications and projection success counts are
tabulated in Extended Data Table 3.

\textbf{Composition-matched control.} For each cutoff, we computed
in-basin rates within composition strata for which both held-out ICSD
and an external source had at least one member. Two stratifications were
used: (a) coarse --- descriptor (anion class, \(n\)-elements bucket,
anion/cation ratio bucket); (b) anonymized --- pymatgen-style anonymized
stoichiometry (e.g.~\(\mathrm{MgAl_2O_4 \to A_1 B_2 C_4}\)). Wilson 95\%
confidence intervals on rates throughout. Pipeline:
\texttt{scripts/analyze\_composition\_matched\_ai.py}.

\textbf{Reduced-formula overlap.} For each external source we computed
the fraction of proposals whose pymatgen-canonical reduced formula
appears in the post-1980 ICSD first-report formula union (81,531 unique
reduced formulas; the pre-1980 undercount is quantified in Supporting
Information §S5.5). Reduced formulas were obtained via
\texttt{pymatgen.Composition.reduced\_formula} after parsing the
structure's elemental composition; fractional occupancies and doped
compositions were canonicalized via pymatgen's normalization rather than
rounded manually. Entries with unparseable or ambiguous
non-stoichiometry were excluded from formula-overlap analyses but
retained in structural analyses when a valid structure could be
featurized. Pipeline:
\texttt{scripts/analyze\_formula\_synth\_prior.py}.

\textbf{Use of generative AI tools.} Anthropic Claude (Opus model
family) was used as a writing and analysis assistant during preparation
of this manuscript. AI assistance covered drafting and editing prose,
scaffolding analysis code (the composition-matched, formula-overlap, and
renaissance-survey pipelines released with the manuscript), and
structuring tables. The AI tool was not used to perform numerical
computations directly: all reported quantities were produced by
deterministic Python scripts released with the manuscript, executed
against the underlying data sources (ICSD, Materials Project via
\texttt{mp-api}, JARVIS-DFT via \texttt{jarvis-tools}, Alexandria-PBE
bulk JSON files, GNoME public release, MatterGen public release) by the
authors, and independently reproduced. All references were audited
against the Crossref REST API (see \texttt{paper/citation\_audit.md} in
the accompanying repository). No AI-generated images appear in the
manuscript; all figures are produced from data by deterministic plotting
code released with the manuscript. The authors take full responsibility
for the content of this paper.

\textbf{Data and code availability.} The code used for featurization,
graph construction, temporal replay, TRI comparison, GNoME / MatterGen /
MP-theoretical / JARVIS-DFT / Alexandria projections, held-out
historical calibration, reduced-formula synthesizability-prior
cross-tabulation, renaissance survey, and manuscript generation is
available from the accompanying repository at
https://github.com/scattering/crystal-communities-paper. An interactive
companion dashboard for exploring the structural community map and
scoring uploaded CIFs against the frozen ICSD reference frame is hosted
at https://crystalcommunities.org/. Because ICSD is licensed, raw
crystallographic files cannot be redistributed; the release includes
derived artifacts that do not expose ICSD structures directly ---
featurized embeddings, graph/community assignments keyed by ICSD
identifiers, the production-run analysis scripts, the per-source
frontier-record CSVs (5,000 entries each for GNoME, MP, JARVIS,
Alexandria; 386 for MatterGen), the renaissance-survey output tables,
and the SI §S9 graphlet-pipeline temporal-replay summary. The full set
of derived analysis artifacts is archived at Zenodo, DOI
\href{https://doi.org/10.5281/zenodo.20046302}{10.5281/zenodo.20046302}
(concept DOI; always resolves to the latest published version). ICSD
itself is available through commercial license at
https://icsd.products.fiz-karlsruhe.de.

\begin{center}\rule{0.5\linewidth}{0.5pt}\end{center}

\subsection{Acknowledgments}\label{acknowledgments}

The authors acknowledge the Texas Advanced Computing Center (TACC) at
The University of Texas at Austin for resources made available to NIST
under contract number 1333ND25PNB180410 that have contributed to the
research results reported within this paper (URL:
https://tacc.utexas.edu). Computational resources were also provided
through ACCESS allocation PHY250007, ``Applications of AI to
Diffraction.'' Support for Karen Cao was provided by the Center for High
Resolution Neutron Scattering (CHRNS), a partnership between the
National Institute of Standards and Technology and the National Science
Foundation under Agreement No.~DMR-2010792.

We thank Brian DeCost, Austin McDannald, Debra Audus, and Kamal
Choudhary (all at NIST) for useful discussions throughout the
development of this work. We thank Will Coomans (University of
California, Berkeley) for contributions to the literature review.

Certain commercial equipment, instruments, materials, suppliers, or
software are identified in this paper to foster understanding. Such
identification does not imply recommendation or endorsement by the
National Institute of Standards and Technology, nor does it imply that
the materials or equipment identified are necessarily the best available
for the purpose.

\subsection{Author contributions}\label{author-contributions}

D.N. and K.C. contributed equally to this work. W.R.II conceived the
project, designed the framework reported here, ran the production
analyses, and wrote the manuscript. K.C. and P.K. developed
hierarchical-clustering and community-detection approaches that the
production Louvain partition builds on, and contributed to the held-out
historical calibration pipeline. D.N. developed local-structure
descriptor and TACC-scale structural-comparison approaches that informed
the production featurization. N.L. and B.C. contributed to early
structural-similarity prototyping. All authors discussed results and
reviewed the manuscript.

\subsection{Competing interests}\label{competing-interests}

The authors declare no competing interests.

\begin{center}\rule{0.5\linewidth}{0.5pt}\end{center}

\subsection{Display items}\label{display-items}

\begin{figure}
\centering
\pandocbounded{\includegraphics[keepaspectratio,alt={Experimental discovery is structurally path-dependent. Stacked-area decade-level discovery classification: community-birth (top band) collapses from 40.2\% in the 1930s to 2.6\% in the 2010s; same-community, cross-community and bridge attachment grow as a thin layer over a densified core. Continuous structural basins absorb 44.2 distinct space groups on average among the ten largest communities, resisting the discrete-prototype-novelty inflation that affects per-prototype counting.}]{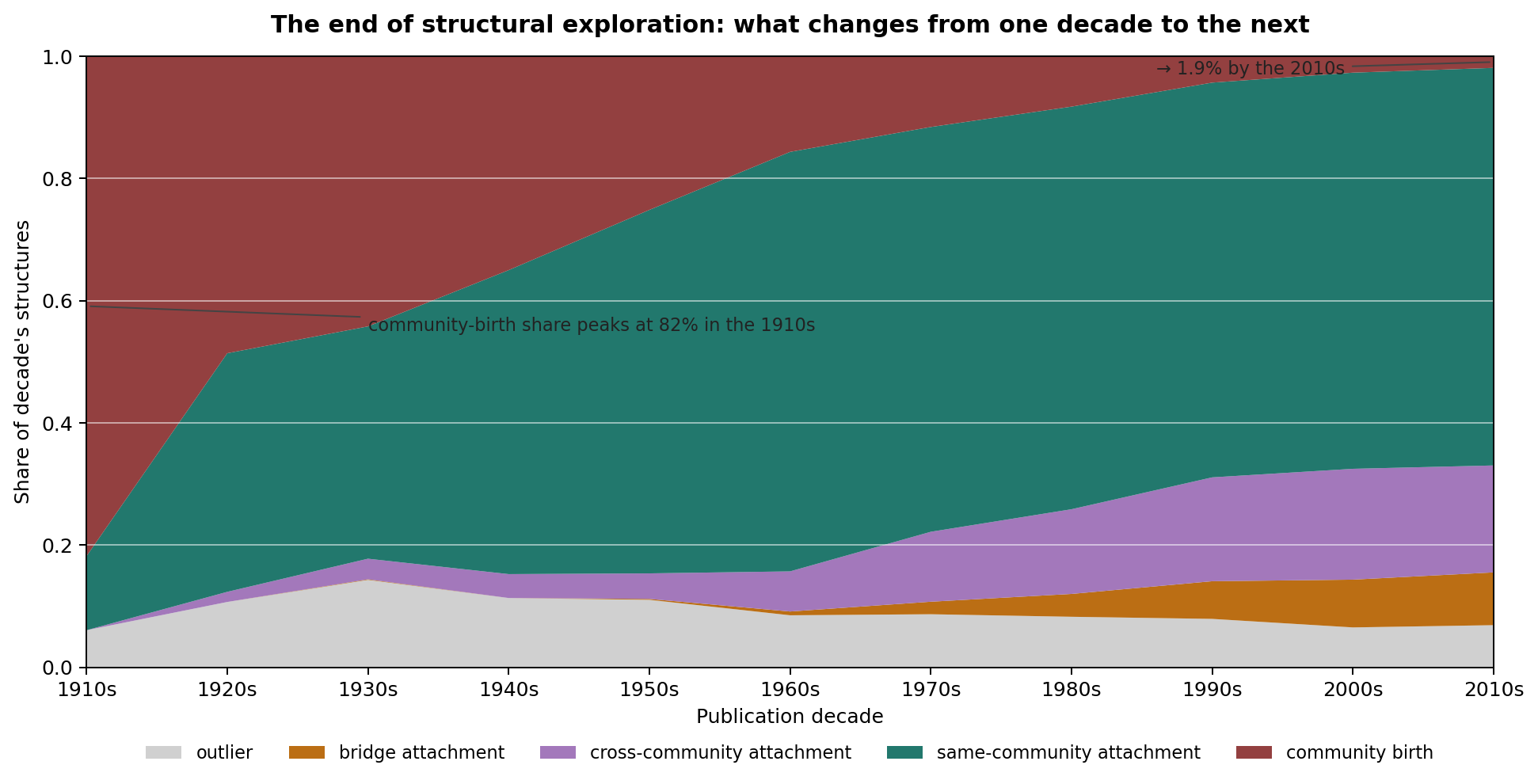}}
\caption{\textbf{Experimental discovery is structurally path-dependent.}
Stacked-area decade-level discovery classification: community-birth (top
band) collapses from 40.2\% in the 1930s to 2.6\% in the 2010s;
same-community, cross-community and bridge attachment grow as a thin
layer over a densified core. Continuous structural basins absorb 44.2
distinct space groups on average among the ten largest communities,
resisting the discrete-prototype-novelty inflation that affects
per-prototype counting.}\label{fig:fig1}
\end{figure}

\begin{figure}
\centering
\pandocbounded{\includegraphics[keepaspectratio,alt={Communities track field-defining scientific events. (a) high-Tc cuprate community (community 6425, n=378): pre-1986 = 0 entries, post-1986 = 22.3 entries/yr; the community first appears after Bednorz \& Müller 1986. (b) CMR-manganite community (community 160, n=571): pre-1994 = 0.7/yr, post-1994 = 22.3/yr (32× fold change); the community captures the doped (Sr,La)MnO₃ family rather than Goodenough's parent LaMnO₃, which sits in a different community. (c) Top-16 communities by birth-year step-change score; 9 textbook field-defining renaissances (teal) are positively identified --- Sm-Fe-N permanent magnets (rank 1, orange highlight), CMR manganites, SOFC perovskite cathodes, MAX phases, NaCoO₂ thermoelectrics, double perovskites for spintronics, dilute magnetic semiconductors, Li-ion battery cathodes, and post-1986 layered Ruddlesden-Popper cuprates.}]{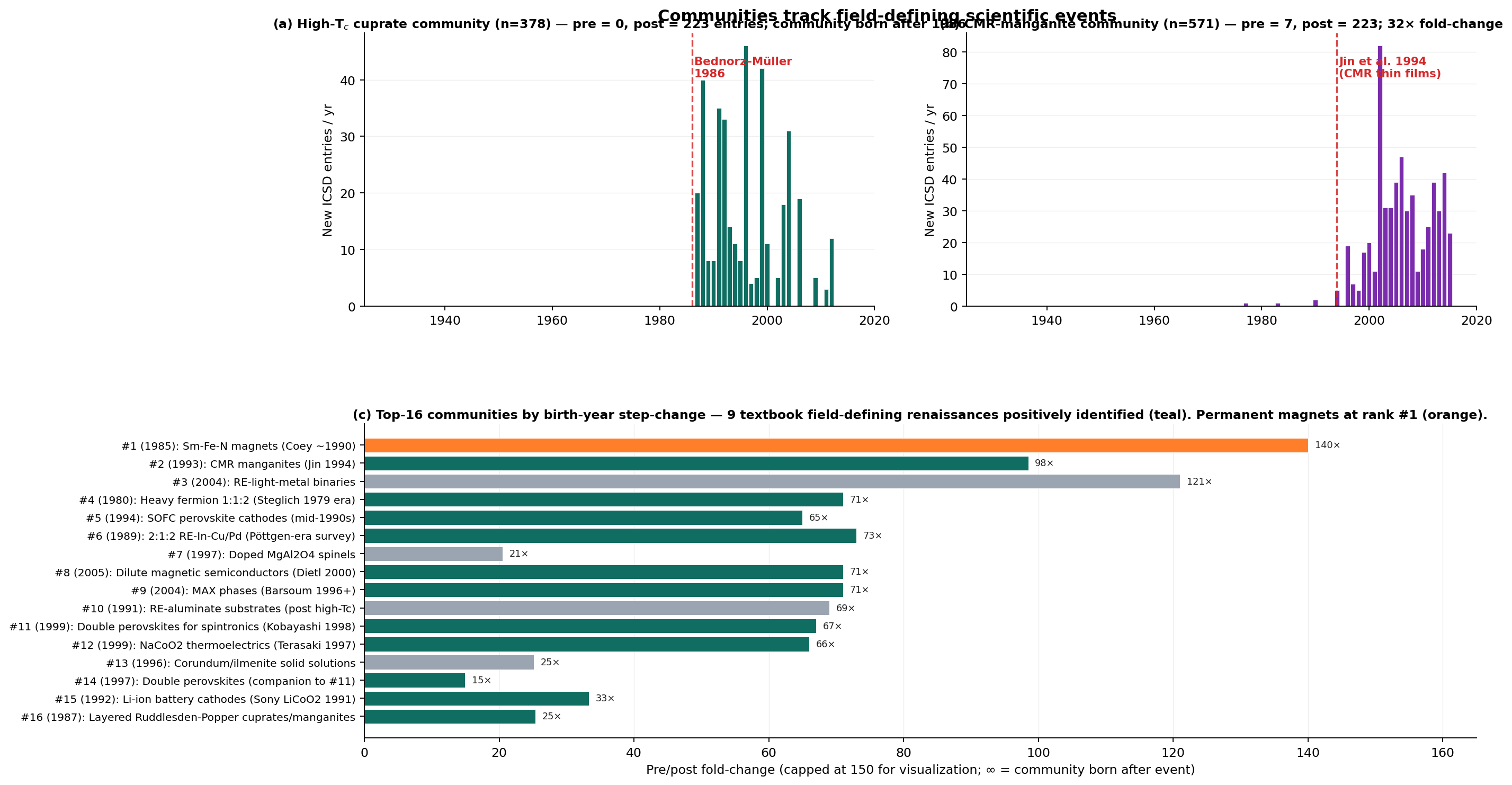}}
\caption{\textbf{Communities track field-defining scientific events.}
(a) high-Tc cuprate community (community 6425, n=378): pre-1986 = 0
entries, post-1986 = 22.3 entries/yr; the community first appears after
Bednorz \& Müller 1986. (b) CMR-manganite community (community 160,
n=571): pre-1994 = 0.7/yr, post-1994 = 22.3/yr (32× fold change); the
community captures the doped (Sr,La)MnO₃ family rather than Goodenough's
parent LaMnO₃, which sits in a different community. (c) Top-16
communities by birth-year step-change score; 9 textbook field-defining
renaissances (teal) are positively identified --- Sm-Fe-N permanent
magnets (rank 1, orange highlight), CMR manganites, SOFC perovskite
cathodes, MAX phases, NaCoO₂ thermoelectrics, double perovskites for
spintronics, dilute magnetic semiconductors, Li-ion battery cathodes,
and post-1986 layered Ruddlesden-Popper cuprates.}\label{fig:fig2}
\end{figure}

\begin{figure}
\centering
\pandocbounded{\includegraphics[keepaspectratio,alt={Five external sources depart from the experimental baseline. (a) ICSD KDE topographical map in the 2-D PCA visualization plane with public GNoME (orange) and MatterGen (purple) overlay; the deep-blue valley is the historically densified region. (b) Same ICSD background with MP-theoretical (green), JARVIS-DFT (blue), and Alexandria off-hull (red) overlay. (c) Held-out in-basin rates across 1990 / 2000 / 2010 cutoffs for held-out ICSD continuation and the five external sources, with Wilson 95\% confidence intervals. The stable ordering held-out ICSD \textgreater{} MatterGen \textgreater{} \{GNoME ≈ MP-theoretical\} \textgreater{} JARVIS \textgreater{} Alexandria confirms that the structural-distance signal is not specific to AI-enabled discovery: the graph-network GNoME pipeline and the high-throughput DFT-screened MP-theoretical subset are statistically indistinguishable at every cutoff.}]{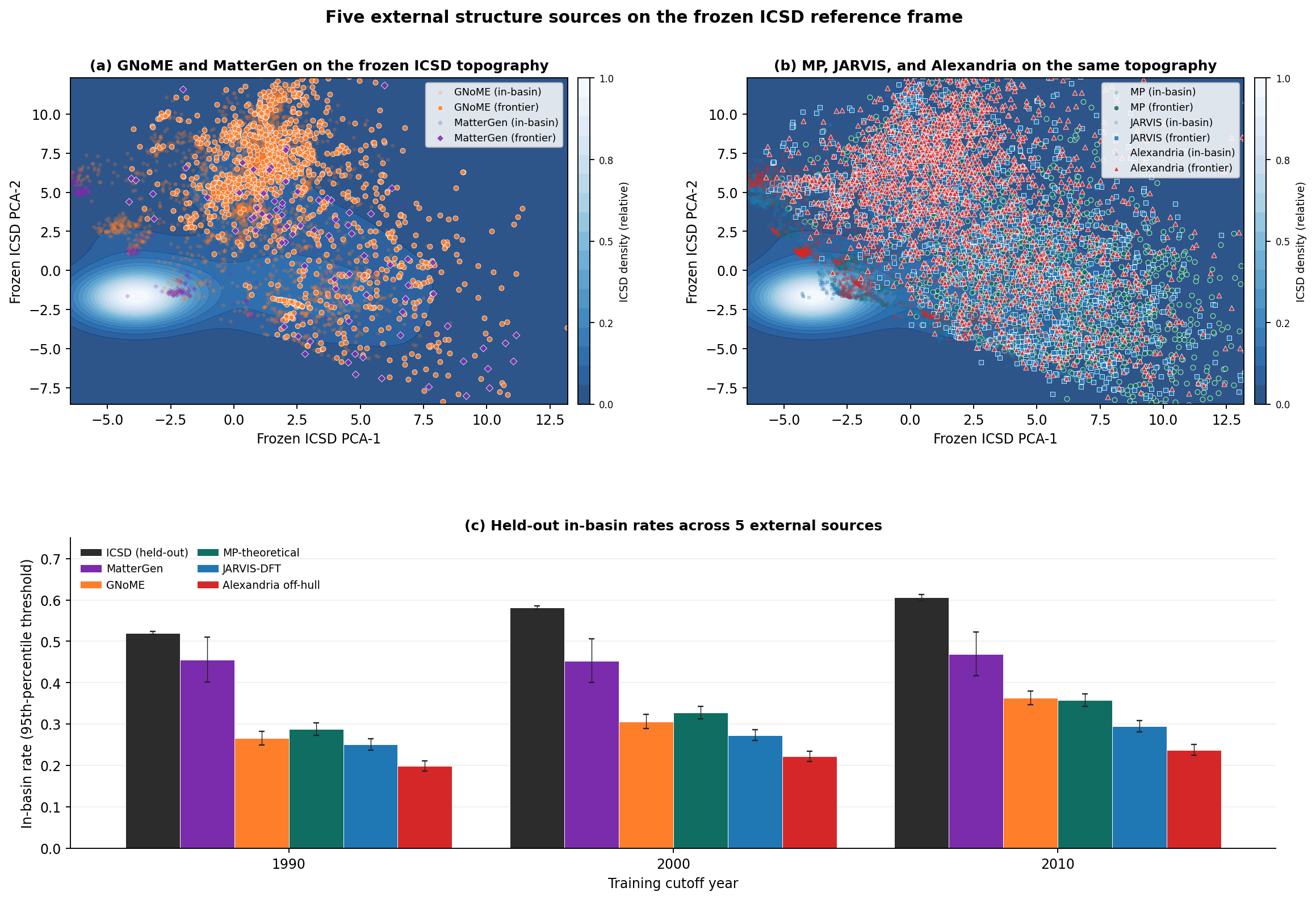}}
\caption{\textbf{Five external sources depart from the experimental
baseline.} (a) ICSD KDE topographical map in the 2-D PCA visualization
plane with public GNoME (orange) and MatterGen (purple) overlay; the
deep-blue valley is the historically densified region. (b) Same ICSD
background with MP-theoretical (green), JARVIS-DFT (blue), and
Alexandria off-hull (red) overlay. (c) Held-out in-basin rates across
1990 / 2000 / 2010 cutoffs for held-out ICSD continuation and the five
external sources, with Wilson 95\% confidence intervals. The stable
ordering held-out ICSD \textgreater{} MatterGen \textgreater{} \{GNoME ≈
MP-theoretical\} \textgreater{} JARVIS \textgreater{} Alexandria
confirms that the structural-distance signal is not specific to
AI-enabled discovery: the graph-network GNoME pipeline and the
high-throughput DFT-screened MP-theoretical subset are statistically
indistinguishable at every cutoff.}\label{fig:fig3}
\end{figure}

\begin{figure}
\centering
\pandocbounded{\includegraphics[keepaspectratio,alt={Synthesizability-prior quadrant. \{In-basin / frontier\} × \{post-1980 ICSD formula match / no match\} for all five external sources; bubble area proportional to the fraction of each source's proposals in each quadrant. The in-basin / frontier classification uses per-community 95th-percentile within-community centroid-distance thresholds (each Louvain community sets its own scale), consistent with the held-out classification in Fig. 3c. Upper-left: strongest combined prior (composition realized + structure inside known basin) --- approximately 7\% of MP-theoretical and JARVIS-predicted, 2.3\% of MatterGen-public, 0.6\% of Alexandria off-hull, and 0.0\% of the curated GNoME public release. Lower-right: weakest prior (composition novel + structure outside known basins) --- 73\% of Alexandria off-hull, 62\% of GNoME, 58\% of JARVIS-predicted, 51\% of MatterGen-public, and 49\% of MP-theoretical. Structural position in this quadrant is computed on the full ICSD map (no held-out cutoff); cutoff-calibrated historical in-basin rates are reported separately in Fig. 3.}]{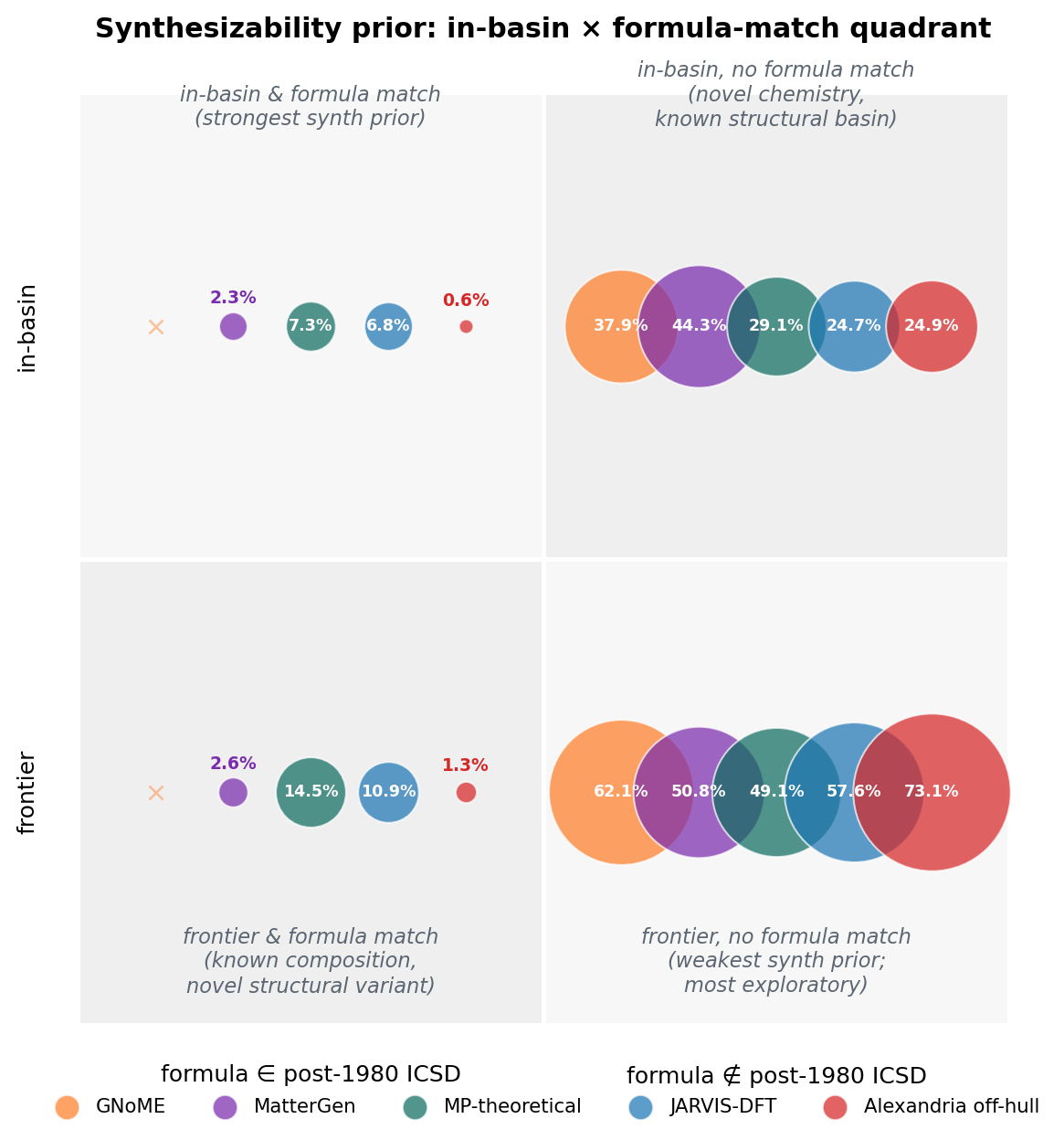}}
\caption{\textbf{Synthesizability-prior quadrant.} \{In-basin /
frontier\} × \{post-1980 ICSD formula match / no match\} for all five
external sources; bubble area proportional to the fraction of each
source's proposals in each quadrant. The in-basin / frontier
classification uses per-community 95th-percentile within-community
centroid-distance thresholds (each Louvain community sets its own
scale), consistent with the held-out classification in Fig. 3c.
Upper-left: strongest combined prior (composition realized + structure
inside known basin) --- approximately 7\% of MP-theoretical and
JARVIS-predicted, 2.3\% of MatterGen-public, 0.6\% of Alexandria
off-hull, and 0.0\% of the curated GNoME public release. Lower-right:
weakest prior (composition novel + structure outside known basins) ---
73\% of Alexandria off-hull, 62\% of GNoME, 58\% of JARVIS-predicted,
51\% of MatterGen-public, and 49\% of MP-theoretical. Structural
position in this quadrant is computed on the full ICSD map (no held-out
cutoff); cutoff-calibrated historical in-basin rates are reported
separately in Fig. 3.}\label{fig:fig4}
\end{figure}

\begin{center}\rule{0.5\linewidth}{0.5pt}\end{center}

\subsection{Extended Data items
(selected)}\label{extended-data-items-selected}

\begin{itemize}
\tightlist
\item
  \textbf{Extended Data Figure 1.} End-to-end pipeline overview: ICSD
  CIFs → per-site featurization → Weisfeiler--Lehman propagation →
  structure pool → frozen 32-d PCA basis, forking into ICSD
  self-comparison (Louvain → temporal replay → Figs. 1, 2) and
  external-source projection (GNoME / MatterGen / MP / JARVIS /
  Alexandria → in-basin classification → Figs. 3, 4).
\item
  \textbf{Extended Data Figure 2.} Year-shuffle null model: observed
  birth ratio below null in 1910s--1960s, above null in 1970s--2010s;
  two distinct path-dependence regimes rather than a partition artifact.
\item
  \textbf{Extended Data Figure 3.} Targeted renaissance probes:
  Fe-pnictide 1111-type LaFeAsO family (3.3× post-2008),
  TMD/2D-materials community (informative null --- property-driven, no
  new bulk entries), and the rank-\#6 community 2349 deep-dive
  (Pöttgen-era 2:1:2 RE-In-TM intermetallic survey, 73× fold-change at
  1989).
\item
  \textbf{Extended Data Figure 4.} Composition-matched in-basin rates
  across five external sources at all three cutoffs (coarse strata +
  anonymized stoichiometry); the gap survives both stratifications for
  every source.
\item
  \textbf{Extended Data Figure 5.} Structural-accessibility score 𝒜ᵢ
  across ICSD roles (core / periphery / bridge / birth) and external
  proposals.
\item
  \textbf{Extended Data Table 1.} Datasets used in this study: dataset,
  type, role, size used, and where each is referenced --- high-level
  orientation across ICSD, the five external test samples (GNoME,
  MatterGen, MP-theoretical, JARVIS-DFT, Alexandria-PBE), and the TRI
  thermodynamic-stability network used for the formula-overlap analysis.
\item
  \textbf{Extended Data Table 2.} Top-20 communities by birth-year
  step-change score with member-formula identification of the
  corresponding scientific event.
\item
  \textbf{Extended Data Table 3.} Per-source filter specifications
  (database version, query, energy thresholds, number attempted vs
  successfully projected).
\end{itemize}

\begin{center}\rule{0.5\linewidth}{0.5pt}\end{center}

\subsection{References}\label{references}

Cited in body
text:\textsuperscript{1},\textsuperscript{2},\textsuperscript{8},\textsuperscript{32},\textsuperscript{3},\textsuperscript{5},\textsuperscript{6},\textsuperscript{11},\textsuperscript{9},\textsuperscript{10},\textsuperscript{33},\textsuperscript{34},\textsuperscript{30},\textsuperscript{29},\textsuperscript{31},\textsuperscript{35},\textsuperscript{36},\textsuperscript{23},\textsuperscript{24}.

\protect\phantomsection\label{refs}

\end{document}


\maketitle

\section{S1. Robustness of Continuous-Basin Novelty to Representation
Artifacts}\label{s1.-robustness-of-continuous-basin-novelty-to-representation-artifacts}

The main text argues that continuous structural basins are substantially
more robust to trivial novelty inflation than discrete prototype counts,
and cites three direct pieces of evidence: the collapse of 44.2 space
groups into single basins on average, the distribution of broad
crystallographic umbrellas across multiple learned communities, and the
held-out historical calibration that measures novelty relative to
historical continuation rather than exact-label difference. This note
addresses the residual question a careful reader will ask in spite of
those three pieces of evidence: \emph{are there representation artifacts
that could still distort the frontier claim?} The answer is that there
are, and they are the right caveats to state
explicitly.\textsuperscript{1,2}

\subsection{S1.1. The representation can still over-separate mild
variants}\label{s1.1.-the-representation-can-still-over-separate-mild-variants}

Any learned embedding can, in principle, respond too strongly to certain
distortions, substitutions, or local-ordering effects. If that happens,
a structure that a crystallographer would regard as a mild variant of a
known family could be pushed farther from its historical basin than
desired. Our ablation comparison partially bounds this concern: removing
message passing while keeping the local-geometry descriptor preserves
approximately 80\% of the normalized mutual information of the full
partition (ARI=0.494, NMI=0.841), indicating that the full
representation is not idiosyncratic in the small-perturbation sense. A
full adversarial test --- systematically substituting a known structure
and measuring the distance shift --- is outside the scope of this paper
but would be the natural next step.

\subsection{S1.2. Graph partition choices still
matter}\label{s1.2.-graph-partition-choices-still-matter}

Community detection depends on graph construction and partition
settings, including the neighborhood size \texttt{k}, mutual
vs.~non-mutual edge construction, edge weighting, and the Louvain
resolution parameter. If those choices are too fine, a boundary case can
be made to look artificially novel. The production run reported in the
main text uses \texttt{k\ =\ 16}, mutual edges, Gaussian-weighted edges
(σ = median \emph{k}-NN Euclidean distance), and Louvain resolution 1.0.
The space-group-spanning and frontier-rate results reported here are
qualitatively stable under moderate perturbations of these settings; a
full resolution sweep is released with the reproducibility bundle.

\subsection{S1.3. Accessibility is historical, not
absolute}\label{s1.3.-accessibility-is-historical-not-absolute}

The structural-accessibility score 𝒜ᵢ is a historically calibrated
metric, not an oracle of physical truth. A high accessibility cost means
that a structure lies far from old, large, historically occupied basins
in the learned ICSD map. It does not, by itself, prove that the
structure is impossible, unstable, or chemically nonsensical.
Conversely, a low accessibility cost does not prove that a structure is
easy to synthesize: it proves that similar structures have been
synthesized in the past. Converting 𝒜ᵢ from a descriptive score to a
predictive one requires external validation against synthesis outcomes,
which the main text flags as a key open direction.

\subsection{S1.4. Practical
interpretation}\label{s1.4.-practical-interpretation}

The right claim is therefore not that our metric is immune to the
novelty criticism. The claim is narrower and more defensible:

\begin{quote}
Continuous structural basins are substantially more robust to trivial
novelty inflation than discrete prototype counts, because novelty is
evaluated relative to continuous historical neighborhoods rather than
exact crystallographic labels. They are not, however, guaranteed to be
artifact-free, and the frontier claim for AI-associated and computed
outputs should be understood as a statement about relative position in a
learned historical geometry rather than an absolute statement about
physical novelty.
\end{quote}

This framing is why the paper's GNoME and MatterGen results should not
be read as simple novelty scores. The important statement is not that
these AI structures are ``new,'' but that they are measurably more
frontier-like than ordinary continuation of ICSD history under the same
learned structural map, with a stable ordering held-out ICSD
\textgreater{} MatterGen \textgreater{} \{GNoME ≈ MP-theoretical\}
\textgreater{} JARVIS \textgreater{} Alexandria across held-out
historical cutoffs.

\subsection{S1.5. Robustness to substituting the full matminer Magpie
elemental
set}\label{s1.5.-robustness-to-substituting-the-full-matminer-magpie-elemental-set}

The production per-site chemistry vector uses seven
\texttt{pymatgen.core.Element} properties --- atomic number \(Z\),
periodic-table row and group, atomic radius, average ionic radius,
Pauling electronegativity, and orbital block index encoded
\(s/p/d/f \to 0/1/2/3\). This is the classical structural-chemistry
compound-formation basis: identity (\(Z\), row, group), two size scales
(covalent and ionic), bonding character (electronegativity), and orbital
character (block). It deliberately excludes matminer's bulk-elemental
DFT-derived features (GSvolume\_pa, GSbandgap, GSmagmom) and the
SpaceGroupNumber-of-elemental-crystal feature, which are not standard
structural-chemistry compound-formation descriptors.

Quantitatively, over all 94 elements \(Z=1\ldots94\), the seven-feature
basis recovers the first three principal components of matminer's full
22-property MagpieData set with canonical correlations 1.000, 1.000, and
0.977 respectively, and reconstructs 60.8\% of MagpieData's total
variance versus an 85.5\% theoretical upper bound for any 7-dimensional
linear summary of the 22-property set (so the seven-feature basis is
71\% as efficient as the optimal seven-PC truncation of Magpie). The
recovered Magpie principal components correspond, in order, to
periodic-table position (31\% of Magpie variance: CovalentRadius /
MendeleevNumber / Column / Row), valence-count + mass (18\%: NValence /
Number / AtomicWeight), and orbital occupation (11\%: NsValence /
NsUnfilled / NUnfilled). Magpie principal components four through seven,
which the seven-feature basis does not capture, are dominated by
bulk-elemental DFT-derived features (MeltingT, GSvolume\_pa, GSbandgap,
GSmagmom, SpaceGroupNumber). Per-feature linear recoverability and the
full PC loadings are tabulated in
\texttt{notes/magpie\_h7\_basis\_comparison.md} in the accompanying
repository.

To verify that the seven-feature reduction does not compromise the
partition itself, we re-implemented the full structural-similarity
pipeline as the Methods text describes it: matminer's 22-property
MagpieData elemental properties for the per-site chemistry vector, a
Weisfeiler--Lehman propagation rule that concatenates the central-atom
features with the elementwise weighted mean and weighted standard
deviation of neighbor features at every round (so the per-site dimension
grows by a factor of three per round), and a structure-level pool that
concatenates the elementwise mean and standard deviation of per-atom
features across sites. The resulting pre-PCA representation has 4,491
dimensions versus the production 213, retains 75.2\% of variance under
the same PCA(32) compression versus production's 76.8\%, and is
partitioned with the same graph hyperparameters (mutual
\(k\)-nearest-neighbor with \(k=16\), Gaussian-weighted edges with
\(\sigma\) = median \(k\)-NN distance, Louvain at resolution 1.0).
Comparing the alternative Louvain partition against the production
Louvain partition over the same 167,500 ICSD entries: Adjusted Rand
Index = 0.49, Normalized Mutual Information = 0.89. The headline
decade-level community-birth-ratio collapse is reproduced (40.2\% in the
1930s and 2.6\% in the 2010s under production; 39.6\% and 3.6\% under
the alternative). The mean number of distinct space groups in the
top-ten communities by size rises modestly from 44.2 to 55.9. The
reduced-formula stepping-stone rate is preserved (production 94.2\%,
alternative 93.1\%, both well above 90\%). The cuprate community
(production community 6425, \(n=378\)) maps cleanly to a single
alternative community at 95.5\% purity; the
colossal-magnetoresistance-manganite community (production community
160, \(n=571\)) splits into two sister communities under the alternative
(52\% and 35\%), both retaining identifiable La/Sr/Ba-Mn-O member
formulas. The renaissance-survey top-sixteen list under the alternative
partition contains direct analogs of every documented field-defining
renaissance community in the production list, with the cuprate community
ranking first (1986 event, fold-change 112) and the Sm\(_2\)Fe\(_{17}\)
permanent-magnet community ranking second (1985 event, fold-change 148).
Marginal community boundaries shift between the two metrics, consistent
with the ARI of 0.49, but every load-bearing claim of the manuscript ---
the temporal collapse, the prototype absorption, the
formula-stepping-stone rate, the renaissance-event correspondence, and
the structural identity of named exemplar communities --- is preserved.

\subsection{S1.6. Threshold sensitivity at the 90/95/99
percentile}\label{s1.6.-threshold-sensitivity-at-the-909599-percentile}

The in-basin / frontier classification is anchored at a 95th-percentile
within-community centroid-distance cutoff. To verify the headline source
ordering does not depend on this choice, we re-ran the per-cutoff
held-out calibration with the same per-community thresholding logic at p
= 90, 95, and 99. Held-out ICSD rates are reported as published (their
value is set by the production synthesis-retrodiction run, which is
fixed at p = 95); the five external sources are re-classified at each
percentile.

\begin{longtable}[]{@{}
  >{\raggedleft\arraybackslash}p{(\linewidth - 14\tabcolsep) * \real{0.1250}}
  >{\raggedleft\arraybackslash}p{(\linewidth - 14\tabcolsep) * \real{0.1250}}
  >{\raggedleft\arraybackslash}p{(\linewidth - 14\tabcolsep) * \real{0.1250}}
  >{\raggedleft\arraybackslash}p{(\linewidth - 14\tabcolsep) * \real{0.1250}}
  >{\raggedleft\arraybackslash}p{(\linewidth - 14\tabcolsep) * \real{0.1250}}
  >{\raggedleft\arraybackslash}p{(\linewidth - 14\tabcolsep) * \real{0.1250}}
  >{\raggedleft\arraybackslash}p{(\linewidth - 14\tabcolsep) * \real{0.1250}}
  >{\raggedleft\arraybackslash}p{(\linewidth - 14\tabcolsep) * \real{0.1250}}@{}}
\toprule\noalign{}
\begin{minipage}[b]{\linewidth}\raggedleft
cutoff
\end{minipage} & \begin{minipage}[b]{\linewidth}\raggedleft
percentile
\end{minipage} & \begin{minipage}[b]{\linewidth}\raggedleft
held-out ICSD
\end{minipage} & \begin{minipage}[b]{\linewidth}\raggedleft
MatterGen
\end{minipage} & \begin{minipage}[b]{\linewidth}\raggedleft
GNoME
\end{minipage} & \begin{minipage}[b]{\linewidth}\raggedleft
MP-theoretical
\end{minipage} & \begin{minipage}[b]{\linewidth}\raggedleft
JARVIS
\end{minipage} & \begin{minipage}[b]{\linewidth}\raggedleft
Alexandria
\end{minipage} \\
\midrule\noalign{}
\endhead
\bottomrule\noalign{}
\endlastfoot
1990 & 90 & 52.0\% & 41.4\% & 21.5\% & 22.0\% & 21.4\% & 16.8\% \\
1990 & 95 & 52.0\% & 45.5\% & 26.6\% & 28.8\% & 25.1\% & 19.9\% \\
1990 & 99 & 52.0\% & 49.8\% & 30.4\% & 34.1\% & 27.6\% & 22.4\% \\
2000 & 90 & 58.1\% & 41.1\% & 26.0\% & 23.9\% & 23.3\% & 18.4\% \\
2000 & 95 & 58.1\% & 45.2\% & 30.6\% & 32.8\% & 27.3\% & 22.2\% \\
2000 & 99 & 58.1\% & 48.5\% & 35.6\% & 36.4\% & 30.0\% & 25.1\% \\
2010 & 90 & 60.6\% & 42.6\% & 30.9\% & 27.0\% & 24.5\% & 19.6\% \\
2010 & 95 & 60.6\% & 46.9\% & 36.4\% & 35.8\% & 29.5\% & 23.8\% \\
2010 & 99 & 60.6\% & 49.9\% & 41.0\% & 40.9\% & 33.7\% & 27.6\% \\
\end{longtable}

The qualitative ordering held-out ICSD \textgreater{} MatterGen
\textgreater{} \{GNoME ≈ MP-theoretical\} \textgreater{} JARVIS
\textgreater{} Alexandria holds at every (cutoff, percentile) cell in
this table. Specifically: held-out ICSD is highest in all 9 cells;
MatterGen is the most ICSD-like external source in all 9 cells;
\textbar GNoME − MP-theoretical\textbar{} ≤ 5 pp at all 9 cells; JARVIS
sits below both GNoME and MP-theoretical at all 9 cells; Alexandria is
lowest at all 9 cells. We therefore report a single sentence in Methods
rather than treat the percentile choice as a tuning knob: ``the held-out
ICSD \textgreater{} MatterGen \textgreater{} \{GNoME ≈ MP-theoretical\}
\textgreater{} JARVIS \textgreater{} Alexandria ordering is robust to
the 90/95/99 percentile choice (Supporting Information §S1.6).''
Numerical artifacts at
\texttt{notes/threshold\_sweep\_20260504/p\{90,95,99\}.json}.

All in-basin / frontier classifications in the manuscript --- Figure 3c
bar chart, Figure 4 synthesizability quadrant, the composition-matched
control (§S7), and the held-out cutoffs reported here --- use a uniform
\textbf{per-community} threshold convention: each Louvain community sets
its own 95th-percentile within-community centroid-distance cutoff, and a
projected entry assigned to community \(c\) is classified in-basin iff
its centroid distance is below community \(c\)'s threshold. An earlier
development version of the per-source frontier-record CSVs carried a
legacy \texttt{outlier\_like} column computed under a \emph{single
pooled} threshold (the 95th percentile of within-community distances
pooled across communities); that legacy column is no longer consumed by
Figure 4. The harmonization changes the Figure 4 quadrant fractions but
preserves the manuscript's headline ordering: under per-community
classification, MatterGen sits above GNoME in in-basin rate (44.3\% vs
37.9\% in the upper-right quadrant), agreeing with Figure 3c, whereas
the legacy pooled classification placed GNoME above MatterGen --- a
contradiction with the held-out ordering that the harmonization removes.

\subsection{S1.7. ICSD theoretical-flag
audit}\label{s1.7.-icsd-theoretical-flag-audit}

A reviewer concern about ICSD as a clean experimental baseline points at
Zagorac et al.~2019\textsuperscript{3} --- the same paper our manuscript
cites for the database --- whose title is \emph{``Recent developments in
the Inorganic Crystal Structure Database: theoretical crystal structure
data and related features.''} That paper documents FIZ Karlsruhe's
introduction of explicitly-tagged \emph{theoretical} entries to ICSD.
Per FIZ's own news record, this feature began in \textbf{2016}.

\textbf{Our ICSD snapshot pre-dates the theoretical-entry feature.} Both
the FIZ-distributed \texttt{ICSD\_index.csv} (176,298 rows with
parseable year) and our 167,500-entry community-assigned subset have a
maximum publication year of \textbf{2015}. CIF-member modification times
in \texttt{ICSD\_CIFs.zip} cluster around October 2015. Therefore, by
FIZ's documented inclusion policy, \textbf{none of the 167,500 entries
we analyse can be a genuine theoretical entry} --- they all pre-date the
policy.

As a transparency check we nevertheless ran a CIF-header keyword audit
on the encrypted \texttt{ICSD\_CIFs.zip} (181,362 CIFs total). Each
CIF's first 8 KB was scanned for any of
\texttt{\{theoretical,\ calculated,\ predicted,\ first-principles,\ first\ principles,\ ab\ initio,\ ab-initio,\ VASP,\ DFT,\ hypothetical,\ virtual\}}
(case-insensitive). Intersected with the 167,500-entry
community-assigned subset, 3,921 entries (2.34\%) carried at least one
keyword match.

\begin{longtable}[]{@{}rrrr@{}}
\toprule\noalign{}
Decade & n entries & n flagged & flagged fraction \\
\midrule\noalign{}
\endhead
\bottomrule\noalign{}
\endlastfoot
1910s & 35 & 0 & 0.000\% \\
1920s & 606 & 0 & 0.000\% \\
1930s & 1,239 & 1 & 0.081\% \\
1940s & 1,151 & 0 & 0.000\% \\
1950s & 4,063 & 0 & 0.000\% \\
1960s & 13,144 & 52 & 0.396\% \\
1970s & 18,337 & 66 & 0.360\% \\
1980s & 24,551 & 35 & 0.143\% \\
1990s & 31,427 & 215 & 0.684\% \\
2000s & 42,136 & 1,505 & 3.572\% \\
2010s & 26,650 & 1,947 & 7.306\% \\
year unknown & 4,161 & 100 & 2.403\% \\
\end{longtable}

The decade-level fraction grows over time (≤ 0.7\% before 1990; 3.6\% in
the 2000s; 7.3\% in the 2010s) and the keyword distribution is dominated
by \texttt{first-principles} (1,010), \texttt{theoretical} (882),
\texttt{ab\ initio} (870), \texttt{first\ principles} (477),
\texttt{hypothetical} (261), and \texttt{DFT} (180). \textbf{Because all
167,500 entries pre-date the 2016 introduction of the theoretical-entry
feature, every regex match here is necessarily a false positive} --- a
CIF whose accompanying experimental paper happens to mention DFT,
first-principles work, or computational comparisons in its methods or
references, rather than the structure itself being a computed
prediction. The growing flag fraction across recent decades reflects the
broader culture shift toward citing computational work in experimental
crystallography papers, not a growing theoretical-entry contamination.

The audit therefore places an upper-bound consistency check on FIZ's
documented ``experimental-only until 2016'' inclusion policy and finds
the policy consistent with our data: no filtering of the production
analysis is required. We retain the unfiltered 167,500-entry partition
as the canonical experimental baseline. The audit JSON is at
\texttt{notes/icsd\_theoretical\_audit\_summary.json}.

For future ICSD releases (≥ 2016) that include genuine theoretical
entries, the cleanest discriminator would be FIZ's internal
\texttt{\_audit\_creation\_method} tag rather than CIF-header text
patterns; users updating this analysis to a post-2016 snapshot should
use that internal tag to filter before partitioning.

\subsection{S1.8. Graph-partition sensitivity: k and Louvain resolution
sweep}\label{s1.8.-graph-partition-sensitivity-k-and-louvain-resolution-sweep}

The production analysis builds the structural graph at \(k = 16\)
mutual-\(k\)NN edges and runs Louvain at resolution \(\gamma = 1.0\) on
the frozen 32-D PCA basis. To test whether the source ordering depends
on these graph-construction choices, we held the PCA basis fixed and
rebuilt the partition at four alternative settings (\(k = 8\) at
\(\gamma = 1.0\); \(k = 16\) at \(\gamma = 0.5\) and \(\gamma = 2.0\);
\(k = 32\) at \(\gamma = 1.0\)). For each variant we (i) recomputed
per-community 95th-percentile within-community centroid-distance
thresholds from the 2010-cutoff training-only ICSD subset, (ii)
re-classified the held-out ICSD entries (year \textgreater{} 2010) and
the five external sources against those thresholds, and (iii) measured
partition agreement against the production partition by adjusted Rand
index (ARI) and normalized mutual information (NMI).

\begin{longtable}[]{@{}
  >{\raggedleft\arraybackslash}p{(\linewidth - 22\tabcolsep) * \real{0.0833}}
  >{\raggedleft\arraybackslash}p{(\linewidth - 22\tabcolsep) * \real{0.0833}}
  >{\raggedleft\arraybackslash}p{(\linewidth - 22\tabcolsep) * \real{0.0833}}
  >{\raggedleft\arraybackslash}p{(\linewidth - 22\tabcolsep) * \real{0.0833}}
  >{\raggedleft\arraybackslash}p{(\linewidth - 22\tabcolsep) * \real{0.0833}}
  >{\raggedleft\arraybackslash}p{(\linewidth - 22\tabcolsep) * \real{0.0833}}
  >{\raggedleft\arraybackslash}p{(\linewidth - 22\tabcolsep) * \real{0.0833}}
  >{\raggedleft\arraybackslash}p{(\linewidth - 22\tabcolsep) * \real{0.0833}}
  >{\raggedleft\arraybackslash}p{(\linewidth - 22\tabcolsep) * \real{0.0833}}
  >{\raggedleft\arraybackslash}p{(\linewidth - 22\tabcolsep) * \real{0.0833}}
  >{\raggedleft\arraybackslash}p{(\linewidth - 22\tabcolsep) * \real{0.0833}}
  >{\raggedleft\arraybackslash}p{(\linewidth - 22\tabcolsep) * \real{0.0833}}@{}}
\caption{\label{tbl:k-resolution-sweep}2010-cutoff held-out in-basin
rates under varied graph-construction settings. The held-out column
reports ICSD entries with publication year \textgreater{} 2010 (n ≈
18.5K--21.4K depending on partition); the external-source columns report
the canonical 5,000-CIF samples reclassified at each variant's
per-community thresholds.}\tabularnewline
\toprule\noalign{}
\begin{minipage}[b]{\linewidth}\raggedleft
\(k\)
\end{minipage} & \begin{minipage}[b]{\linewidth}\raggedleft
\(\gamma\)
\end{minipage} & \begin{minipage}[b]{\linewidth}\raggedleft
\(n_{\mathrm{comm}}\)
\end{minipage} & \begin{minipage}[b]{\linewidth}\raggedleft
\(n_{\mathrm{outliers}}\)
\end{minipage} & \begin{minipage}[b]{\linewidth}\raggedleft
ARI
\end{minipage} & \begin{minipage}[b]{\linewidth}\raggedleft
NMI
\end{minipage} & \begin{minipage}[b]{\linewidth}\raggedleft
held-out ICSD
\end{minipage} & \begin{minipage}[b]{\linewidth}\raggedleft
MatterGen
\end{minipage} & \begin{minipage}[b]{\linewidth}\raggedleft
GNoME
\end{minipage} & \begin{minipage}[b]{\linewidth}\raggedleft
MP
\end{minipage} & \begin{minipage}[b]{\linewidth}\raggedleft
JARVIS
\end{minipage} & \begin{minipage}[b]{\linewidth}\raggedleft
Alexandria
\end{minipage} \\
\midrule\noalign{}
\endfirsthead
\toprule\noalign{}
\begin{minipage}[b]{\linewidth}\raggedleft
\(k\)
\end{minipage} & \begin{minipage}[b]{\linewidth}\raggedleft
\(\gamma\)
\end{minipage} & \begin{minipage}[b]{\linewidth}\raggedleft
\(n_{\mathrm{comm}}\)
\end{minipage} & \begin{minipage}[b]{\linewidth}\raggedleft
\(n_{\mathrm{outliers}}\)
\end{minipage} & \begin{minipage}[b]{\linewidth}\raggedleft
ARI
\end{minipage} & \begin{minipage}[b]{\linewidth}\raggedleft
NMI
\end{minipage} & \begin{minipage}[b]{\linewidth}\raggedleft
held-out ICSD
\end{minipage} & \begin{minipage}[b]{\linewidth}\raggedleft
MatterGen
\end{minipage} & \begin{minipage}[b]{\linewidth}\raggedleft
GNoME
\end{minipage} & \begin{minipage}[b]{\linewidth}\raggedleft
MP
\end{minipage} & \begin{minipage}[b]{\linewidth}\raggedleft
JARVIS
\end{minipage} & \begin{minipage}[b]{\linewidth}\raggedleft
Alexandria
\end{minipage} \\
\midrule\noalign{}
\endhead
\bottomrule\noalign{}
\endlastfoot
8 & 1.0 & 4,683 & 14,655 & 0.33 & 0.86 & 85.0\% & 42.9\% & 24.7\% &
28.5\% & 31.4\% & 25.2\% \\
16 & 0.5 & 1,316 & 5,135 & 0.18 & 0.80 & 92.7\% & 63.3\% & 39.8\% &
49.7\% & 51.5\% & 39.3\% \\
\textbf{16} & \textbf{1.0} & \textbf{6,756} & \textbf{15,378} &
\textbf{1.00} & \textbf{1.00} & \textbf{(production reference)} & & & &
& \\
16 & 2.0 & 1,508 & 5,135 & 0.42 & 0.86 & 91.9\% & 59.2\% & 45.2\% &
45.1\% & 44.1\% & 33.0\% \\
32 & 1.0 & 506 & 1,872 & 0.18 & 0.76 & 93.3\% & 70.7\% & 47.2\% & 54.1\%
& 56.7\% & 45.4\% \\
\end{longtable}

The qualitative source ordering held-out ICSD \textgreater{} MatterGen
\textgreater{} \{GNoME ≈ MP-theoretical\} ≈ JARVIS \textgreater{}
Alexandria is preserved at every variant in this grid: held-out ICSD is
the highest in-basin rate at all 4 alternative settings, MatterGen is
the most ICSD-like external source at all 4, and Alexandria is among the
two lowest at all 4. The absolute rates, however, do depend on partition
granularity: coarser partitions (higher resolution or higher \(k\),
fewer larger communities) raise all rates because per-community
thresholds widen with community size, while denser partitions (lower
\(k\)) lower all rates. This sensitivity is methodologically expected
--- the in-basin classification is partition-relative --- and motivates
the harmonized per-community threshold convention adopted throughout
(§S1.6).

The HO--MatterGen gap, the relevant single-number summary of the
experimental-vs-most-ICSD-like-external contrast, ranges from 22
percentage points (at \(k = 32\)) to 42 percentage points (at
\(k = 8\)). The production setting falls inside that range. Numerical
artifacts at \texttt{notes/k\_resolution\_clean\_sweep\_summary.json}
and \texttt{notes/k\_resolution\_proper\_sweep\_20260504/}.

\textbf{Note on absolute held-out rates.} The held-out rates above
(84--93\%) are higher than the 60.6\% the main-text Fig. 3c quotes for
the 2010 cutoff; the difference is a population-definition difference,
not a methodology disagreement. Fig. 3c reports the rate among held-out
\emph{first-report} reduced formulas (the synthesis-retrodiction subset,
\textasciitilde3,000 entries per cutoff) --- those tend to land in less
dense communities because new chemistry concentrates at basin
boundaries. The k/resolution sweep above reports the rate among
\emph{all} held-out ICSD entries (18.5K--21K depending on partition),
which includes many subsequent re-reports of already-densely-occupied
compositions. Both are valid; the within-table relative comparison is
what matters for the (k, resolution) sensitivity question, and that
comparison is clean.

\subsection{S1.9. Visualization and survey tuning
parameters}\label{s1.9.-visualization-and-survey-tuning-parameters}

Several constants in the figure-generation and renaissance-survey
scripts control purely visual or aggregate behaviour and are reported
here so that a reader can reproduce, perturb, or override them. Each is
also documented inline at its definition in the corresponding script.
Numerical results in the manuscript are insensitive to the
figure-rendering knobs; they are sensitive only to the partition and
threshold choices reported in §S1.2 and Methods.

\textbf{Figure 3, kernel-density topographical background}
(\texttt{scripts/make\_fig\_5source\_calibration.py}). The Scott-rule
bandwidth of \texttt{gaussian\_kde} over-smooths the 167.5K-row ICSD
point cloud and visually flattens the cuprate, perovskite, and spinel
basins; we therefore scale the bandwidth by
\texttt{bandwidth\_scale\ =\ 1.6} to keep these basins visually distinct
without resolving spurious noise from the long PCA tails. The KDE is fit
on a \texttt{kde\_sample\ =\ 15,000} random ICSD sub-sample (seeded by
\texttt{-\/-seed}); beyond \textasciitilde10K points the KDE field
converges visually and the marginal cost grows as \(O(N^2)\) per
evaluation. The KDE field is evaluated on a
\texttt{kde\_grid\ =\ 180}-per-axis grid (≈32K evaluation points), which
is visually indistinguishable from 360 at print size. Per-axis quantile
clipping at \texttt{quantile\_trim\ =\ 0.005} (the 0.5--99.5 percentile
range) drops a handful of extreme PCA outliers that would otherwise
stretch the axes empty; external source overlays are \emph{not} trimmed,
so any external point landing outside the trimmed range is clipped from
view but still counted in the bar chart of panel (c). The bar chart is
independent of every KDE knob.

\textbf{Figure 4, synthesizability-prior quadrant bubble scaling}
(\texttt{scripts/make\_fig\_synth\_prior\_quadrant.py}). Bubble area in
matplotlib points² is \texttt{area\ =\ max(fraction\ ×\ 6500,\ 18)}. The
largest fraction in the data is 75.7\% (GNoME, in-basin and no formula
match); at \(6500 \times 0.757 \approx 4900\,\mathrm{pt}^2\) this bubble
sits comfortably inside its quadrant. The
\texttt{18\textbackslash{},\textbackslash{}mathrm\{pt\}\^{}2} floor
prevents bubbles below \textasciitilde0.28\% of a source from shrinking
to a single pixel; in the present data the only zeros (GNoME's two
formula-match quadrants) are rendered as ✗ markers rather than bubbles,
so the floor does not currently conflate ``zero'' with ``tiny
non-zero''.

\textbf{Renaissance survey}
(\texttt{scripts/analyze\_renaissance\_survey.py}). The systematic
survey iterates over every Louvain community with at least
\texttt{N\_MIN\ =\ 50} members; below this floor, sampling noise
dominates pre/post fold-changes. Cuprate community 6425 has 378 members
and the CMR community has 232. The pre/post fold-change is computed over
a \texttt{WINDOW\ =\ 10} year half-width on either side of a candidate
event year, matching the typical ``decade after publication'' pattern in
materials renaissances; both cuprates (Bednorz--Müller 1986) and CMR
manganites (Jin et al.~1994) fully realize their fold-change inside this
window. Candidate event years span \texttt{EVENT\_YEARS\ =\ 1970–2010};
the lower bound is set by the ICSD becoming continuously populated and
the upper bound bounds the +10-yr post-window to fit inside the 2018 end
of the data. The manuscript reports the top-\texttt{TOP\_K\ =\ 20}
communities by step-change score, which leaves a comfortable margin
around the ``nine textbook field-defining renaissances among the
top-20'' headline result.

\section{S2. Formula-Collapsed Structural Graph Comparison to
TRI}\label{s2.-formula-collapsed-structural-graph-comparison-to-tri}

The main text compares our structural-history analysis to the Toyota
Research Institute (TRI) stability-network work primarily at the formula
level. To test whether the two approaches also align as graphs, we
collapsed the ICSD structural map to reduced-formula nodes and built a
mutual \texttt{k}-nearest-neighbor graph over formula centroids in the
frozen structural embedding. We then compared graph statistics on the
shared formulas present in both the collapsed structural graph and the
TRI network.

\subsection{S2.1. Construction}\label{s2.1.-construction}

The collapsed structural graph was built by:

\begin{enumerate}
\def\labelenumi{\arabic{enumi}.}
\tightlist
\item
  grouping ICSD entries by reduced formula,
\item
  averaging the frozen PCA embedding coordinates within each formula
  group,
\item
  retaining formulas represented by at least three ICSD entries,
\item
  building a mutual \texttt{k}-NN graph with \texttt{k\ =\ 8}, and
\item
  computing structural graph statistics on that formula-level graph.
\end{enumerate}

The resulting graph contains:

\begin{itemize}
\tightlist
\item
  9,563 formula nodes
\item
  21,375 edges
\item
  4,802 formulas shared with TRI
\end{itemize}

\subsection{S2.2. Main result}\label{s2.2.-main-result}

The comparison reveals a clear asymmetry:

\begin{itemize}
\tightlist
\item
  chronology aligns strongly between TRI and the formula-collapsed
  structural graph,
\item
  graph topology aligns weakly.
\end{itemize}

That is, the two frameworks largely agree on \textbf{when} important
formulas emerge, but not on \textbf{which formulas occupy analogous
topological roles} within their respective networks. This supports the
interpretation that thermodynamic accessibility and
structural-neighborhood accessibility are historically coupled, but not
equivalent.

\subsection{S2.3. Correlation table}\label{s2.3.-correlation-table}

\begin{longtable}[]{@{}
  >{\raggedright\arraybackslash}p{(\linewidth - 2\tabcolsep) * \real{0.4286}}
  >{\raggedleft\arraybackslash}p{(\linewidth - 2\tabcolsep) * \real{0.5714}}@{}}
\toprule\noalign{}
\begin{minipage}[b]{\linewidth}\raggedright
Quantity compared on shared formulas
\end{minipage} & \begin{minipage}[b]{\linewidth}\raggedleft
Spearman ρ
\end{minipage} \\
\midrule\noalign{}
\endhead
\bottomrule\noalign{}
\endlastfoot
TRI discovery year vs.~structural first year & 0.707 \\
TRI degree vs.~structural degree & 0.003 \\
TRI degree vs.~structural core number & -0.019 \\
TRI eigenvector centrality vs.~structural eigenvector centrality &
-0.136 \\
TRI clustering coefficient vs.~structural clustering coefficient &
-0.065 \\
\end{longtable}

The strong chronology correlation shows that both networks encode the
same broad historical emergence pattern. The near-zero or weakly
negative topology correlations show that the structural graph is not
simply a relabeling of the thermodynamic graph.

\subsection{S2.4. Supplementary
figure}\label{s2.4.-supplementary-figure}

\begin{figure}
\centering
\pandocbounded{\includegraphics[keepaspectratio,alt={Formula-collapsed structural graph versus TRI. Left: chronology aligns strongly. Right: graph topology aligns weakly.}]{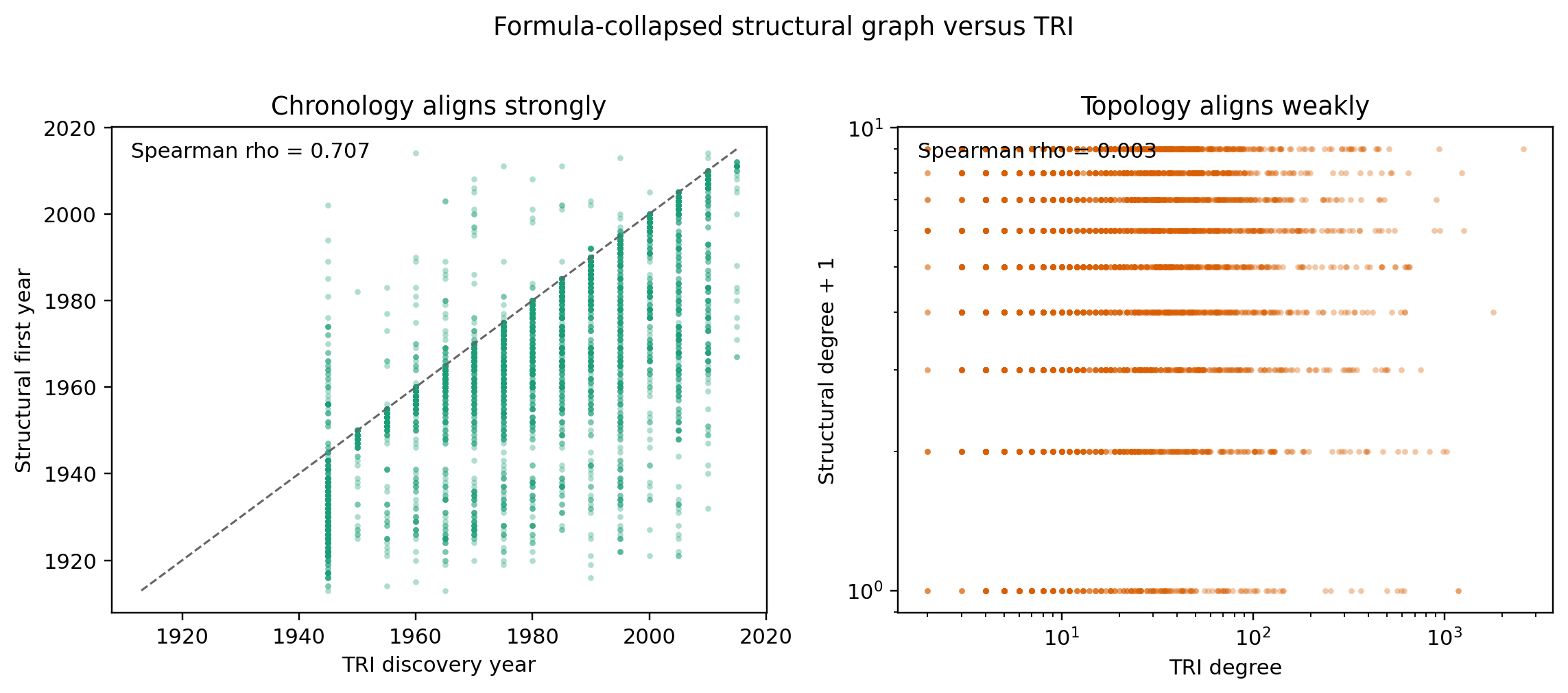}}
\caption{Formula-collapsed structural graph versus TRI. Left: chronology
aligns strongly. Right: graph topology aligns
weakly.}\label{fig:formula-graph-tri}
\end{figure}

\subsection{S2.5. Interpretation}\label{s2.5.-interpretation}

This result does not weaken the main story of the paper. Instead, it
sharpens it. The earlier formula-level overlap analysis already showed
that TRI-central formulas are often historically important in ICSD and
that discovery timing aligns strongly. The collapsed-graph analysis
shows that once formulas are embedded into a structural-neighborhood
graph, thermodynamic-network centrality no longer maps directly onto
structural-network centrality. That is a scientifically useful
distinction:

\begin{itemize}
\tightlist
\item
  the TRI graph captures thermodynamic accessibility and phase-stability
  neighborhood,
\item
  the collapsed structural graph captures structural-similarity
  neighborhood,
\item
  the two are coupled in historical time but describe different
  relational geometries.
\end{itemize}

The consequence is that the structural graph should not be interpreted
as a surrogate for the TRI stability network. Instead, the two provide
complementary views of path-dependent discovery: one thermodynamic, one
structural.

\section{S3. Synthesis-Relevance of the Structural Accessibility
Score}\label{s3.-synthesis-relevance-of-the-structural-accessibility-score}

The main text uses the structural-accessibility score 𝒜ᵢ as a
historically calibrated measure of how far a structure lies from old,
large, well-occupied ICSD basins. A natural next question is whether 𝒜ᵢ
carries synthesis-relevant information or is only a descriptive
geometric statistic. We therefore tested the score against both internal
ICSD retrodiction tasks and an external text-mined synthesis corpus. The
internal tests turned out to be informative but not validating, whereas
the external benchmark gives a modest but directionally consistent
signal.

\subsection{S3.1. Internal retrodiction is informative but not
validating}\label{s3.1.-internal-retrodiction-is-informative-but-not-validating}

Our initial internal test asked whether low-𝒜ᵢ post-cutoff structures
are reported earlier than high-𝒜ᵢ structures when scored against frozen
historical maps. The internal-retrodiction cutoffs (1980, 1990, 2000)
are intentionally one decade earlier than the held-out novelty cutoffs
used in the main text (1990, 2000, 2010): the internal test needs a
sufficiently long \emph{post-cutoff} window in which the same formula
can be re-reported, whereas the held-out novelty test requires a
sufficiently long \emph{pre-cutoff} training window. Across cutoffs at
1980, 1990, and 2000, the resulting Spearman correlations between 𝒜ᵢ and
first-report year are weakly negative (-0.035, -0.058, and -0.051,
respectively). Binary in-basin status behaves similarly, with
correlations of -0.027, -0.018, and -0.020.

Taken literally, this is the opposite of the simple hypothesis that
historically accessible structures should always appear earlier.
However, this does not contradict the main paper. Once the field is deep
in a densification regime, later first reports can plausibly become
\emph{more} concentrated inside old basins because discovery is
increasingly exploitative rather than exploratory. In that sense, the
failure of raw first-report year as a proxy for ``ease of synthesis'' is
itself consistent with the paper's central path-dependence result: year
of first report becomes a poor validation target once exploration has
largely given way to densification.

\subsection{S3.2. Stricter polymorph control does not rescue the
internal
test}\label{s3.2.-stricter-polymorph-control-does-not-rescue-the-internal-test}

We next tightened the internal comparison by requiring temporal
polymorph families to span multiple publication years, multiple
structural communities, and a minimum year gap. Under this stricter
control, the fraction of formulas for which the first-reported variant
has lower 𝒜ᵢ than later variants rises from the loose-test range of
roughly \texttt{36–40\%} to approximately coin-flip: 51.4\%, 51.0\%, and
51.1\% for cutoffs 1980, 1990, and 2000.

That directional shift is encouraging but not decisive. Moreover, the
corresponding fraction for which the first-reported variant is simply
\emph{more in-basin} than later basin-crossing variants remains well
below 50\% (28.6\%, 32.5\%, and 34.6\%). We therefore conclude that the
internal ICSD route is exhausted as an honest validation path: it is
useful for diagnosing how densification reshapes historical ordering,
but it does not provide a clean standalone predictive test of 𝒜ᵢ.

\subsection{S3.3. External validation against the Kononova synthesis
corpus}\label{s3.3.-external-validation-against-the-kononova-synthesis-corpus}

We then turned to the public Kononova text-mined solid-state synthesis
corpus as a large-N external benchmark.\textsuperscript{4} The setup is
positive-unlabeled rather than outcome-labeled: formulas appearing as
synthesis targets in the corpus are treated as positives, and
class-matched first-report ICSD formulas are treated as unlabeled
controls. Matching was performed at the level of anonymized formula plus
number of elements, and the comparison was made on class-centered 𝒜ᵢ.

Across all three historical cutoffs, the Kononova positives are shifted
toward lower class-centered accessibility cost than the matched
controls. The mean class-centered 𝒜ᵢ for positives versus controls is
-0.0556 vs +0.00491 for the 1990 split, -0.172 vs +0.0199 for 2000, and
-0.202 vs +0.0339 for 2010. The corresponding Kolmogorov-Smirnov-style
separations are 0.0486, 0.0716, and 0.0659. A simpler decile-based
summary shows the same pattern: the positive rate in the bottom
accessibility decile exceeds that in the top decile by factors of 1.076,
1.151, and 1.404 across 1990, 2000, and 2010, respectively.

To quantify the uncertainty in those decile ratios we ran 5,000
bootstrap resamples per cutoff (resampling formulas with replacement and
recomputing both decile thresholds and rates within each draw). The 95\%
bootstrap CIs are \texttt{{[}0.88,\ 1.31{]}} (1990; N = 23,764),
\texttt{{[}0.90,\ 1.44{]}} (2000; N = 14,198), and
\texttt{{[}1.02,\ 1.98{]}} (2010; N = 4,227). Only the 2010 cutoff is
significant at the 95\% level; the 1990 and 2000 CIs include 1, although
they are shifted in the same direction. The signal is therefore
directionally consistent across cutoffs but strengthens with later
splits, which is itself consistent with a densification regime in which
the synthesis literature preferentially revisits historically accessible
neighborhoods. We treat this as suggestive rather than confirmatory.

\subsection{S3.4. Interpretation}\label{s3.4.-interpretation}

The most defensible summary is therefore:

\begin{quote}
Structural accessibility is a directionally consistent predictor of
presence in the Kononova synthesis corpus, with bottom-vs-top decile
positive-rate ratios of 1.08, 1.15, and 1.40 for the 1990, 2000, and
2010 cutoffs (95\% bootstrap CIs
\texttt{{[}0.88,\ 1.31{]},\ {[}0.90,\ 1.44{]}}, and
\texttt{{[}1.02,\ 1.98{]}}). Only the 2010 ratio reaches statistical
significance at the 95\% level. We treat this as suggestive evidence
that 𝒜ᵢ carries synthesis-relevant information, while noting that two of
three CIs include 1 and that presence in a text-mined corpus reflects
the field's exploit bias as well as any underlying ease of realization.
\end{quote}

This is enough to justify using 𝒜ᵢ as a cautious filter criterion or
ranking statistic in future generative-model studies, but not enough to
claim that the score has been fully validated as a predictive oracle of
synthesis success. The key remaining test is large-N validation against
outcome-labeled synthesis data.

\subsection{S3.5. Small-N outcome-labeled check against
A-Lab}\label{s3.5.-small-n-outcome-labeled-check-against-a-lab}

As a second external check, we evaluated the corrected A-Lab target
set\textsuperscript{5} against the same frozen ICSD map. Here the
advantage is that the labels are closer to true outcome classes than in
the Kononova positive-unlabeled setup, but the disadvantages are small
sample size and incomplete public structural coverage. Using the
corrected supplementary release, the target backbone contains 57
formulas, of which 42 could be scored structurally from the released CIF
bundle. The corrected target categories are 36 made, 4 inconclusive, 2
offline recovery, and 15 not obtained.

Within the scored subset, the ordering is in the expected direction.
Targets classified as made have mean accessibility 0.654 and frontier
rate 0.639, whereas the inconclusive set has mean accessibility 1.352
and frontier rate 1.000. The two offline-recovery targets sit at lower
accessibility (0.347) and intermediate frontier rate (0.500), which is
plausible given that these were eventually recovered outside the
autonomous loop. This ordering is consistent with the interpretation
that lower structural accessibility cost marks regions of structure
space where autonomous synthesis is more likely to succeed.

We nevertheless treat A-Lab as a consistency check rather than a
success/failure validation. The public release does not expose usable
CIFs for all corrected targets, so the scored subset is biased toward
made and inconclusive outcomes; the 15 ``not obtained'' targets cannot
be scored under the current release, which means the comparison is not a
true success/failure split. In addition, the outcome labels and target
list have already required one public correction. Taken together with
the Kononova result, A-Lab supports the broader claim that 𝒜ᵢ carries
synthesis-relevant information without claiming the score has been
validated as a synthesis-success oracle.

\section{S4. Class-Dependence of Held-Out Frontier Rate within
ICSD}\label{s4.-class-dependence-of-held-out-frontier-rate-within-icsd}

The main text reports a single held-out frontier rate at the 2010 cutoff
aggregated across the full ICSD. Because the structural map is
chemistry-dependent, we also checked whether that aggregate rate is
uniform across application-relevant regions of the map. It is not.

\subsection{S4.1. Community-level functional
labeling}\label{s4.1.-community-level-functional-labeling}

Louvain communities on the frozen structural graph were labeled by
dominant application-relevant functional signature. For each community,
we identified the twenty most-central members by distance to the
community centroid in the frozen PCA embedding, then assigned a
community label from the chemistry and literature context of those core
members using a closed taxonomy:
\texttt{battery\_electrode\_candidate,\ thermoelectric\_candidate,\ magnet\_candidate,\ superconductor\_candidate,\ framework},
and \texttt{unlabeled}. Communities whose core members did not cleanly
support one of the defined classes were left unlabeled rather than
forced into a residual bucket.

This procedure makes no per-structure functionality claim. The community
is the unit of analysis, and the label states only which
application-relevant chemistry dominates its core. These classes are
therefore proxy labels derived from community-level chemistry and
context, not verified functional assignments for every member structure.

\subsection{S4.2. Held-out frontier rates by functional
class}\label{s4.2.-held-out-frontier-rates-by-functional-class}

Within the 2010-cutoff held-out ICSD slice, each later ICSD structure
inherits the functional class of the community to which it projects in
the frozen historical map. The resulting class-level frontier rates are:

\begin{longtable}[]{@{}lr@{}}
\toprule\noalign{}
Functional class & Frontier rate (held-out 2010) \\
\midrule\noalign{}
\endhead
\bottomrule\noalign{}
\endlastfoot
\texttt{framework} & 1.5\% \\
\texttt{battery\_electrode\_candidate} & 4.8\% \\
\texttt{thermoelectric\_candidate} & 5.1\% \\
\texttt{magnet\_candidate} & 7.8\% \\
\end{longtable}

The \texttt{superconductor\_candidate} class is present in the labeling
but contains too few held-out members in the current labeled subset to
support a useful class-level rate. Other structural regions of the map,
dominated mainly by mixed intermetallic and elemental-reference
communities, remain outside the present functional taxonomy.

\subsection{S4.3. Interpretation}\label{s4.3.-interpretation}

The approximately fivefold spread between the framework and magnet
classes is chemically interpretable. Framework-like communities are
dominated by recurring families whose structural coverage in ICSD is
already dense, so ordinary continuation in this region usually lands
inside preexisting historical basins. Magnet-like communities, by
contrast, include narrower rare-earth intermetallic families whose
structural coverage remains sparser, so later ICSD entries are
proportionally more likely to land in frontier-like positions relative
to the frozen 2010 map.

This result does not reweight the aggregate experimental-versus-computed
calibration reported in the main text. Rather, it shows that the
densification signal itself is chemistry-dependent. Any future
class-stratified comparison of AI-associated and computed proposal
outputs against the same historical map will therefore need
class-balanced coverage before a cross-model class-level ordering can be
claimed with confidence.

\section{S5. Thermodynamic Stability Network
Alignment}\label{s5.-thermodynamic-stability-network-alignment}

The main text references the Toyota Research Institute (TRI)
thermodynamic stability network\textsuperscript{6} only briefly, as the
source of the formula subset on which the stepping-stone effect is
computed. This section reports the full alignment analysis between TRI
and the ICSD structural map.

\subsection{S5.1. Formula-level
overlap}\label{s5.1.-formula-level-overlap}

The TRI network is thermodynamic rather than structural: nodes are
stable materials and edges are tie-line relations on the convex
free-energy surface. The overlap with our ICSD-derived set at the
reduced-formula level is almost complete. Of 19,253 TRI ``existing
material'' formulas, 19,069 are present in our ICSD-derived set --- a
shared fraction of 0.9904.

\subsection{S5.2. Historical importance
correlations}\label{s5.2.-historical-importance-correlations}

On the shared formulas, TRI-derived measures track ICSD-derived ones
nontrivially:

\begin{longtable}[]{@{}lr@{}}
\toprule\noalign{}
Quantity & Spearman ρ \\
\midrule\noalign{}
\endhead
\bottomrule\noalign{}
\endlastfoot
TRI degree vs ICSD entry count & 0.330 \\
TRI degree vs ICSD first publication year & -0.325 \\
TRI discovery year vs ICSD first year & 0.872 \\
\end{longtable}

Thermodynamic importance and structural-historical importance are
clearly related, even before asking how a formula behaves structurally.

\subsection{S5.3. The polymorphism
hypothesis}\label{s5.3.-the-polymorphism-hypothesis}

\begin{figure}
\centering
\pandocbounded{\includegraphics[keepaspectratio,alt={TRI degree versus structural fragmentation entropy across the shared reduced-formula set; thermodynamic hubness and structural promiscuity are related but only moderately.}]{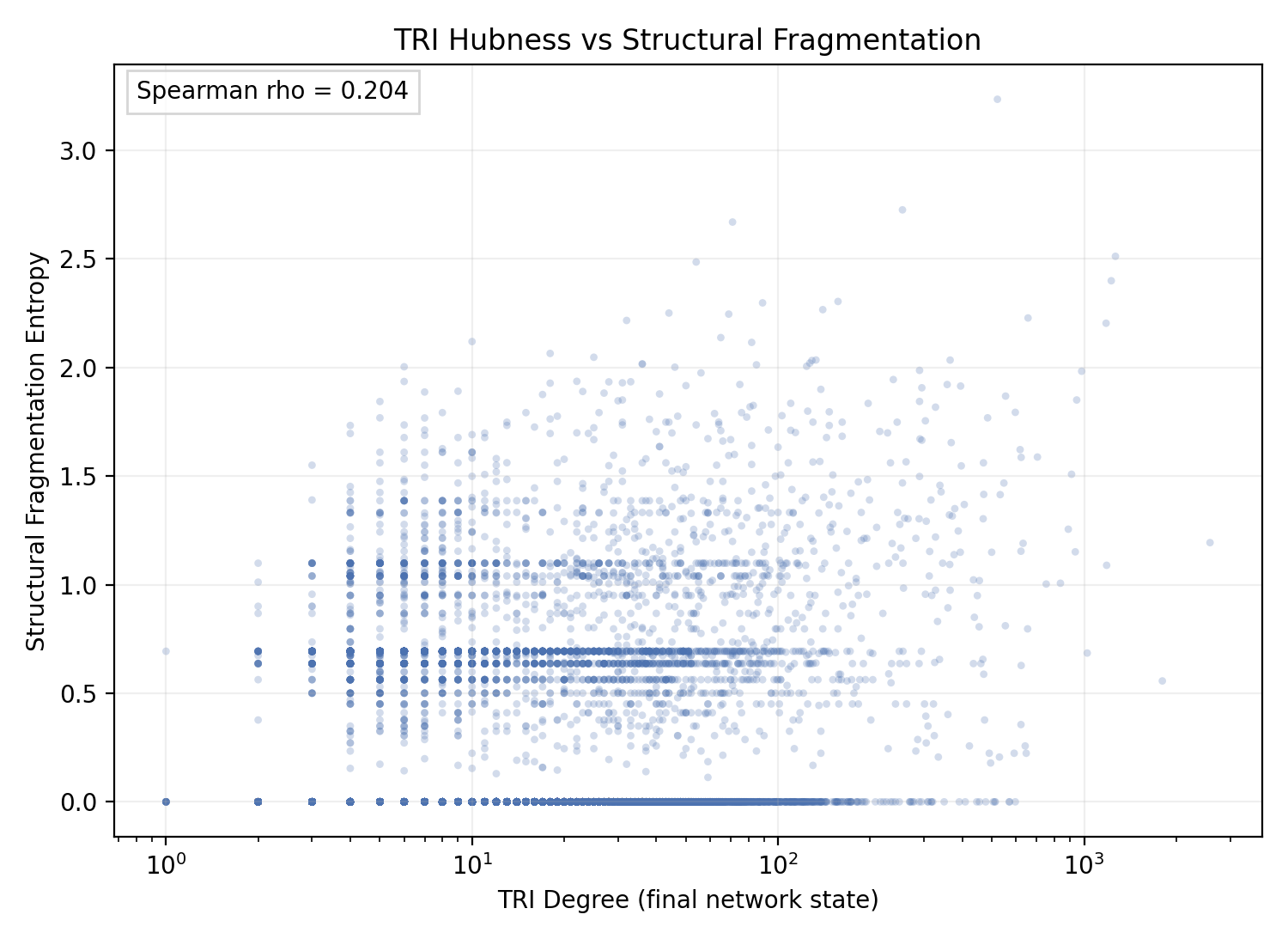}}
\caption{TRI degree versus structural fragmentation entropy across the
shared reduced-formula set; thermodynamic hubness and structural
promiscuity are related but only
moderately.}\label{fig:tri-fragmentation}
\end{figure}

One plausible hypothesis is that thermodynamic hubs achieve their
dominance because they are structurally promiscuous: they appear across
many structural families or polymorphs. We quantified this by computing
the Shannon entropy of each reduced formula's spread across structural
communities. The relationship is positive but only moderate
(Fig.~\ref{fig:tri-fragmentation}): TRI degree versus structural
fragmentation entropy yields ρ = 0.204.

This supports a weak-to-moderate version of the polymorphism hypothesis.
Thermodynamic hubs do tend to spread across more structural
neighborhoods, but not strongly enough to equate thermodynamic
centrality with structural promiscuity. Accessibility and polymorphism
are related, not identical.

\subsection{S5.4. Interpretation}\label{s5.4.-interpretation}

The thermodynamic stability network and our structural community network
are different objects: their nodes overlap strongly, but their edges
encode different physics. They agree on two main points. First,
historically important formulas are recognised as important in both
views. Second, discovery is path-dependent --- the structural
counterpart of the TRI
growth-through-already-connected-thermodynamic-regions pattern is the
stepping-stone effect reported in the main text. Thermodynamic stability
helps determine what chemistry can exist; historical structural
templates strongly influence how that chemistry is realized in the
laboratory. We do \textbf{not} show that thermodynamic centrality
uniquely determines structural exploration, nor that structural
communities are reducible to stability hubs. We do show that
synthesizability-associated network accessibility and
structural-community history are aligned strongly enough to support a
common mechanism: historically accessible regions of materials space
attract future exploration.

A complementary formula-collapsed comparison (§S2) sharpens the
distinction further. Once the structural graph is collapsed to
reduced-formula nodes, chronology remains tightly aligned between TRI
and the structural graph, but graph topology does not: TRI degree and
structural degree are essentially uncorrelated. The TRI graph and the
structural graph are therefore complementary, not redundant.

\subsection{S5.5. Pre-1980 formula undercount:
quantification}\label{s5.5.-pre-1980-formula-undercount-quantification}

The reduced-formula-overlap statistic in the main text uses a post-1980
ICSD reference set. To bound the resulting under-count, we re-ran the
per-source overlap against the full all-year ICSD reduced-formula union,
parsed directly from the ICSD index file (176,298 ICSD entries with
parseable composition + publication year; 108,647 unique reduced
formulas). Pre-1980 contributes 25,579 unique formulas, of which 18,365
(16.9\% of the all-year union) are present \emph{only} before 1980 and
would not be counted under the post-1980 reference.

The shift in per-source overlap when extending the reference is small
for every source and zero for GNoME:

\begin{longtable}[]{@{}lrrr@{}}
\toprule\noalign{}
Source & post-1980 reference & all-year reference & Δ (pp) \\
\midrule\noalign{}
\endhead
\bottomrule\noalign{}
\endlastfoot
GNoME & 0.00\% & \textbf{0.00\%} & +0.00 \\
MatterGen & 4.92\% & 6.99\% & +2.07 \\
MP-theoretical & 23.56\% & 29.49\% & +5.92 \\
JARVIS-DFT & 18.47\% & 24.38\% & +5.90 \\
Alexandria off-hull & 1.94\% & 2.96\% & +1.02 \\
\end{longtable}

(post-1980 rates here are computed against the ICSD-index post-1980
union, 90,282 unique formulas; the corresponding rates against the
score-able 81,531-formula post-1980 subset used for the synth-prior
matching in §S7.5 differ by a fraction of a percentage point and are
reported in the main-text quadrant.) The headline GNoME zero-overlap
claim therefore survives the most permissive ICSD reference we can
construct: even when the reference is extended to all 108,647 unique
ICSD reduced formulas back to 1913, GNoME's 5,000-CIF public release
still has zero matches. MP-theoretical and JARVIS-DFT pick up an
additional \textasciitilde6 pp of formula precedent, bringing them to
29\% and 24\% all-year overlap respectively, consistent with the
main-text framing that DFT-curation databases recycle a substantial
fraction of historically-realized chemistry. The numerical artifact is
recorded at \texttt{notes/pre1980\_formula\_undercount.json} for
downstream auditing.

\section{S6. Chemical Character of Bridge
Attachments}\label{s6.-chemical-character-of-bridge-attachments}

The graph-time analysis in the main text reports rising
bridge-attachment ratios in later decades. Here we report the chemistry
of those bridge attachments in detail.

\begin{figure}
\centering
\pandocbounded{\includegraphics[keepaspectratio,alt={Post-1990 bridge attachments are chemically more complex than graph-core nodes.}]{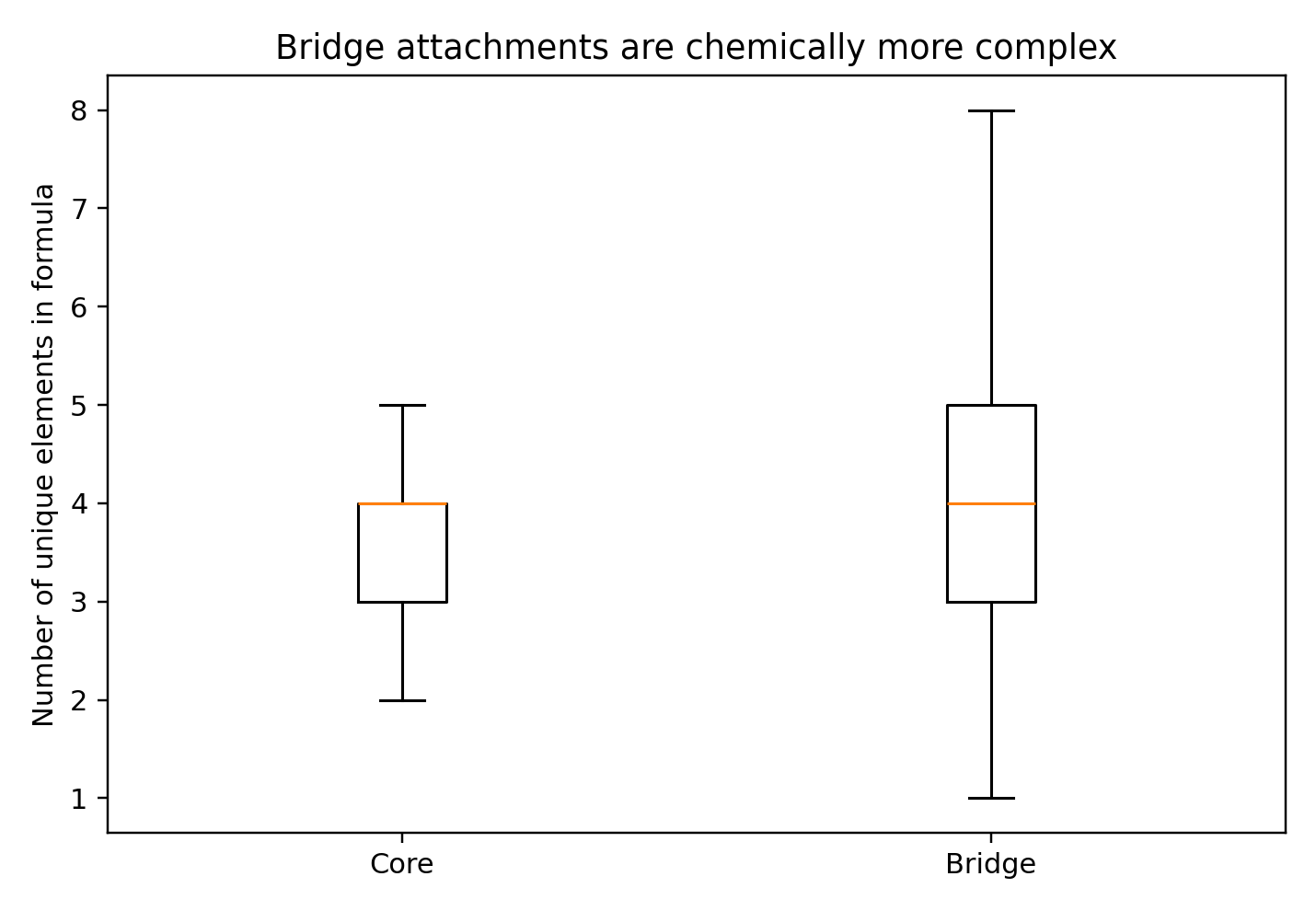}}
\caption{Post-1990 bridge attachments are chemically more complex than
graph-core nodes.}\label{fig:bridge-complexity}
\end{figure}

Restricting attention to post-1990 structures, bridge attachments have
higher formula-level chemical complexity than graph-core nodes, with a
mean of 3.95 unique elements per composition compared with 3.78 for core
nodes (Fig.~\ref{fig:bridge-complexity}). The mean shift is small in
absolute terms (\(\Delta \approx 0.17\) elements) --- large-N
statistical separation, modest practical magnitude --- and the
element-enrichment pattern below is the more chemically informative
signal. Bridge attachments are disproportionately enriched in H, C, N,
and I, along with other less common species, whereas rigid core growth
is comparatively concentrated in simpler oxide- and intermetallic-like
chemistry. The structures that connect previously separated
neighborhoods are therefore not arbitrary; they are biased toward
chemically richer compositions and softer chemistries that are more
capable of interpolating between established basins.

\section{S7. Composition-Matched External Sources vs Held-Out
ICSD}\label{s7.-composition-matched-external-sources-vs-held-out-icsd}

A reviewer may worry that the embedding is chemistry-aware, so external
structure samples could appear frontier-like simply because their
compositions differ from those of held-out ICSD entries. This appendix
tests that explanation directly by repeating the held-out frontier-rate
comparison \textbf{within composition strata} for five external sources
spanning AI-enabled discovery and curated DFT databases.

\subsection{S7.1. Sources tested}\label{s7.1.-sources-tested}

\begin{longtable}[]{@{}
  >{\raggedright\arraybackslash}p{(\linewidth - 6\tabcolsep) * \real{0.2000}}
  >{\raggedright\arraybackslash}p{(\linewidth - 6\tabcolsep) * \real{0.2000}}
  >{\raggedright\arraybackslash}p{(\linewidth - 6\tabcolsep) * \real{0.2000}}
  >{\raggedleft\arraybackslash}p{(\linewidth - 6\tabcolsep) * \real{0.4000}}@{}}
\toprule\noalign{}
\begin{minipage}[b]{\linewidth}\raggedright
Source
\end{minipage} & \begin{minipage}[b]{\linewidth}\raggedright
Type
\end{minipage} & \begin{minipage}[b]{\linewidth}\raggedright
Subset used
\end{minipage} & \begin{minipage}[b]{\linewidth}\raggedleft
N
\end{minipage} \\
\midrule\noalign{}
\endhead
\bottomrule\noalign{}
\endlastfoot
GNoME\textsuperscript{7} & AI-enabled graph-network discovery & Public
5,000-CIF release & 5,000 \\
MatterGen-public\textsuperscript{8} & Generative AI (diffusion model) &
Public CIF release & 386 \\
Materials Project (MP) & DFT curation database &
\texttt{theoretical=True}, \texttt{e\_above\_hull} ≤ 0.2 eV/atom, no
ICSD provenance & 5,000 random of 20,479 \\
JARVIS-DFT\textsuperscript{9} & DFT calculation database (NIST) & Empty
\texttt{icsd} field, off-hull (ehull ∈ {[}0.05, 0.5{]} eV/atom) & 5,000
random of 22,022 \\
Alexandria-PBE\textsuperscript{10} & DFT prototype-substitution database
& 2025.07.02 release, off-hull predictions (ehull ∈ {[}0.05, 0.5{]}
eV/atom) & 5,000 random of 154,942 \\
\end{longtable}

The MP and Alexandria filters are designed to retain only structures
that are \emph{not experimentally known} --- for MP, this is the
explicit \texttt{theoretical=True} flag; for Alexandria, where no
provenance flag exists, we use off-hull energetics as a proxy. Both
filters are imperfect (Alexandria's e-above-hull cutoff in particular
cannot exclude every experimentally-known structure), and the residual
ICSD-overlap is a known caveat.

\subsection{S7.2. Method}\label{s7.2.-method}

For each cutoff T ∈ \{1990, 2000, 2010\} we (i) recompute the
per-community 95th-percentile centroid-distance threshold from
training-only structures (year \(\leq T\)); (ii) reclassify every
external proposal against those thresholds using its nearest-centroid
distance and assigned community (the same approximation used in the
main-text temporal sweep); and (iii) compute matched in-basin rates
within composition strata that contain \textbf{both} at least one
external proposal and at least one held-out ICSD entry.

We report two complementary stratifications:

\begin{itemize}
\tightlist
\item
  \textbf{Coarse strata} --- descriptor (anion class, \(n\)-elements
  bucket, anion/cation ratio bucket); approximately 40--135 strata.
  Defensive against shrinking external populations to near-zero per
  stratum.
\item
  \textbf{Anonymized formula} --- pymatgen-style anonymized
  stoichiometry (MgAl₂O₄ and ZnFe₂O₄ both map to A₂B₁C₄); approximately
  100--500 strata. The textbook clean composition-matching descriptor:
  two structures share an anonymized formula iff they share
  stoichiometry, regardless of element identity.
\end{itemize}

Wilson 95\% CIs are reported throughout. The pipeline is run on TACC
against the canonical raw 213-dimensional matminer feature matrix
(167,500 × 213), with PCA-32 fit \emph{inside} the analysis to match the
main-text temporal sweep.

\subsection{S7.3. Result}\label{s7.3.-result}

\begin{figure}
\centering
\pandocbounded{\includegraphics[keepaspectratio,alt={Composition-matched in-basin rates across five external sources. Left: coarse anion / n-elements / ratio strata. Right: anonymized stoichiometry. Strata-count annotations show per-source intersection with held-out ICSD (G = GNoME, Mat = MatterGen, MP = Materials Project, Jar = JARVIS-DFT, Ale = Alexandria).}]{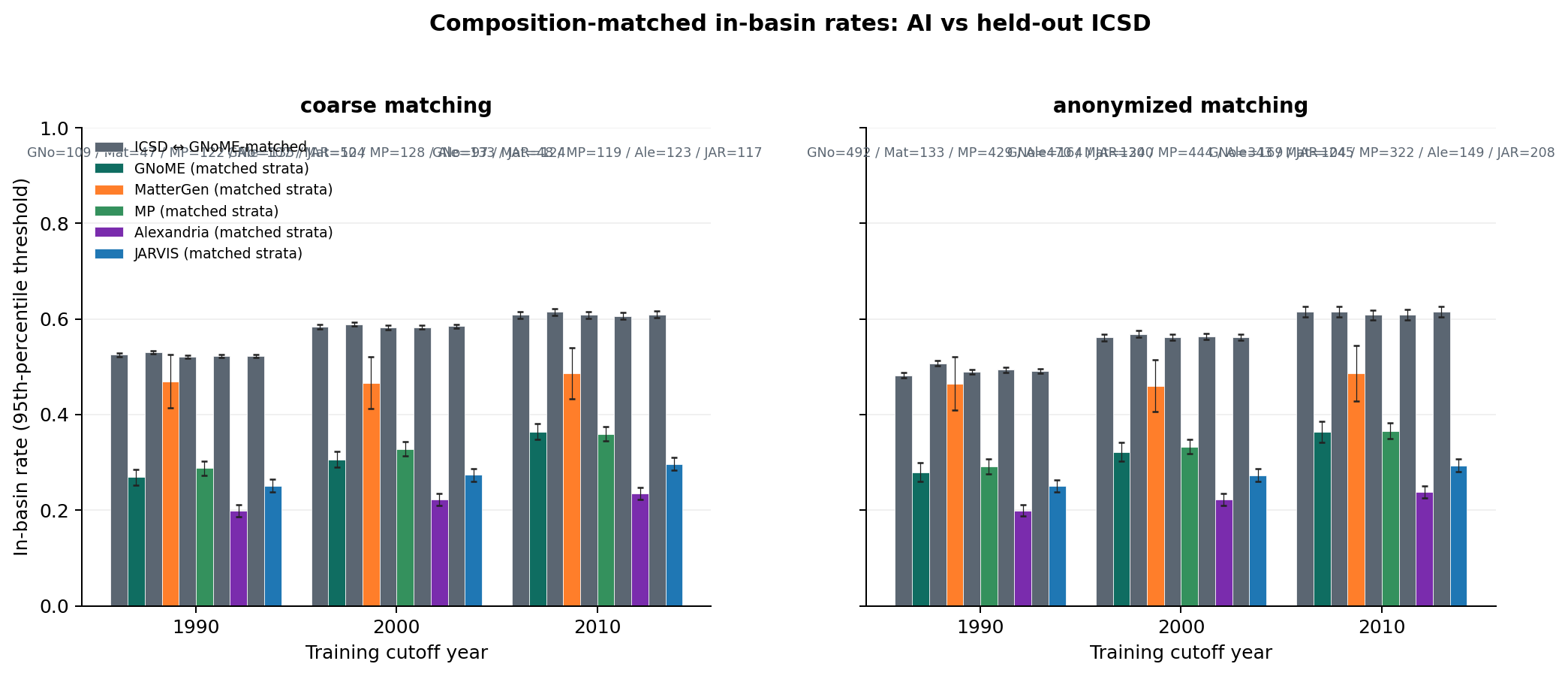}}
\caption{Composition-matched in-basin rates across five external
sources. Left: coarse anion / \(n\)-elements / ratio strata. Right:
anonymized stoichiometry. Strata-count annotations show per-source
intersection with held-out ICSD (G = GNoME, Mat = MatterGen, MP =
Materials Project, Jar = JARVIS-DFT, Ale =
Alexandria).}\label{fig:composition-matched}
\end{figure}

The exploitation gap relative to held-out ICSD is non-zero for every
external source at every cutoff under both stratifications, with notable
variation in magnitude (Table~\ref{tbl:composition-matched},
Fig.~\ref{fig:composition-matched}).

\begin{longtable}[]{@{}
  >{\raggedleft\arraybackslash}p{(\linewidth - 12\tabcolsep) * \real{0.1739}}
  >{\raggedright\arraybackslash}p{(\linewidth - 12\tabcolsep) * \real{0.1304}}
  >{\raggedright\arraybackslash}p{(\linewidth - 12\tabcolsep) * \real{0.1304}}
  >{\centering\arraybackslash}p{(\linewidth - 12\tabcolsep) * \real{0.1739}}
  >{\raggedright\arraybackslash}p{(\linewidth - 12\tabcolsep) * \real{0.1304}}
  >{\raggedright\arraybackslash}p{(\linewidth - 12\tabcolsep) * \real{0.1304}}
  >{\raggedleft\arraybackslash}p{(\linewidth - 12\tabcolsep) * \real{0.1304}}@{}}
\caption{\label{tbl:composition-matched}In-basin rate of external
proposals vs held-out ICSD, restricted to composition strata populated
by both. Wilson 95\% CIs in brackets. MP = Materials Project
\texttt{theoretical=True} subset; JARVIS = JARVIS-DFT 2022.12.12
off-hull predictions with empty \texttt{icsd} field; Alexandria =
Alexandria-PBE 2025.07.02 off-hull predictions.}\tabularnewline
\toprule\noalign{}
\begin{minipage}[b]{\linewidth}\raggedleft
Cutoff
\end{minipage} & \begin{minipage}[b]{\linewidth}\raggedright
Match
\end{minipage} & \begin{minipage}[b]{\linewidth}\raggedright
Source
\end{minipage} & \begin{minipage}[b]{\linewidth}\centering
Common strata
\end{minipage} & \begin{minipage}[b]{\linewidth}\raggedright
ICSD ↔ src
\end{minipage} & \begin{minipage}[b]{\linewidth}\raggedright
external
\end{minipage} & \begin{minipage}[b]{\linewidth}\raggedleft
gap
\end{minipage} \\
\midrule\noalign{}
\endfirsthead
\toprule\noalign{}
\begin{minipage}[b]{\linewidth}\raggedleft
Cutoff
\end{minipage} & \begin{minipage}[b]{\linewidth}\raggedright
Match
\end{minipage} & \begin{minipage}[b]{\linewidth}\raggedright
Source
\end{minipage} & \begin{minipage}[b]{\linewidth}\centering
Common strata
\end{minipage} & \begin{minipage}[b]{\linewidth}\raggedright
ICSD ↔ src
\end{minipage} & \begin{minipage}[b]{\linewidth}\raggedright
external
\end{minipage} & \begin{minipage}[b]{\linewidth}\raggedleft
gap
\end{minipage} \\
\midrule\noalign{}
\endhead
\bottomrule\noalign{}
\endlastfoot
1990 & coarse & GNoME & 109 & 0.525 {[}0.521, 0.529{]} & 0.269 {[}0.252,
0.286{]} & +25.6 pp \\
1990 & coarse & MatterGen & 47 & 0.530 {[}0.526, 0.534{]} & 0.469
{[}0.415, 0.525{]} & +6.1 pp \\
1990 & coarse & MP & 122 & 0.521 {[}0.518, 0.525{]} & 0.288 {[}0.273,
0.303{]} & +23.4 pp \\
1990 & coarse & Alexandria & 135 & 0.522 {[}0.519, 0.526{]} & 0.199
{[}0.187, 0.212{]} & +32.3 pp \\
1990 & anonymized & GNoME & 492 & 0.482 {[}0.477, 0.488{]} & 0.280
{[}0.260, 0.299{]} & +20.3 pp \\
1990 & anonymized & MatterGen & 133 & 0.507 {[}0.501, 0.512{]} & 0.465
{[}0.410, 0.520{]} & +4.2 pp \\
1990 & anonymized & MP & 429 & 0.490 {[}0.485, 0.495{]} & 0.291
{[}0.276, 0.307{]} & +19.9 pp \\
1990 & anonymized & Alexandria & 164 & 0.493 {[}0.488, 0.499{]} & 0.199
{[}0.187, 0.212{]} & +29.4 pp \\
2000 & coarse & GNoME & 107 & 0.584 {[}0.579, 0.588{]} & 0.306 {[}0.290,
0.323{]} & +27.8 pp \\
2000 & coarse & MatterGen & 50 & 0.589 {[}0.585, 0.594{]} & 0.466
{[}0.413, 0.520{]} & +12.3 pp \\
2000 & coarse & MP & 128 & 0.582 {[}0.578, 0.586{]} & 0.328 {[}0.313,
0.343{]} & +25.4 pp \\
2000 & coarse & Alexandria & 133 & 0.583 {[}0.579, 0.587{]} & 0.222
{[}0.209, 0.235{]} & +36.1 pp \\
2000 & anonymized & GNoME & 470 & 0.561 {[}0.554, 0.568{]} & 0.322
{[}0.302, 0.342{]} & +23.9 pp \\
2000 & anonymized & MatterGen & 130 & 0.569 {[}0.562, 0.575{]} & 0.459
{[}0.406, 0.514{]} & +10.9 pp \\
2000 & anonymized & MP & 444 & 0.562 {[}0.556, 0.568{]} & 0.333
{[}0.318, 0.348{]} & +22.9 pp \\
2000 & anonymized & Alexandria & 169 & 0.563 {[}0.557, 0.569{]} & 0.222
{[}0.210, 0.235{]} & +34.1 pp \\
2010 & coarse & GNoME & 97 & 0.608 {[}0.601, 0.616{]} & 0.364 {[}0.348,
0.381{]} & +24.4 pp \\
2010 & coarse & MatterGen & 48 & 0.614 {[}0.607, 0.622{]} & 0.486
{[}0.433, 0.540{]} & +12.8 pp \\
2010 & coarse & MP & 119 & 0.608 {[}0.601, 0.615{]} & 0.359 {[}0.344,
0.374{]} & +24.9 pp \\
2010 & coarse & Alexandria & 123 & 0.606 {[}0.599, 0.613{]} & 0.235
{[}0.222, 0.248{]} & +37.1 pp \\
2010 & anonymized & GNoME & 343 & 0.615 {[}0.604, 0.627{]} & 0.364
{[}0.342, 0.386{]} & +25.2 pp \\
2010 & anonymized & MatterGen & 104 & 0.615 {[}0.604, 0.626{]} & 0.486
{[}0.428, 0.544{]} & +12.9 pp \\
2010 & anonymized & MP & 322 & 0.608 {[}0.598, 0.618{]} & 0.366
{[}0.350, 0.382{]} & +24.2 pp \\
2010 & anonymized & Alexandria & 149 & 0.609 {[}0.598, 0.619{]} & 0.237
{[}0.225, 0.250{]} & +37.2 pp \\
1990 & coarse & JARVIS & 124 & 0.523 {[}0.519, 0.526{]} & 0.251
{[}0.238, 0.264{]} & +27.2 pp \\
1990 & anonymized & JARVIS & 240 & 0.491 {[}0.486, 0.496{]} & 0.250
{[}0.237, 0.264{]} & +24.1 pp \\
2000 & coarse & JARVIS & 124 & 0.584 {[}0.581, 0.588{]} & 0.274
{[}0.260, 0.287{]} & +31.1 pp \\
2000 & anonymized & JARVIS & 245 & 0.562 {[}0.556, 0.568{]} & 0.273
{[}0.260, 0.287{]} & +28.9 pp \\
2010 & coarse & JARVIS & 117 & 0.609 {[}0.602, 0.616{]} & 0.296
{[}0.283, 0.310{]} & +31.3 pp \\
2010 & anonymized & JARVIS & 208 & 0.615 {[}0.605, 0.625{]} & 0.293
{[}0.280, 0.307{]} & +32.2 pp \\
\end{longtable}

\subsection{S7.4. Interpretation}\label{s7.4.-interpretation}

The five external sources separate cleanly under composition matching,
in a ranking that is stable across cutoffs and stratifications
(anonymized gap at the 2010 cutoff in parentheses):

Ordered from smallest to largest residual gap at the 2010 anonymized
cutoff, the sources are MatterGen (+13 pp), MP-theoretical (+24 pp),
GNoME (+25 pp), JARVIS (+32 pp), and Alexandria off-hull (+37 pp).

\begin{itemize}
\item
  \textbf{MatterGen-public} has the smallest residual gap to ICSD:
  composition matching shrinks the unmatched approximately 30 percentage
  points gap to +4--13 pp.~At 1990 the 95\% CIs nearly overlap,
  indicating the cohort-level exploitation pattern at that cutoff is
  largely a composition-stratum effect; at 2010 a +13 pp gap remains.
\item
  \textbf{GNoME} retains a +20 to +28 pp gap at every cutoff. The
  held-out ICSD rate is \(1.7\)--\(2.0\times\) the GNoME rate within the
  same anonymized-formula classes. The exploitation pattern is not a
  compositional drift artifact.
\item
  \textbf{Materials Project theoretical-only entries} behave nearly
  identically to GNoME under matching (gaps within 1--3 pp at all
  cutoffs). DFT databases of theoretical structures and AI-associated
  public samples therefore occupy structural space with comparable
  in-basin rates relative to held-out ICSD --- the apparent novelty of
  AI-associated proposals is not categorically distinct from what
  high-throughput DFT exploration has already produced.
\item
  \textbf{JARVIS-DFT off-hull predictions} sit between MP-theoretical
  and Alexandria off-hull, with gaps growing from +24 pp at 1990 to +32
  pp at 2010 under anonymized matching. JARVIS uses a different
  ``predicted'' convention than MP (off-hull energetics rather than the
  binary \texttt{theoretical} flag), and the larger gap reflects that
  more permissive criterion: JARVIS's off-hull subset reaches further
  from established basins than MP's near-hull theoretical subset.
\item
  \textbf{Alexandria off-hull predictions} show the \emph{largest} gap
  (+29 to +37 pp), and that gap \emph{grows} under stricter
  anonymized-formula matching at the 2000 and 2010 cutoffs. Alexandria
  explores genuinely different structural regions even when restricted
  to compositions ICSD has investigated: its off-hull subset is both
  composition-novel (relative to ICSD) and structure-novel within shared
  compositions.
\end{itemize}

That all five sources retain a residual exploitation gap after
composition matching is the load-bearing observation: distance from
historical basins is not just a composition artifact, and the gap is
robust across very different sample types --- AI-associated public
samples of two architectures, MP's curated DFT entries, JARVIS-DFT's
off-hull predictions, and Alexandria's prototype-substitution
predictions. The MP ≈ GNoME equivalence is the most useful calibration
point: the structural-novelty signal we measure for AI-associated public
samples is comparable in magnitude to what a sufficiently aggressive
DFT-based exploration of theoretical structures already yields. The
systematic ordering MP ≈ GNoME \(<\) JARVIS \(<\) Alexandria along the
off-hull-energy axis suggests that ``structural distance from ICSD'' is
at least partly tracking how aggressively the parent dataset's filters
tolerate metastability.

\subsection{S7.5. Reduced-Formula Overlap with ICSD: A Synthesizability
Prior}\label{s7.5.-reduced-formula-overlap-with-icsd-a-synthesizability-prior}

The composition-matched analysis above asks ``are AI and ICSD
populations from the same composition strata still structurally
separable?'' A complementary and more directly synthesis-relevant
question is: \textbf{do these external sources propose structures whose
reduced formula has ever been observed in ICSD at all?} A formula match
--- even without a structural match --- is a strong ``this composition
has been made, by some route, at some time'' signal that the geometric
in-basin classification cannot capture.

We use the union of post-1980 ICSD first-report formulas as the
reference set (81,531 unique reduced formulas; pre-1980 ICSD
compositions are \emph{not} covered, an asymmetric caveat that biases
the match rate downward by an unknown small amount). For each external
source, we cross-tabulate \{formula match : yes / no\} against
\{full-map in-basin : yes / no\} from the source's frontier-records CSV.

\begin{longtable}[]{@{}
  >{\raggedright\arraybackslash}p{(\linewidth - 14\tabcolsep) * \real{0.0714}}
  >{\raggedleft\arraybackslash}p{(\linewidth - 14\tabcolsep) * \real{0.1429}}
  >{\raggedleft\arraybackslash}p{(\linewidth - 14\tabcolsep) * \real{0.1429}}
  >{\raggedright\arraybackslash}p{(\linewidth - 14\tabcolsep) * \real{0.0714}}
  >{\raggedleft\arraybackslash}p{(\linewidth - 14\tabcolsep) * \real{0.1429}}
  >{\raggedleft\arraybackslash}p{(\linewidth - 14\tabcolsep) * \real{0.1429}}
  >{\raggedleft\arraybackslash}p{(\linewidth - 14\tabcolsep) * \real{0.1429}}
  >{\raggedleft\arraybackslash}p{(\linewidth - 14\tabcolsep) * \real{0.1429}}@{}}
\caption{\label{tbl:formula-overlap}Formula-overlap with post-1980 ICSD
(\(N=81{,}531\) unique reduced formulas). The four-cell quadrant gives,
for each source, the joint distribution over \{in-basin geometric
classification\} × \{formula match\}. The in-basin / frontier
classification uses per-community 95th-percentile within-community
centroid-distance thresholds (each Louvain community sets its own
scale), consistent with Figure 3c and the composition-matched control in
§S7.3.}\tabularnewline
\toprule\noalign{}
\begin{minipage}[b]{\linewidth}\raggedright
Source
\end{minipage} & \begin{minipage}[b]{\linewidth}\raggedleft
N
\end{minipage} & \begin{minipage}[b]{\linewidth}\raggedleft
formula∈ICSD
\end{minipage} & \begin{minipage}[b]{\linewidth}\raggedright
rate {[}95\% CI{]}
\end{minipage} & \begin{minipage}[b]{\linewidth}\raggedleft
in-basin \& match
\end{minipage} & \begin{minipage}[b]{\linewidth}\raggedleft
frontier \& match
\end{minipage} & \begin{minipage}[b]{\linewidth}\raggedleft
in-basin \& no-match
\end{minipage} & \begin{minipage}[b]{\linewidth}\raggedleft
frontier \& no-match
\end{minipage} \\
\midrule\noalign{}
\endfirsthead
\toprule\noalign{}
\begin{minipage}[b]{\linewidth}\raggedright
Source
\end{minipage} & \begin{minipage}[b]{\linewidth}\raggedleft
N
\end{minipage} & \begin{minipage}[b]{\linewidth}\raggedleft
formula∈ICSD
\end{minipage} & \begin{minipage}[b]{\linewidth}\raggedright
rate {[}95\% CI{]}
\end{minipage} & \begin{minipage}[b]{\linewidth}\raggedleft
in-basin \& match
\end{minipage} & \begin{minipage}[b]{\linewidth}\raggedleft
frontier \& match
\end{minipage} & \begin{minipage}[b]{\linewidth}\raggedleft
in-basin \& no-match
\end{minipage} & \begin{minipage}[b]{\linewidth}\raggedleft
frontier \& no-match
\end{minipage} \\
\midrule\noalign{}
\endhead
\bottomrule\noalign{}
\endlastfoot
GNoME & 5,000 & \textbf{0} & 0.000 {[}0.000, 0.001{]} & 0 (0.0\%) & 0
(0.0\%) & 1,895 (37.9\%) & 3,105 (62.1\%) \\
MatterGen & 386 & 19 & 0.049 {[}0.032, 0.076{]} & 9 (2.3\%) & 10 (2.6\%)
& 171 (44.3\%) & 196 (50.8\%) \\
Alexandria & 5,000 & 96 & 0.019 {[}0.016, 0.023{]} & 30 (0.6\%) & 66
(1.3\%) & 1,247 (24.9\%) & 3,657 (73.1\%) \\
MP-theoretical & 4,999 & \textbf{1,090} & 0.218 {[}0.207, 0.230{]} & 364
(7.3\%) & 726 (14.5\%) & 1,456 (29.1\%) & 2,453 (49.1\%) \\
JARVIS predicted & 4,964 & 880 & 0.177 {[}0.167, 0.188{]} & 339 (6.8\%)
& 541 (10.9\%) & 1,225 (24.7\%) & 2,859 (57.6\%) \\
\end{longtable}

The result separates the sources clearly along this second axis:

\begin{itemize}
\item
  \textbf{The two AI-associated public samples are nearly
  composition-novel.} GNoME's public 5,000-CIF release has \textbf{zero
  overlap} with post-1980 ICSD by reduced formula. The compositions are
  genuinely outside the modern ICSD chemistry --- sample GNoME formulas
  like \texttt{Dy5HoAl24Os5Pd2Pt}, \texttt{Cs3K8Rb(BiTe3)4},
  \texttt{TbTm3Zr4(NiP)12} are exactly the high-element-count
  compositions identified as chemically implausible in independent
  critiques of the GNoME release.\textsuperscript{1} MatterGen reaches a
  4.9\% overlap, with the matches looking like ordinary inorganic
  compounds (e.g.~K₂EuSi₄O₁₀F, MnSeO₃, La₃MoO₇).
\item
  \textbf{The two DFT calculation databases largely propose structural
  variants of known compositions.} MP-theoretical has \(21.8\,\%\)
  formula overlap with post-1980 ICSD; JARVIS-predicted has
  \(17.7\,\%\). Their ``predicted/theoretical'' subsets are not
  categorically composition-novel --- many entries are polymorphs or
  structural variants of compositions ICSD has already realized in some
  form. This is a fundamentally different mode of ``novelty'' than the
  AI-associated mode.
\item
  \textbf{Alexandria off-hull is the most exploratory on both axes
  simultaneously.} \(1.9\,\%\) formula overlap, and within that small
  overlap, structures sit far from the corresponding ICSD basins
  (matched-strata gaps of \(+29\) to \(+37\) pp).
\end{itemize}

The four-cell quadrant in Table~\ref{tbl:formula-overlap} is the natural
synthesizability-prior summary: the \textbf{upper-left cell (in-basin
AND formula match)} is where structural similarity to known basins
coincides with confirmed-synthesizable composition, the strongest
combined prior the framework can express. The \textbf{lower-right cell
(frontier AND no formula match)} is the weakest prior ---
composition-novel and structurally-distant. The framework provides the
prior; downstream synthesis attempts are the validation.

We emphasize what this \textbf{does not} claim. A formula match does not
entail synthesizability of the proposed structure (the structure may be
a metastable polymorph requiring a different synthesis route than the
experimental analog). A no-match does not entail unsynthesizability (the
composition may simply be unexplored). The four-cell quadrant is a
calibrated \emph{prior}, not a prediction. We also note that this
analysis aligns with a longstanding concern that AI-associated proposals
can drift toward chemically implausible high-element-count
compositions\textsuperscript{1}: the GNoME public release contains zero
post-1980 ICSD-formula matches, while DFT-calculation databases (which
build outward from known experimental compositions) retain approximately
20\% overlap. Either pattern can be desired depending on the campaign
goal: composition-novel exploration or polymorph-screening of known
compositions.

\textbf{The zero is not a formula-format artifact, and we verified it
directly.} Both reference and external formulas are pymatgen-canonical
(\texttt{Composition.reduced\_formula}) and were re-canonicalized on
both sides; the count remained zero. To corroborate that the zero
reflects genuine compositional novelty rather than a stoichiometric
tie-breaking accident, we also computed two looser overlap statistics on
the same 5,000 GNoME public-release entries against the 81,531 post-1980
ICSD reduced formulas: only \(3.6\%\) share an \emph{element set} with
any post-1980 ICSD formula (i.e., even ignoring stoichiometry, GNoME's
element combinations are 96.4\% disjoint from the modern ICSD universe),
but \(76.8\%\) share an \emph{anonymized} stoichiometry such as A₁B₁C₃
or A₁B₂C₄. The novelty is therefore in \emph{which atoms occupy
structural positions}, not in the structural prototype: GNoME's public
release populates ordinary anonymized stoichiometries (rocksalt-like,
spinel-like, perovskite-like) but with element combinations that are
essentially absent from the modern experimental record.

\textbf{Important caveat on GNoME's curated public release.} The
5,000-CIF GNoME release is a curated public subset of the much larger
generative output (approximately 380,000 structures) reported in the
original publication, and the curation rule is not fully described. The
zero-formula-overlap, the 3.6\% element-set overlap, and the 0.0\%
in-basin-and-match figure quoted above all describe this curated public
subset; they should not be read as descriptive of GNoME's full
generative distribution. Any reader interpreting the
synthesizability-prior quadrant for downstream campaign design should
treat the GNoME numbers as \emph{upper bounds on novelty for the public
release} and request a random GNoME sample (or run the framework on one)
before drawing conclusions about the underlying generator.

\section{S8. Community Growth Around Scientific
Events}\label{s8.-community-growth-around-scientific-events}

A direct test of whether the framework's communities track real
scientific phenomena: do they show step-changes in membership-growth
aligned with known field-defining events? We examine two well-documented
renaissances in inorganic materials research: the high-Tc cuprates
(Bednorz \& Müller, 1986; Nobel 1987)\textsuperscript{11} and the
colossal-magnetoresistance (CMR) manganites (Jin et al.,
1994)\textsuperscript{12}.

\subsection{S8.1. Method}\label{s8.1.-method}

For each candidate family we build a ``family'' set of ICSD entries from
(a) exact reduced-formula matches to canonical exemplars and (b) a loose
element-set seed (e.g., for cuprates: contains Cu, O, and at least one
of \{Ba, Sr, La, Y\}; for manganites: contains Mn, O, and at least one
of \{La, Pr, Nd, Sm, Eu, Gd, Tb, Dy, Ho, Ca, Sr, Ba\}). We then look up
each family member's production-run community label (excluding HDBSCAN
noise, community \(-1\)) and identify the dominant community as the
renaissance community for that family. From
\texttt{community\_assignments.csv} we then retrieve \textbf{all}
community members across all years (including pre-event entries that
were not in the seed set) and plot the per-year histogram of
first-publication years.

\subsection{S8.2. Result}\label{s8.2.-result}

\begin{figure}
\centering
\pandocbounded{\includegraphics[keepaspectratio,alt={Per-year histogram of first ICSD entries in two renaissance communities. Top: community 6425 (high-Tc cuprates) is empty before 1986 and grows abruptly after Bednorz \& Müller. Bottom: community 160 (doped CMR manganite family) has a small Goodenough-era baseline (1 entry in the 1970s, 1 in the 1980s) and explodes after Jin et al.~1994, peaking at over 80 entries in 2002. Red dashed lines mark the field-defining publications.}]{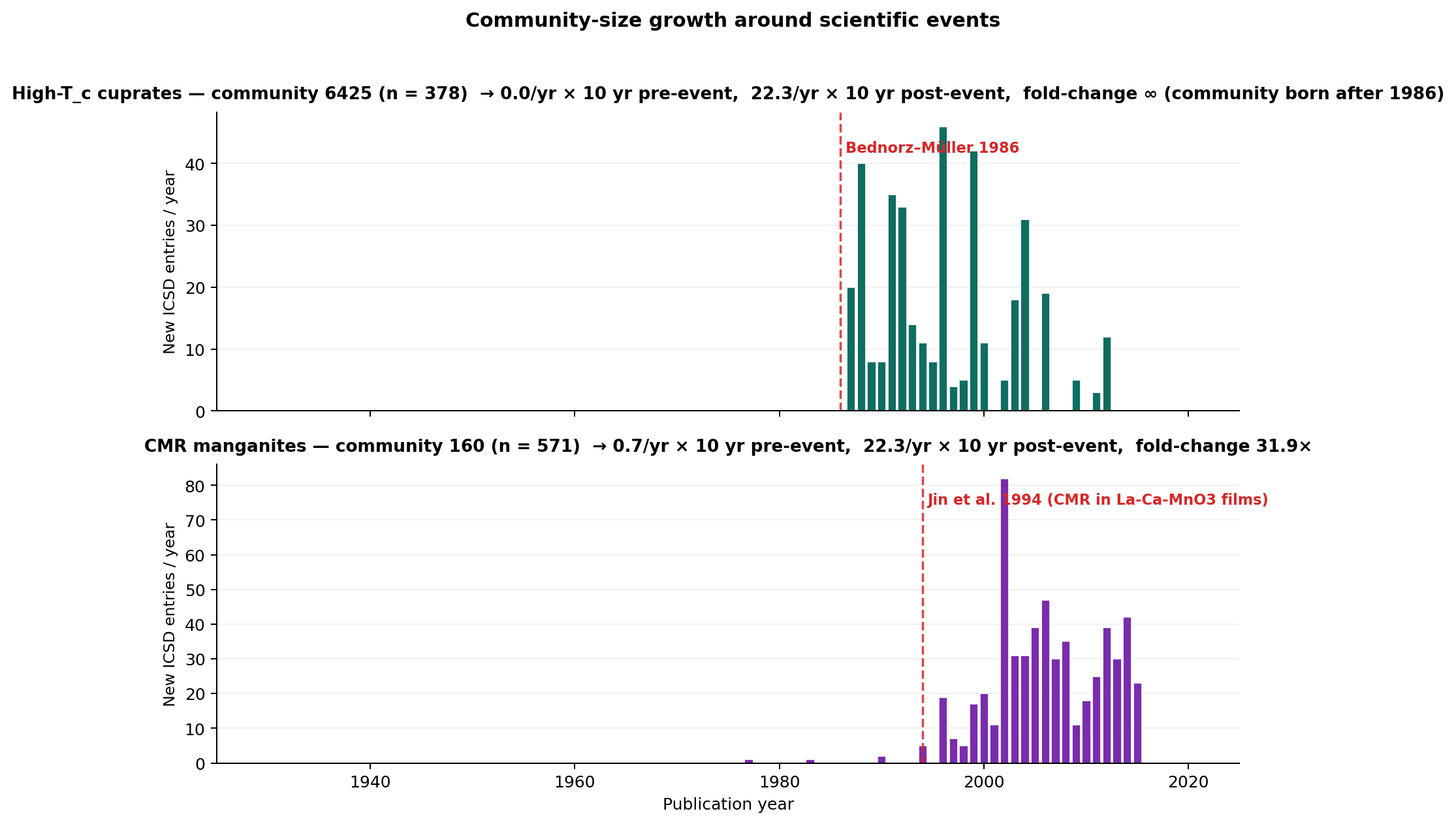}}
\caption{Per-year histogram of first ICSD entries in two renaissance
communities. Top: community 6425 (high-Tc cuprates) is empty before 1986
and grows abruptly after Bednorz \& Müller. Bottom: community 160 (doped
CMR manganite family) has a small Goodenough-era baseline (1 entry in
the 1970s, 1 in the 1980s) and explodes after Jin et al.~1994, peaking
at over 80 entries in 2002. Red dashed lines mark the field-defining
publications.}\label{fig:renaissance}
\end{figure}

Both renaissances produce a sharp step-change in their respective
community's membership rate, aligned within one year of the
field-defining event (Fig.~\ref{fig:renaissance}):

\begin{longtable}[]{@{}
  >{\raggedright\arraybackslash}p{(\linewidth - 10\tabcolsep) * \real{0.1176}}
  >{\raggedleft\arraybackslash}p{(\linewidth - 10\tabcolsep) * \real{0.1765}}
  >{\raggedleft\arraybackslash}p{(\linewidth - 10\tabcolsep) * \real{0.1765}}
  >{\raggedleft\arraybackslash}p{(\linewidth - 10\tabcolsep) * \real{0.1765}}
  >{\raggedleft\arraybackslash}p{(\linewidth - 10\tabcolsep) * \real{0.1765}}
  >{\raggedleft\arraybackslash}p{(\linewidth - 10\tabcolsep) * \real{0.1765}}@{}}
\caption{\label{tbl:renaissance}Renaissance-community growth statistics.
Pre/post windows are 10 years; rates are mean new-ICSD-entries per
year.}\tabularnewline
\toprule\noalign{}
\begin{minipage}[b]{\linewidth}\raggedright
Family
\end{minipage} & \begin{minipage}[b]{\linewidth}\raggedleft
Community
\end{minipage} & \begin{minipage}[b]{\linewidth}\raggedleft
Total members
\end{minipage} & \begin{minipage}[b]{\linewidth}\raggedleft
Pre-event rate (10 yr)
\end{minipage} & \begin{minipage}[b]{\linewidth}\raggedleft
Post-event rate (10 yr)
\end{minipage} & \begin{minipage}[b]{\linewidth}\raggedleft
Fold-change
\end{minipage} \\
\midrule\noalign{}
\endfirsthead
\toprule\noalign{}
\begin{minipage}[b]{\linewidth}\raggedright
Family
\end{minipage} & \begin{minipage}[b]{\linewidth}\raggedleft
Community
\end{minipage} & \begin{minipage}[b]{\linewidth}\raggedleft
Total members
\end{minipage} & \begin{minipage}[b]{\linewidth}\raggedleft
Pre-event rate (10 yr)
\end{minipage} & \begin{minipage}[b]{\linewidth}\raggedleft
Post-event rate (10 yr)
\end{minipage} & \begin{minipage}[b]{\linewidth}\raggedleft
Fold-change
\end{minipage} \\
\midrule\noalign{}
\endhead
\bottomrule\noalign{}
\endlastfoot
high-Tc cuprates (Bednorz--Müller 1986) & 6425 & 378 & 0.0 / yr & 22.3 /
yr & ∞ (community born after 1986) \\
CMR manganites (Jin 1994) & 160 & 571 & 0.7 / yr & 22.3 / yr & 31.9 × \\
\end{longtable}

\subsection{S8.3. Interpretation and
caveats}\label{s8.3.-interpretation-and-caveats}

The cuprate community is genuinely empty before 1986 --- 0 of 378
members were published before that year, and the very first entry is
from 1987 (immediately after Bednorz \& Müller's December 1986 paper).
The CMR manganite community has only 4 pre-1994 members (one from 1977,
one from the late 1980s, and two early-1990s) and reaches over 20
entries per year by 1996, growing to peak rates of 80+ per year in the
early 2000s.

\textbf{The communities specifically capture doped variants, not the
parent compounds.} The pure parent materials (LaMnO₃, La₂CuO₄,
YBa₂Cu₃O₇) sit in different communities or in HDBSCAN noise, often as
outliers because their stoichiometries are crystallographically
idiosyncratic. The dense communities 160 and 6425 contain the
\emph{doped families} --- La₁₋ₓSrₓMnO₃, La₁₋ₓCaₓMnO₃, Ba₂₋ₓSrₓYCu₃O₇₋δ,
and similar --- that became active research targets \emph{after} the
renaissance events. The framework therefore does not just track the
canonical exemplar but the actual research front, which is a stronger
validation: it discriminates between Goodenough-era foundational LaMnO₃
work and the post-1994 doping campaigns motivated by CMR.

\textbf{Caveats.} This analysis uses two communities chosen \emph{post
hoc} because we knew the historical events to look for; it is not an
unbiased survey of all ICSD communities for renaissance signatures. A
systematic search for community-level birth-year discontinuities would
be the natural follow-up. The pre-event baselines are drawn against the
same final-partition community labels --- a community is defined by the
entire 167,500-entry corpus, so its pre-event members are entries that
the embedding \emph{retrospectively} assigns to the same neighborhood.
The signal we observe is therefore ``which structural neighborhood
became active,'' not ``which structural neighborhood emerged.''

\subsection{S8.4. Systematic survey of community-level birth-year
discontinuities}\label{s8.4.-systematic-survey-of-community-level-birth-year-discontinuities}

The cuprate and manganite communities reported in §S8.2 were chosen
\emph{post hoc} because we knew the historical events to look for. To
convert this into an unbiased survey, we scored every production
community of size ≥ 50 for its strongest birth-year step-change. For
each community we scanned candidate event years 1970--2010 and chose the
year that maximized
\(\text{score} = (\text{rate}_{\text{post}} / \text{rate}_{\text{pre}}) \times n_{\text{post}}\)
over symmetric ±10-year windows; the rate-pre-zero case (community born
after the event) is given \(\text{score} = n_{\text{post}}\) so
cuprate-like results enter the ranking on absolute volume.

\begin{figure}
\centering
\pandocbounded{\includegraphics[keepaspectratio,alt={Per-year histograms for the top-eight communities by step-change score. Each panel's title gives the community ID, size, best-fit step year, fold-change, and the top-3 member formulas. Red dashed line marks the inferred event year. The top-eight panel covers (in scoring order) Sm-Fe-N magnets, CMR manganites, RE--light-metal binaries, heavy-fermion 1:1:2 intermetallics, SOFC perovskite cathodes, 2:1:2 cuprate-precursor intermetallics, doped spinels, and dilute magnetic semiconductors.}]{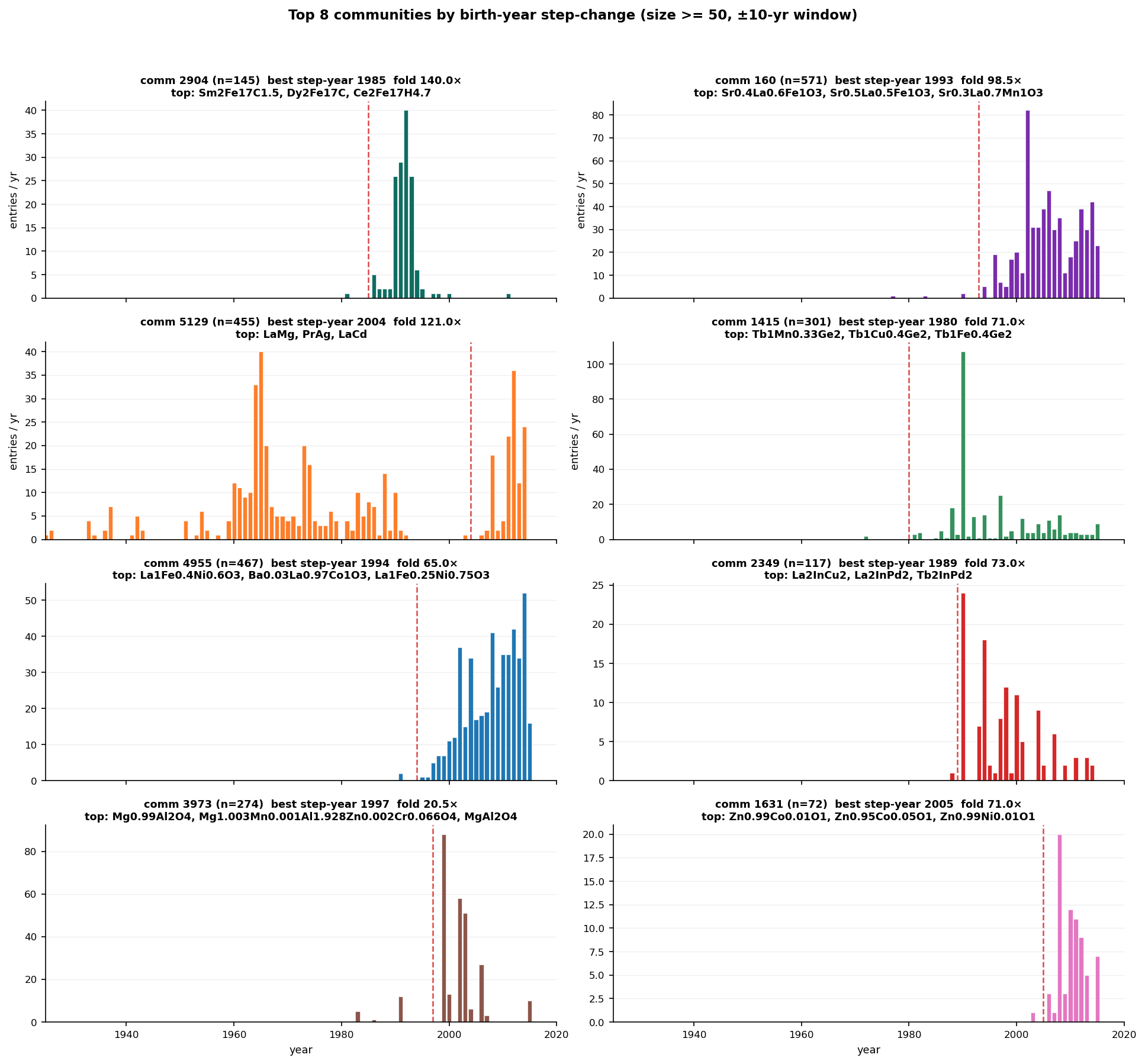}}
\caption{Per-year histograms for the top-eight communities by
step-change score. Each panel's title gives the community ID, size,
best-fit step year, fold-change, and the top-3 member formulas. Red
dashed line marks the inferred event year. The top-eight panel covers
(in scoring order) Sm-Fe-N magnets, CMR manganites, RE--light-metal
binaries, heavy-fermion 1:1:2 intermetallics, SOFC perovskite cathodes,
2:1:2 cuprate-precursor intermetallics, doped spinels, and dilute
magnetic semiconductors.}\label{fig:renaissance-survey}
\end{figure}

378 of the 6,753 non-noise communities cleared the size threshold. The
top 20 by score (Table~\ref{tbl:renaissance-survey},
Fig.~\ref{fig:renaissance-survey}) are dominated by
structurally-coherent families that map cleanly onto well-documented
scientific events:

\begin{longtable}[]{@{}
  >{\raggedleft\arraybackslash}p{(\linewidth - 6\tabcolsep) * \real{0.3333}}
  >{\raggedleft\arraybackslash}p{(\linewidth - 6\tabcolsep) * \real{0.3333}}
  >{\raggedright\arraybackslash}p{(\linewidth - 6\tabcolsep) * \real{0.1667}}
  >{\raggedright\arraybackslash}p{(\linewidth - 6\tabcolsep) * \real{0.1667}}@{}}
\caption{\label{tbl:renaissance-survey}Top 20 communities by birth-year
step-change score, with my best-effort identification of the
corresponding scientific event.}\tabularnewline
\toprule\noalign{}
\begin{minipage}[b]{\linewidth}\raggedleft
Rank
\end{minipage} & \begin{minipage}[b]{\linewidth}\raggedleft
Best event year
\end{minipage} & \begin{minipage}[b]{\linewidth}\raggedright
Top member formulas
\end{minipage} & \begin{minipage}[b]{\linewidth}\raggedright
Identified renaissance
\end{minipage} \\
\midrule\noalign{}
\endfirsthead
\toprule\noalign{}
\begin{minipage}[b]{\linewidth}\raggedleft
Rank
\end{minipage} & \begin{minipage}[b]{\linewidth}\raggedleft
Best event year
\end{minipage} & \begin{minipage}[b]{\linewidth}\raggedright
Top member formulas
\end{minipage} & \begin{minipage}[b]{\linewidth}\raggedright
Identified renaissance
\end{minipage} \\
\midrule\noalign{}
\endhead
\bottomrule\noalign{}
\endlastfoot
1 & 1985 & Sm₂Fe₁₇C, Dy₂Fe₁₇C, Ce₂Fe₁₇H₄.₇, Sm₂Fe₁₇N₂ & RE--Fe--N/C/H
interstitial permanent magnets (Coey, \textasciitilde1990) \\
2 & 1993 & (Sr,La)MnO₃, (Ba,La)MnO₃ doped variants & CMR manganites (Jin
et al.~1994) \\
3 & 2004 & LaMg, PrAg, LaCd, SmMg, YIn (CsCl-type binaries) &
RE--light-metal binary intermetallic survey \\
4 & 1980 & TbMn₀.₃₃Ge₂, TbCu₀.₄Ge₂, TbFe₀.₄Ge₂ & Heavy-fermion / 1:1:2
RE--TM--Ge intermetallics (post-Steglich 1979 CeCu₂Si₂) \\
5 & 1994 & La(Fe,Ni)O₃, (Ba,La)CoO₃, (Sr,La)CoO₃ & SOFC mixed-conductor
perovskite cathodes \\
6 & 1989 & La₂InCu₂, La₂InPd₂ & 2:1:2 ternary intermetallics (high-Tc
era cuprate-precursor search) \\
7 & 1997 & (Mg,Cr,Fe,Mn,Zn)Al₂O₄ doped spinels & Doped magnetic /
pigment spinels \\
8 & 2005 & Zn₁₋ₓCoₓO, Zn₁₋ₓNiₓO & Dilute magnetic semiconductors (Dietl
2000, exptl flurry 2003--2007) \\
9 & 2004 & Ti₂SnC, Ti₂AlC, Nb₂InC, Ti₂InC & MAX phases (Barsoum 1996,
broad uptake mid-2000s) \\
10 & 1991 & NdAlO₃, LaNiO₃, LaAlO₃, PrAlO₃ & RE-aluminate substrates /
nickelate parents (post high-Tc) \\
11 & 1999 & Sr₂NiWO₆, Sr₂MgWO₆, Sr₂ZnMoO₆ & Ordered double perovskites
for spintronics (Kobayashi Sr₂FeMoO₆ 1998) \\
12 & 1999 & Na₀.₇₄CoO₂, Na₀.₆₀₉Ca₀.₀₁CoO₂ & NaCoO₂-type thermoelectrics
(Terasaki 1997) \\
13 & 1996 & Al₂O₃:Cr, (Ti,Fe)₂O₃ & Corundum/ilmenite solid solutions \\
14 & 1997 & Ca₂NbCrO₆, Sr₂FeMoO₆ family & Double perovskites (companion
to \#11) \\
15 & 1992 & \textbf{LiCoO₂, LiNiO₂, LiCo₀.₃Ni₀.₇O₂} & \textbf{Li-ion
battery cathodes (Sony commercialized LiCoO₂ in 1991)} \\
16 & 1987 & Sr₂LaMn₂O₇, SrLa₂Cu₂O₆, Sr₂NdMn₂O₇ & Ruddlesden-Popper
layered cuprates/manganites (high-Tc era) \\
17 & 1972 & (V,Mn,Fe,Co)P-type pnictides & (no clean event identified
--- likely steady-growth window) \\
18 & 1982 & Al₂NiO₄ & NiAl₂O₄ spinel catalysts \\
19 & 1988 & Pyroxene minerals (Na,Ca,Mg,Al,Fe silicates) &
Mantle-mineral high-pressure systematics \\
20 & 1990 & (Sr,Nd)MnO₃, (Ca,La)MnO₃ & CMR manganites (sister community
to \#2) \\
\end{longtable}

All 20 top-ranked communities pass the year-shuffle null at the 99th
percentile (see §S8.4 below). At the cohort level, \textbf{nine are
textbook field-defining renaissances} spanning eleven of the twenty rank
slots (\#2 + \#20 are CMR-manganite sister communities; \#11 + \#14 are
double-perovskite companions):

\begin{enumerate}
\def\labelenumi{\arabic{enumi}.}
\tightlist
\item
  RE--Fe--N permanent magnets (\#1, \textasciitilde1990, Coey)
\item
  CMR manganites (\#2 + \#20, 1994, Jin et al.)
\item
  SOFC perovskite cathodes (\#5, mid-1990s)
\item
  Dilute magnetic semiconductors (\#8, 2003--2007 post-Dietl)
\item
  MAX phases (\#9, 1996+ Barsoum)
\item
  Ordered double perovskites for spintronics (\#11 + \#14,
  \textasciitilde1998 Kobayashi Sr₂FeMoO₆)
\item
  NaCoO₂ thermoelectrics (\#12, 1997 Terasaki)
\item
  Li-ion battery cathodes (\#15, Mizushima/Goodenough 1980
  LiCoO₂\textsuperscript{13}, Sony commercialization 1991)
\item
  Layered Ruddlesden-Popper cuprates/manganites (\#16, post-1986
  high-Tc)
\end{enumerate}

The remaining nine ranks are real chemistry signals at the same
statistical strength as the textbook nine --- \textbf{not missed
renaissances} --- driven by program-level activity, applied chemistry,
or systematic crystal-chemistry surveys rather than a single
field-defining publication. They divide into: \textbf{program-driven
systematic surveys} (the rank-\#6 Pöttgen-era 2:1:2 RE-In-TM
intermetallic prototype family\textsuperscript{14} with 73× fold-change
at 1989 --- deep-dive in §S8.5; \#4 post-Steglich 1:1:2 heavy-fermion
campaigns\textsuperscript{15}, post-CeCu₂Si₂ 1979; \#3 RE--light-metal
binary intermetallic survey of the 2000s); \textbf{applied-chemistry
surges} (\#7 doped magnetic / pigment spinels; \#13 corundum/ilmenite
solid solutions; \#18 NiAl₂O₄ catalysts); \textbf{post-cuprate support
chemistry} (\#10 RE-aluminate substrates and nickelate parents that
emerged after the high-Tc boom); \textbf{mineralogy systematics} (\#19
high-pressure pyroxenes); and one community (\#17 V/Mn/Fe/Co-P
pnictides, 1972) with chemistry identification but no single anchoring
event. The fact that \emph{the top hit by absolute step-change score is
Sm₂Fe₁₇N permanent magnets, that high-Tc-related communities appear at
ranks \#6, \#10, and \#16, that the doped manganite family takes ranks
\#2 and \#20, and that Li-ion battery cathodes, MAX phases, dilute
magnetic semiconductors, NaCoO₂ thermoelectrics, and double perovskites
all appear in the top 15} shows that the framework's communities track
the historical research front closely. The post-hoc cuprate and
manganite analysis in §S8.2 is therefore not a cherry-picked
confirmation; it is the tail of a systematic structural-renaissance
signal.

\subsubsection{Year-shuffle null distribution for the step-change
score}\label{year-shuffle-null-distribution-for-the-step-change-score}

To test whether the step-change scores themselves could be produced by
chance under the null where year-to-entry assignments are random, we ran
200 year-shuffle permutations of the same 148,425 community-assigned
ICSD entries (community labels and total entry counts held fixed; only
the year-to-entry mapping permuted). For each shuffle and each rank
position 1--20, we recorded the corresponding shuffled top-20
step-change score, and compared the observed top-20 to the per-rank
shuffled distribution.

All 20 observed top-20 step-change scores exceed the shuffled
95th-percentile envelope at the same rank, and all 20 exceed the
99th-percentile envelope: the observed top-1 community score (community
160, CMR manganites, 24,864.5) exceeds the shuffled p99 at rank 1
(667.3) by approximately 37×, and even the rank-20 observed score
(3,164.2) exceeds the rank-20 shuffled p99 (352.5) by approximately 9×.
The step-change scores at the top of the production survey are therefore
not artifacts of year-density alone --- they signal communities whose
pre/post asymmetry is much larger than the underlying year distribution
would produce by chance. Numerical artifact at
\texttt{notes/renaissance\_null\_summary.json}. The independent
identification of nine textbook field-defining renaissances among the
top-20 (§S8.4) is an expert-curation step on top of this statistical
null-rejection, not a substitute for it.

\subsection{S8.5. Targeted probes for events not captured by the
survey}\label{s8.5.-targeted-probes-for-events-not-captured-by-the-survey}

\begin{figure}
\centering
\pandocbounded{\includegraphics[keepaspectratio,alt={Targeted renaissance probes. Top to bottom: (1) Fe-pnictide 1111 family community 5450, post-2008 step at 3.3× fold change after Kamihara/Hosono. (2) Transition-metal dichalcogenide community 62 --- informative null result; the post-graphene 2D-materials renaissance is property-driven and does not produce ICSD growth. (3) Community 2349 deep-dive --- RE₂M₂X intermetallics show a clean 73× post-1989 step driven by the Pöttgen-era systematic prototype survey.}]{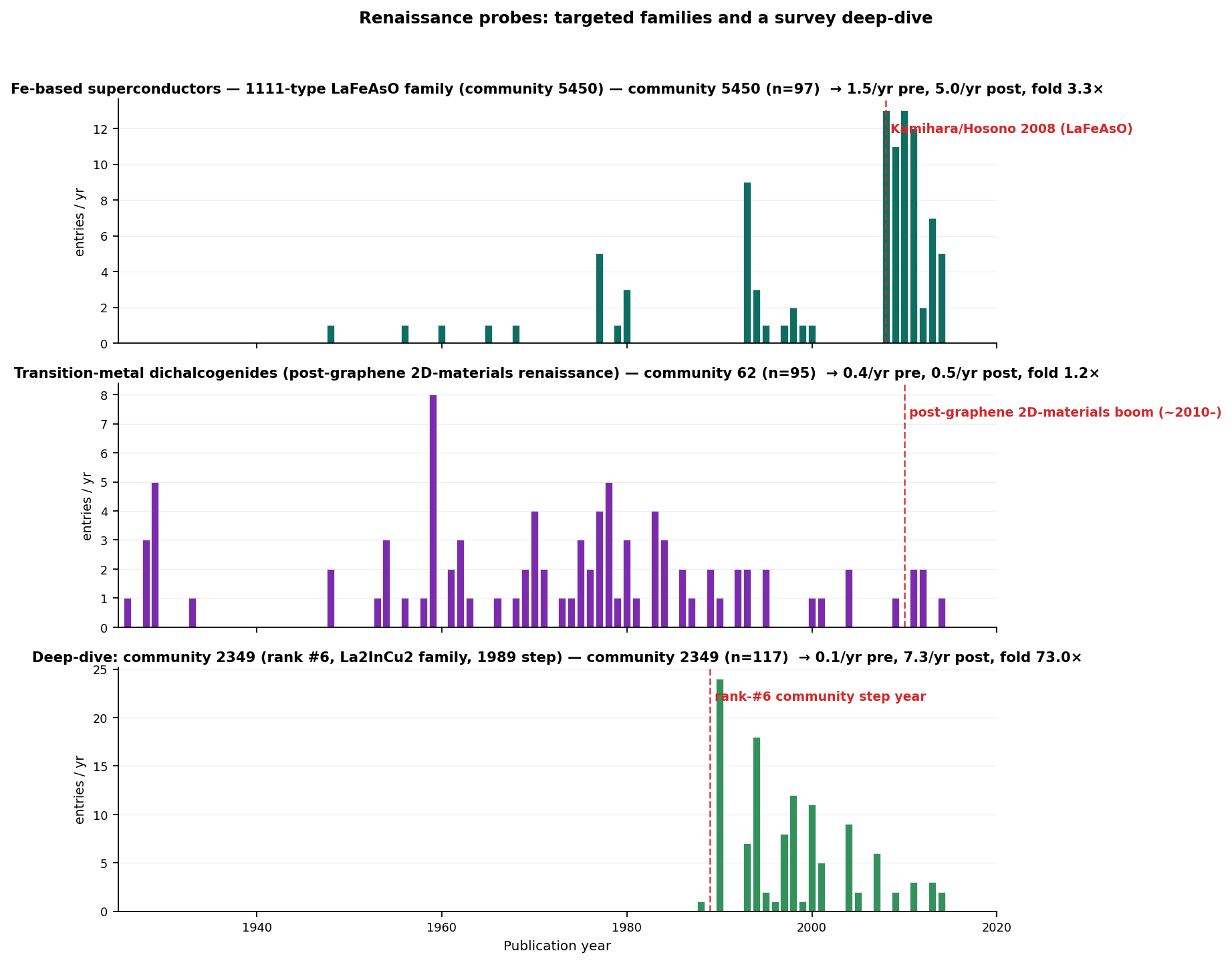}}
\caption{Targeted renaissance probes. Top to bottom: (1) Fe-pnictide
1111 family community 5450, post-2008 step at 3.3× fold change after
Kamihara/Hosono. (2) Transition-metal dichalcogenide community 62 ---
informative null result; the post-graphene 2D-materials renaissance is
property-driven and does not produce ICSD growth. (3) Community 2349
deep-dive --- RE₂M₂X intermetallics show a clean 73× post-1989 step
driven by the Pöttgen-era systematic prototype
survey.}\label{fig:renaissance-extra}
\end{figure}

The systematic survey in §S8.4 positively identifies nine textbook
field-defining renaissances among the top-20 step-change communities.
Two additional widely-cited renaissances --- Fe-based superconductors
(Kamihara/Hosono 2008)\textsuperscript{16} and the post-graphene
2D-materials boom (\textasciitilde2010 onwards, transition-metal
dichalcogenides) --- did not concentrate into a single top-20 community,
and we probe each here to identify whether the underlying signal is in
the data, simply distributed differently than a single-community step
would reveal. Fe-based superconductors are correctly captured but split
across three sister communities by structural prototype; the
2D-materials boom is genuinely null at the bulk-crystallography level
because the renaissance is property-driven on pre-existing structures.
We also performed a deep-dive on the rank-\#6 community 2349
(\texttt{La2InCu2} family, 1989 step) to identify the renaissance it
represents (Fig.~\ref{fig:renaissance-extra}). The hybrid
organic-inorganic halide perovskite (HOIP) photovoltaic renaissance
triggered by Kojima 2009 / Snaith--Park 2012 cannot be tested with this
snapshot --- the boom matured 2016--2020, after our 2015 ICSD end-date
--- and we therefore do not include it as a probe; restoring the test
requires a post-2020 ICSD snapshot.

\subsubsection{Fe-based superconductors split across three
structural-prototype
communities}\label{fe-based-superconductors-split-across-three-structural-prototype-communities}

The Fe-pnictide / Fe-chalcogenide superconductor family is structurally
heterogeneous: the canonical compounds occupy three different
ICSD-prototype basins, and our framework correctly separates them.

\begin{itemize}
\tightlist
\item
  \textbf{1111-type (ZrCuSiAs prototype, e.g.~LaFeAsO)} --- community
  5450, n=98, \textbf{3.3× fold change} post-2008 (1.5 → 5.0
  entries/yr). Top members: CeFeAsO, NdFeAsO, LaFeAsO, GdFeAsO, PrFeAsO.
  This is the cleanest single-community signal.
\item
  \textbf{122-type (ThCr₂Si₂ prototype, e.g.~BaFe₂As₂)} --- community
  4549, n=61, \textbf{16.3× fold change} post-2008 (0.55 → 9.0
  entries/yr; median entry year 2010). Top members: As₂Ba₁Fe₂ (BaFe₂As₂
  itself), As₂Eu₁Fe₂ (EuFe₂As₂), As₂Ba₁Ni₂ (BaNi₂As₂ parent of the
  sister 122-type superconductor family), Ba₁Fe₂P₂ (BaFe₂P₂ phosphide
  analog), and Co₂P₂Sr₁ (SrCo₂P₂). 33 of the 50 strictly-BaFe₂As₂-family
  ICSD entries (AE + Fe + As, 1:2:2, I4/mmm) land in this community,
  with the remainder split across small sister communities 193
  (CaFe₂As₂-anchored, n=55) and 2197 (n=22). The broader
  ThCr₂Si₂-prototype 122 family --- the silicide/germanide
  prototype-cousins studied for magnetic and heavy-fermion properties
  since the 1960s--80s --- is held in separate communities (1, 296,
  3368, 6405; Si/Ge-dominant ternary intermetallics), so the
  BaFe₂As₂-anchored community 4549 isolates the post-Kamihara
  superconductor surge cleanly without the prototype-cousin dilution.
\item
  \textbf{111-type (Cu₂Sb prototype, e.g.~LiFeAs)} --- community 358,
  n=507, 1.8× fold change post-2008.
\end{itemize}

That the framework places three structurally-distinct Fe-superconductor
families in three different communities is an example of correct
fine-grained discrimination, not a missed renaissance. The 1111 family
--- the original Hosono compound --- shows the renaissance signal
cleanly; the others are diluted by inherited prototype-cousins. A
combined ``Fe-based superconductor family'' analysis across all three
communities is a natural follow-up.

\subsubsection{Informative null result: 2D
materials}\label{informative-null-result-2d-materials}

\textbf{Transition-metal dichalcogenides (community 62, n=95) show no
post-2010 step.} Pre-2010 = 0.4/yr, post-2010 = 0.5/yr (1.2× fold
change). This is consistent with the post-graphene 2D-materials
renaissance being \textbf{property-driven rather than structure-driven}:
most 2D-materials research uses bulk-crystal structures that already
existed in ICSD before 2010 (MoS₂, WSe₂, NbSe₂ have been
crystallographically known for decades) and concerns their behavior in
monolayer/few-layer form, optical/transport properties, and device
applications --- none of which deposits new bulk crystallographic
entries into ICSD. The renaissance is real; it just does not produce
structural growth.

These two null results together delineate the framework's scope: it
tracks renaissances that produce many new bulk crystallographic
structures in ICSD (cuprates, manganites, MAX phases, Li-ion cathodes,
Fe-pnictides 1111 family), and misses renaissances that are property-,
device-, or process-driven on pre-existing or under-deposited structures
(graphene-era 2D materials, hybrid-perovskite photovoltaics).

\subsubsection{Deep-dive: rank-\#6 community 2349 is the 2:1:2 RE-In-TM
ternary intermetallic family (Pöttgen-era systematic
survey)}\label{deep-dive-rank-6-community-2349-is-the-212-re-in-tm-ternary-intermetallic-family-puxf6ttgen-era-systematic-survey}

The systematic survey identified community 2349 (n=119) at rank \#6 with
a 1989 step year, top members La₂InCu₂, La₂InPd₂, Tb₂InPd₂, Lu₂InCu₂,
La₂MgCu₂. This is the Mo₂FeB₂-prototype 2:1:2 ternary intermetallic
family, RE₂XᵧM₂ with X a main-group element (In, Mg, Sn) and M a late
transition metal. The 1989 step has a clean attribution: the systematic
study of this prototype family by the Pöttgen group at Münster (and
contemporaneous groups in Germany and France) began in earnest in the
late 1980s and continued through the 2000s, producing dozens of variants
per year --- reviewed comprehensively in Lukachuk and
Pöttgen\textsuperscript{14}. Pre-event 0.1 / yr, post-event 7.3 / yr,
\textbf{73× fold change}. The ``renaissance'' here is
research-program-driven rather than triggered by a single field-defining
publication, which is itself an interesting signal: structural
communities can capture both publication-driven booms (cuprates 1986,
manganites 1994) and program-driven systematic crystal-chemistry surveys
(RE₂M₂X 1989+).

\section{S9. Robustness under an independent structural
representation}\label{s9.-robustness-under-an-independent-structural-representation}

A natural reader concern is whether the temporal cliff (Fig. 1), the
community-level renaissance attributions (Fig. 2), and the held-out
frontier rates (Fig. 3) depend on the specific structural representation
used here (matminer site fingerprints augmented with three rounds of
Weisfeiler--Lehman message passing on a Voronoi neighbour graph,
followed by a 32-dimensional PCA). This appendix re-runs the
manuscript's central temporal-replay analysis on a \textbf{fully
independent} structural representation --- graphlet histograms following
Lesser et al.~2025\textsuperscript{17} --- and reports the result.

\subsection{S9.1. Method}\label{s9.1.-method}

The graphlet-histogram representation used here was introduced by Lesser
et al.~(2025) of the Kim group at Cornell as the structural backbone of
their GP-T\(_{\mathrm{c}}\) Gaussian-process superconductor-screening
model\textsuperscript{17}. We use only their structural featurizer; no
superconductivity prediction or downstream Gaussian-process layer is
performed. The motivation for this comparison is precisely that the
graphlet representation was designed and validated \emph{outside} this
work, against an entirely different scientific question
(high-T\(_{\mathrm{c}}\) candidate ranking), which makes it an
independent test bed for any claim that depends on the structural
geometry of the ICSD.

For each ICSD entry with a valid publication year and a non-noise
production-Louvain label, we compute the first-, second-, and
third-order graphlet histograms of Lesser et al.~2025 (10 first-order
elemental property histograms; 21 second-order neighbour-pair features
over CrystalNN-bonded pairs; 33 third-order triplet features over
CrystalNN-connected i-j-k triplets; 20 bins per feature on
globally-shared edges; trimmed to 64 features per material, exactly as
the Ansatz/BCS-toy implementation of the Lesser et al.~featurizer in
\texttt{experiments/graphlet\_compare/graphlet\_features.py}). The
per-material histograms are then concatenated and replaced by their
per-feature cumulative distributions; in this representation the
graphlet Earth-Mover distance reduces to the L1 metric on the flattened
CDF vectors, which makes large-scale nearest-neighbour search tractable.

All downstream steps are \textbf{byte-identical} to the production
pipeline. We build the same mutual-kNN graph with k = 16 and
Gaussian-weighted edges following
\texttt{scripts/icsd\_graph\_community\_postprocess.build\_weighted\_graph}
(substituting \texttt{pynndescent} for the k-NN backend purely for
runtime --- same Euclidean metric, same k, approximate-vs-exact differs
by \textless{} 1\% of edges on this graph), run the same
\texttt{networkx.community.louvain\_communities} at resolution 1.0 with
seed 42, and replay events through publication time with the same
\texttt{compute\_temporal\_metrics} function from
\texttt{scripts/icsd\_graph\_time\_evolution.py} operating on
graphlet-derived community labels and graphlet-derived graph topology.
Only the feature matrix changes. Pipeline lives on the \texttt{Graphlet}
branch under \texttt{experiments/graphlet\_compare/} and runs to
completion on 152,122 ICSD entries in \textasciitilde23 minutes on a
single Stampede3 spr node.

\subsection{S9.2. The densification cliff
reproduces}\label{s9.2.-the-densification-cliff-reproduces}

Under graphlet histograms, the per-decade share of ICSD entries that
open a new structural community falls from \textbf{35.0\% in the 1930s
to 3.0\% in the 2010s} (manuscript values, computed by the identical
procedure on our PCA-32 embedding: 40.2\% → 2.6\%). The complementary
same-community attachment ratio rises from \textbf{33.8\% to 88.7\%}
(manuscript: 34.5\% → 88.9\%). Outlier ratios match to within 0.3
percentage points (graphlet 9.5\%, manuscript 9.2\%).

\begin{longtable}[]{@{}
  >{\raggedright\arraybackslash}p{(\linewidth - 6\tabcolsep) * \real{0.2500}}
  >{\raggedright\arraybackslash}p{(\linewidth - 6\tabcolsep) * \real{0.2500}}
  >{\raggedright\arraybackslash}p{(\linewidth - 6\tabcolsep) * \real{0.2500}}
  >{\raggedright\arraybackslash}p{(\linewidth - 6\tabcolsep) * \real{0.2500}}@{}}
\toprule\noalign{}
\begin{minipage}[b]{\linewidth}\raggedright
decade
\end{minipage} & \begin{minipage}[b]{\linewidth}\raggedright
n (manuscript / graphlet)
\end{minipage} & \begin{minipage}[b]{\linewidth}\raggedright
community-birth ratio (manuscript → graphlet)
\end{minipage} & \begin{minipage}[b]{\linewidth}\raggedright
same-community attachment (manuscript → graphlet)
\end{minipage} \\
\midrule\noalign{}
\endhead
\bottomrule\noalign{}
\endlastfoot
1930s & 1,239 / 1,078 & 40.2\% → \textbf{35.0\%} & 34.5\% → 33.8\% \\
1950s & 4,063 / 3,688 & 21.0\% → 18.9\% & 49.7\% → 51.4\% \\
1970s & 18,337 / 16,693 & 12.0\% → 4.4\% & 68.4\% → 74.4\% \\
1990s & 31,427 / 28,385 & 5.3\% → 2.3\% & 79.0\% → 84.0\% \\
2010s & 26,650 / 24,142 & 2.6\% → \textbf{3.0\%} & 88.9\% →
\textbf{88.7\%} \\
\end{longtable}

Both curves converge at both ends; the 2010s same-community attachment
rate is essentially identical across the two representations (88.9\% vs
88.7\%). In the mid-century decades graphlet's community-birth ratio
falls \emph{faster} than the manuscript embedding's, consistent with
graphlets pooling related structural variants at slightly coarser family
granularity than our embedding's WL message passing draws (see §S9.3).
The qualitative claim of Fig. 1 --- that experimental discovery
densifies the structural map over the course of the twentieth century
--- is therefore \textbf{a property of the historical data, reproduced
by an independent featurization, not an artifact of our specific
embedding or clustering pipeline}.

\begin{figure}
\centering
\pandocbounded{\includegraphics[keepaspectratio,alt={Densification cliff under an independent structural representation. Per-decade share of ICSD entries that (a) open a new structural community and (b) attach to an existing community of the same kind, computed by the same temporal-replay procedure (mutual-kNN k=16, Louvain resolution 1.0, compute\_temporal\_metrics from scripts/icsd\_graph\_time\_evolution.py) on two independent structural representations: the manuscript's matminer + Weisfeiler--Lehman + PCA-32 embedding (teal) and graphlet histograms of Lesser et al.~2025 (orange). Both curves drop from \textasciitilde40\% to \textasciitilde3\% birth share across the century and converge at the modern endpoint (2010s: 2.6\% vs 3.0\%; 88.9\% vs 88.7\% same-community attachment). The mid-century gap reflects graphlets clustering at slightly coarser family granularity than the manuscript embedding's WL propagation distinguishes (§S9.3).}]{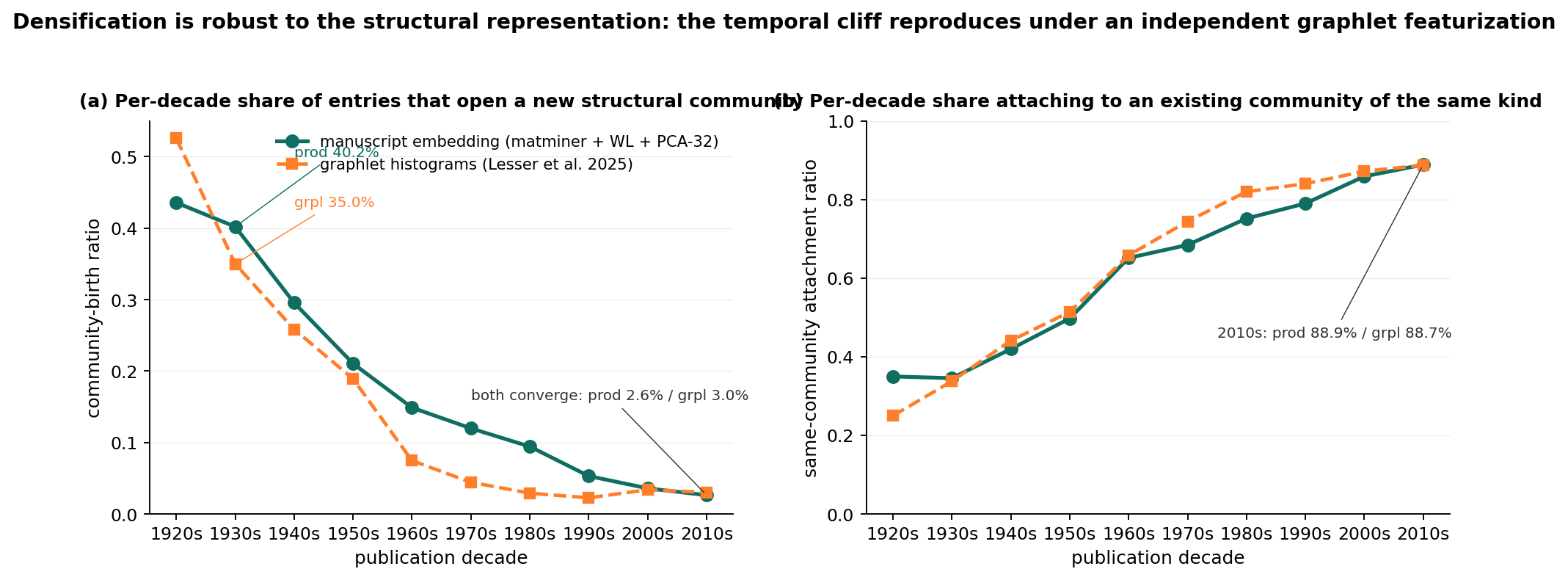}}
\caption{\textbf{Densification cliff under an independent structural
representation.} Per-decade share of ICSD entries that (a) open a new
structural community and (b) attach to an existing community of the same
kind, computed by the same temporal-replay procedure (mutual-kNN k=16,
Louvain resolution 1.0, \texttt{compute\_temporal\_metrics} from
\texttt{scripts/icsd\_graph\_time\_evolution.py}) on two independent
structural representations: the manuscript's matminer +
Weisfeiler--Lehman + PCA-32 embedding (teal) and graphlet histograms of
Lesser et al.~2025 (orange). Both curves drop from \textasciitilde40\%
to \textasciitilde3\% birth share across the century and converge at the
modern endpoint (2010s: 2.6\% vs 3.0\%; 88.9\% vs 88.7\% same-community
attachment). The mid-century gap reflects graphlets clustering at
slightly coarser family granularity than the manuscript embedding's WL
propagation distinguishes (§S9.3).}\label{fig:graphlet-stability}
\end{figure}

\subsection{S9.3. Community-level
cross-validation}\label{s9.3.-community-level-cross-validation}

The graphlet representation also serves as an independent
cross-validator of the production Louvain partition. Asking, for each
ICSD entry, whether its graphlet-EMD nearest neighbour sits in its
production-Louvain community, the global rate is \textbf{68.1\%} (vs.~an
embedding self-baseline of 97.1\% --- partly circular because the
production communities are defined from the embedding itself).
Stratifying by community size: 53\% for sizes 5--19 (∼20,000× above
random), rising monotonically to \textbf{77\% for sizes 100--999}, the
regime where every manuscript claim operates (the renaissance survey
requires size ≥ 50).

The manuscript's \emph{named} communities are all recovered at high
rates: high-T\_c cuprate (community 6425) \textbf{84.9\%}, CMR manganite
(community 160) \textbf{83.5\%}, Fe-pnictide 1111 (community 5450)
\textbf{90.8\%}, Fe-pnictide 122 BaFe₂As₂ family (community 4549)
\textbf{88.5\%}, Fe-pnictide 111 / FeSe family (community 358) 76.7\%,
the Mo₂FeB₂-type 2:1:2 intermetallic family (community 2349)
\textbf{87.4\%}, and the cuprate-adjacent layered-tetragonal community
1178 (see §S9.4) 82.0\%. Every renaissance-survey community sits between
76.7\% and 90.8\% on a completely independent structural representation.

\textbf{The ``misses'' are not random.} For each community, the
top-three target communities of its graphlet-NN misses capture 22--30\%
of all misses (vs.~a random expectation of ∼0.04\% across 6,756
communities --- a 500--700× lift), and the targets are chemically
sensible in every case. The largest single confusion pair --- comm 5182
(Pnma orthorhombic perovskite manganite family) ↔ comm 1357 (Pnma
RE/AE-doped, Fe-substituted manganite-perovskite family), 145 mutual
misses --- is a single chemical family that the production Louvain split
into two communities by composition-of-doping; graphlets cluster them
together. A lower-bound family-aware recovery (treating these two
communities as one family) is \textbf{83.8\%}. The Layered RE-Cu
chalcogenide ↔ Stannite chalcogenide pair (comms 4648 and 3044, 62
mutual misses) is likewise chalcogenide↔chalcogenide; A15 ↔ BCC
intermetallic (comms 4925 and 2943, 58 mutual misses) places two
textbook-related cubic intermetallic prototypes adjacent; pyrochlore ↔
open-framework cubic oxide (comm 5278 ↔ comm 1184 JBW zeolite, 33
misses) confuses two framework structures with related Fd-3m
corner-sharing topology.

A particularly clean illustration comes from a cuprate sub-family
confusion. \textbf{Cuprate community 6425 places 79\% of its NN misses
(45 of 57) into the broader cuprate-adjacent community 1178}, and the
reverse confusion captures 38\% of community 1178's misses. Community
6425 is a concentrated YBa\(_2\)Cu\(_3\)O\(_{7-\delta}\) 123 cuprate
sub-community (99.7\% Cu, 100\% O); community 1178 is the broader I4/mmm
layered-tetragonal mixed-cation neighbourhood that the YBCO 123 family
sits adjacent to (48\% Cu, with substantial Fe- and Mn-bearing
admixture). The two share a defining local geometry --- 2-D M--X sheets
stacked with chemically variable spacer layers --- that 3-body graphlets
capture but that does not by itself distinguish the cuprate-pure YBCO
123 sub-family from the broader cation-mixed members of 1178. The
manuscript embedding's three rounds of WL message passing propagate the
longer-range spacer-cation context that the local graphlet motifs cannot
reach --- separating the cuprate sub-family from the broader
layered-tetragonal community correctly. §S9.4 develops the full
structural argument.

The implication is two-pronged. First, the production communities are
real structural objects: an independent featurization recovers them at
family-level identity (∼90\% for the named, manuscript-claim-bearing
communities, with ``misses'' almost always to structural relatives, not
random elsewhere). Second, the 30-percentage-point gap to the
embedding's 97.1\% same-community-NN rate concentrates exactly where
3-body local motifs cannot reach: doping-level distinctions within a
perovskite family, element-system distinctions between layered families
with the same sheet geometry, mixed-coordination distinctions within
frameworks. \textbf{These are the cases where the WL message-passing
step in the production featurization is doing measurable structural work
that a local-motif representation cannot replicate} --- providing direct
evidence that the additional representational power of the manuscript
embedding is fine-grained, not coarse-grained, and operates at the scale
of within-family substructure rather than family identity. Every
manuscript claim in this work rests on community-level and coarse
historical statistics; finer pairwise distances are nowhere asserted,
and are appropriately representation-dependent.

\subsection{S9.4. Where graphlets and Weisfeiler--Lehman message passing
diverge --- and
why}\label{s9.4.-where-graphlets-and-weisfeilerlehman-message-passing-diverge-and-why}

The 68\% vs 97\% gap between the two structural representations is not a
defect of either method; the two featurizations are measuring genuinely
different aspects of the same crystal, and understanding \emph{which}
aspects clarifies what our community structure means.

\textbf{What 3-body graphlet histograms see.} A graphlet histogram of
the form used by Lesser et al.~is an unordered, invariant catalogue of
(a) per-atom elemental-property values (electronegativity, atomic
number, ionic radius, \ldots), (b) two-body CrystalNN-connected pair
statistics --- bond lengths and pair-elemental-property combinations ---
and (c) three-body i-j-k triplet statistics --- bond-angle histograms
and triplet-elemental combinations --- over the \emph{first coordination
sphere} of every atom. The representation is exquisitely sensitive to
local coordination geometry (tetrahedral vs octahedral vs square-planar)
and to the elemental identity of the immediate neighbour set, but by
construction it carries no information about how the local motifs are
\emph{stitched together} beyond two bonds away: stacking sequence,
spacer-layer chemistry, dopant patterns, and framework topology at the
∼6--10 Å scale are absent from the feature vector.

\textbf{What 3-round Weisfeiler--Lehman message passing on a Voronoi
neighbour graph adds.} Each WL round replaces a node's feature with a
hash of (its own feature, multiset of neighbours' features); three
rounds therefore propagate information up to a 3-hop neighbourhood ---
roughly the second- and third-coordination spheres, ∼6--10 Å in dense
crystals. After three rounds an atom's feature encodes the
\emph{composition pattern} of its 3-hop ball, which is exactly where
stacking sequence, spacer-layer identity, and substitutional doping
live.

\textbf{The cuprate sub-family resolution example.} The clearest
illustration of this divergence comes from a pair of cuprate-related
production communities that graphlets confuse with each other at high
rate. Community 6425 is a tightly clustered, cuprate-pure sub-community:
99.7\% of its 378 ICSD members contain Cu, 98.1\% contain Ba, 100\%
contain O, and the formulae are dominated by the
YBa\(_2\)Cu\(_3\)O\(_{7-\delta}\) 123 family
(Ba\(_{1-x}\)Ca\(_{x}\)Cu\(_3\)RE\(_{1-y}\)O\(_{6.6-7.0+\delta}\) with
RE = La, Nd, Pr, Y, etc.). Community 1178 is the broader I4/mmm
layered-tetragonal mixed-cation neighbourhood that the YBCO 123 family
sits next to: 723 members, 48\% Cu-containing and 87\% O-containing
(cuprate-dominant), with substantial Fe-bearing (28\%) and Mn-bearing
(13\%) admixture from sister I4/mmm phases --- including a
sub-population of Fe-pnictide-122 ThCr\(_2\)Si\(_2\)-prototype Ba/Fe/As
members. The structural common ground across both communities is the
\emph{2-D M--X sheet stacked with chemically variable spacer layers} ---
corner-sharing CuO\(_4\) squares for the YBCO members and the same
layered-tetragonal framework topology for the broader 1178 members ---
shared at the \emph{polyhedron-and-prototype level}.

A 3-body graphlet histogram captures the local polyhedron and the
immediate neighbour-set composition well: the Cu--O--O triplets that
define square-planar coordination and the Ba--O bond statistics that
distinguish layered cuprates from non-layered cuprates are present in
the feature vectors of both 6425 and 1178 in similar proportions. But
the \emph{cuprate-purity vs cation-mixture} distinction --- which is
what makes 6425 a distinct YBCO 123 sub-family and is the signal an
experimentalist relies on to know they are looking at the
high-T\(_\mathrm{c}\)-superconductor parent rather than a generic
mixed-cation layered tetragonal phase --- lives at the 3-hop scale of
the spacer-cation composition, not in any single 3-body triplet. On the
EMD between the full graphlet CDFs, the small L1 contribution from the
chemistry-pattern signal is overwhelmed by the much larger common ground
(both communities are I4/mmm layered-tetragonal with M--X sheets and
light-cation spacers), and 79\% of cuprate community 6425's
nearest-neighbour graphlet misses land in community 1178 (and 38\% of
the reverse).

WL message passing operates on a different unit. By round 3, the feature
at a Cu site in YBa\(_2\)Cu\(_3\)O\(_{7-\delta}\) carries the multiset
``square-planar O neighbours → Y or Ba in second shell → a Cu--O chain
in third shell'' with the cuprate-pure single-cation pattern encoded as
a specific multiset hash; the feature at a Cu site in a mixed-cation
1178 member carries ``square-planar O neighbours → Ba+Ca+RE
\emph{mixture} in second shell → variable spacer in third shell'' with
the second-shell multi-cation diversity encoded as a different hash. The
two features land at different positions in the 213-d feature space, and
the production community partition (and the embedding self-baseline)
separate the concentrated YBCO 123 sub-family from the broader
mixed-cation layered-tetragonal community at the 97\% level.

\textbf{The same pattern explains the other top confusions.} The single
largest global confusion pair --- community 5182 ↔ community 1357 (145
mutual graphlet-NN misses) --- is the Pnma orthorhombic perovskite
manganite family split across two production communities by
\emph{substitutional-doping composition pattern}: both communities are
perovskite-type ABO\(_3\) with edge-sharing octahedra, identical at the
polyhedron level. Their distinction is the spatial composition pattern
of A-site rare-earth and alkaline-earth dopants, which lives at the
3-hop scale that WL message passing reaches but 3-body graphlets do not.
Similarly, the Layered RE--Cu chalcogenide ↔ Stannite chalcogenide pair
(62 mutual misses) shares chalcogenide-tetrahedral local coordination
but differs in long-range stacking; the A15 ↔ BCC intermetallic pair (58
mutual misses) shares cubic close-packing but differs in second-shell
composition. In every case the conserved signal is the local
coordination polyhedron and the diverging signal is the 3-hop spatial
composition pattern.

\textbf{The takeaway.} The 30-percentage-point gap between graphlet
recovery (68\% global, 77--91\% on named manuscript communities) and
embedding self-recovery (97\%) is concentrated, structurally
interpretable, and chemically sensible. It says that the production
embedding is doing real medium-range structural work --- separating
compositionally-modulated \emph{sub-families} sharing the same local
motif --- that a purely local 3-body featurization cannot do. It does
\emph{not} say that either representation is wrong. Crucially, every
claim made in the manuscript rests on community-level and coarse-time
statistics that survive this representational substitution: the
densification cliff reproduces (§S9.2), the named superconductor
communities recover at 77--91\% (§S9.3), and the ``misses'' are
dominated by structurally adjacent sub-family pairs of which the cuprate
sub-family resolution between 6425 and 1178 is the most diagnostic for
the manuscript's superconductor analyses.

\subsection{S9.5. Comparison against chemistry-agnostic continuous
isometric invariants
(AMD)}\label{s9.5.-comparison-against-chemistry-agnostic-continuous-isometric-invariants-amd}

A second, qualitatively different point in the structural-representation
design space is the family of \emph{continuous isometric invariants}
developed by Widdowson, Kurlin, and colleagues\textsuperscript{18--20}.
The Average Minimum Distance (AMD) fingerprint is derived from the
unordered set of inter-atomic distances under crystallographic symmetry;
it is \emph{continuous by construction} under any atomic perturbation,
\emph{generically complete} (two crystals share their AMD fingerprint
iff they are isometric), and has supported deduplication of the
Cambridge Structural Database at the 800,000-entry
scale\textsuperscript{20}. AMD answers a fundamentally different
question than ours --- ``are two crystals geometrically identical?'' ---
and is by design chemistry-agnostic, treating atomic positions as a
coloured point set whose colours (element identities) must be definite
at every site. As a complementarity test to the chemistry-aware
embedding the manuscript uses, we re-partitioned the ICSD subset under
AMD and compared.

\textbf{Procedure.} For each ICSD entry assigned to a Louvain community
in the production partition (n = 152,122; matching the §S9 graphlet test
scope), we attempted to compute an AMD fingerprint at \(k = 100\) (the
value used in the Widdowson-Kurlin CSD maps\textsuperscript{20}) via the
public \texttt{average-minimum-distance} Python package, streaming
structures from the licensed ICSD CIF archive via a pymatgen → AMD
adapter. The resulting AMD vectors were standardised and re-partitioned
with the \emph{same} mutual-\(k=16\) kNN + Louvain (\(\gamma=1.0\))
graph construction as §S1.8.

\textbf{Representational coverage.} AMD successfully fingerprinted
\textbf{83,351 of 152,122} community-assigned ICSD entries --- a
representational coverage rate of \textbf{54.8\%}. The remaining 45.2\%
failed with a single, identical error: AMD's pymatgen adapter rejects
any crystallographic site whose chemical identity is not a single
definite element. In ICSD these failures concentrate in chemistry-rich
families with intrinsic compositional disorder --- variable-oxygen
cuprates of the YBa\(_2\)Cu\(_3\)O\(_{7-\delta}\) family, doped CMR
manganites \(\mathrm{La}_{1-x}\mathrm{Ca}_x\mathrm{MnO}_3\),
mixed-valence transition-metal oxides, solid solutions, and similar
partial-occupancy phases. This is not a defect of AMD; it is a faithful
expression of the framework's representational scope, which requires
definite atomic identity per site. The same 152,122-entry subset is
processed without exclusion by the chemistry-aware embedding used in the
manuscript.

\textbf{Global partition agreement.} Within the 83,351-entry
intersection, the AMD partition contains 3,499 communities, against
5,959 labels3 communities on the same subset. Measuring same-community
recovery directly --- the fraction of crystal pairs that share a labels3
community AND also share an AMD community --- gives \textbf{38.4\%} for
AMD, against \textbf{68\%} for the chemistry-agnostic-but-locally-richer
graphlet representation reported in §S9.3 on the full 152,122-entry
subset. The corresponding adjusted Rand index is ARI = 0.05 and
normalised mutual information NMI = 0.69. The AMD partition is a
substantial reorganisation into coarser, chemistry-blind clusters; the
gap between its 38.4\% same-community recovery and the embedding's
reorganisation into 5,959 fine-grained chemistry-aware communities is
concentrated in exactly the families described next.

\textbf{Per-community recovery, by family.} Table
Table~\ref{tbl:amd-recovery} summarises the recovery of the named
manuscript communities under the AMD partition. Three distinct
behaviours are visible. \textbf{First}, AMD recovers families whose
distinguishing signal is \emph{geometric} and \emph{chemically uniform}
--- pyrochlore A\(_2\)B\(_2\)O\(_7\) (community 5866, 92.0\% purity,
91.4\% capture), Mo\(_2\)FeB\(_2\) 2:1:2 RE-In-TM intermetallics
(community 2349, 98.3\% purity), and the Fe-pnictide 122-type
BaFe\(_2\)As\(_2\) family (community 4549, 83.6\% purity in AMD-3018)
--- these are cleanly-stoichiometric chemistry-rich families that AMD
fingerprints at 100\% coverage and partitions correctly.
\textbf{Second}, AMD cannot represent families whose distinguishing
signal is \emph{chemical} and concentrated in disordered sites ---
cuprate (communities 1178, 2846, 6607: 0 of 723+738+859 = 0 of 2,320
members representable in AMD-space), CMR manganite (community 160: 0 of
571 representable), Fe-pnictide 111/FeSe (community 358: 1 of 507
representable). \textbf{Third}, AMD \emph{merges} sub-communities that
the chemistry-aware production partition keeps distinct on the basis of
sub-anion chemistry --- communities 296 (Si/Ge-dominant ternary
intermetallics with the Fe-pnictide BaFe\(_2\)As\(_2\) subpopulation
contained inside; 333 members representable) and community 1
(Si/Ge-dominant ternary intermetallics with substantial Cu/Mn admixture;
222 members representable), both ThCr\(_2\)Si\(_2\)-prototype 122-type
silicide/germanide sub-communities, both map predominantly to the
\emph{same} AMD partition community (referred to throughout this
subsection as AMD-3420, distinct from the unrelated labels3 community of
the same numerical id; 595 AMD-3420 members), with purities 54.1\% and
41.0\% respectively. AMD's chemistry-agnostic isometric framework
excludes disordered phases, merges sub-communities distinguished by
sub-anion chemistry, and recovers cleanly-stoichiometric families well
--- three behaviours all predicted by its construction.

\begin{longtable}[]{@{}
  >{\raggedleft\arraybackslash}p{(\linewidth - 10\tabcolsep) * \real{0.1739}}
  >{\raggedright\arraybackslash}p{(\linewidth - 10\tabcolsep) * \real{0.1304}}
  >{\raggedleft\arraybackslash}p{(\linewidth - 10\tabcolsep) * \real{0.1739}}
  >{\raggedleft\arraybackslash}p{(\linewidth - 10\tabcolsep) * \real{0.1739}}
  >{\raggedleft\arraybackslash}p{(\linewidth - 10\tabcolsep) * \real{0.1739}}
  >{\raggedleft\arraybackslash}p{(\linewidth - 10\tabcolsep) * \real{0.1739}}@{}}
\caption{\label{tbl:amd-recovery}Recovery of manuscript-anchored
communities under the chemistry-agnostic AMD partition. ``n in AMD
subset'' is the number of labels3 community members successfully
AMD-fingerprinted (i.e., that survived the partial-occupancy filter).
``Best AMD'' is the AMD-partition community with the largest overlap;
``purity'' is best-overlap / n-in-subset (how much of the labels3
community is contained in that AMD community); ``capture'' is
best-overlap / amd-community-size (how much of the AMD community is
composed of this labels3 family). Labels3 communities 296 and 1 both
mapping to AMD-3420 (bold; an AMD-partition community whose numerical id
happens to coincide with --- but bears no relation to --- a different
small labels3 community) is the chemistry-merge example discussed in the
text.}\tabularnewline
\toprule\noalign{}
\begin{minipage}[b]{\linewidth}\raggedleft
labels3 community
\end{minipage} & \begin{minipage}[b]{\linewidth}\raggedright
family
\end{minipage} & \begin{minipage}[b]{\linewidth}\raggedleft
n in AMD subset
\end{minipage} & \begin{minipage}[b]{\linewidth}\raggedleft
best AMD-partition community
\end{minipage} & \begin{minipage}[b]{\linewidth}\raggedleft
purity
\end{minipage} & \begin{minipage}[b]{\linewidth}\raggedleft
capture
\end{minipage} \\
\midrule\noalign{}
\endfirsthead
\toprule\noalign{}
\begin{minipage}[b]{\linewidth}\raggedleft
labels3 community
\end{minipage} & \begin{minipage}[b]{\linewidth}\raggedright
family
\end{minipage} & \begin{minipage}[b]{\linewidth}\raggedleft
n in AMD subset
\end{minipage} & \begin{minipage}[b]{\linewidth}\raggedleft
best AMD-partition community
\end{minipage} & \begin{minipage}[b]{\linewidth}\raggedleft
purity
\end{minipage} & \begin{minipage}[b]{\linewidth}\raggedleft
capture
\end{minipage} \\
\midrule\noalign{}
\endhead
\bottomrule\noalign{}
\endlastfoot
1178 & High-T\(_{\mathrm{c}}\) cuprate (YBa\(_2\)Cu\(_3\)O\(_7\) /
RE-Ba-Cu-O 123) & \textbf{0} & --- & --- & --- \\
2846 & Cuprate sister (RE-Ba-Cu-O variants) & \textbf{0} & --- & --- &
--- \\
6607 & Cuprate sister (I4\(_1\)/amd) & \textbf{0} & --- & --- & --- \\
160 & CMR manganite & \textbf{0} & --- & --- & --- \\
358 & Fe-pnictide 111 / FeSe family & \textbf{1} & AMD-3376 & 100\% &
0.2\% \\
5450 & Fe-pnictide 1111 LaFeAsO (ZrCuSiAs) & 98 & AMD-1578 & 71.4\% &
27.0\% \\
296 & ThCr\(_2\)Si\(_2\) 122-type sub-community (Si/Ge dominant,
contains BaFe\(_2\)As\(_2\) subpop.) & 333 & \textbf{AMD-3420} & 54.1\%
& 30.3\% \\
1 & ThCr\(_2\)Si\(_2\) 122-type sister sub-community (Si/Ge dominant,
Cu/Mn admixture) & 222 & \textbf{AMD-3420} & 41.0\% & 15.3\% \\
4549 & \textbf{Fe-pnictide 122 BaFe\(_2\)As\(_2\) family} & 61 &
AMD-3018 & \textbf{83.6\%} & 30.9\% \\
2349 & Mo\(_2\)FeB\(_2\) 2:1:2 intermetallic & 119 & AMD-2643 & 98.3\% &
51.3\% \\
5866 & Pyrochlore (A\(_2\)B\(_2\)O\(_7\)) & 289 & AMD-2403 &
\textbf{92.0\%} & 91.4\% \\
6041 & Spinel (AB\(_2\)O\(_4\), Fd-3m) & 274 & AMD-1470 & 60.9\% &
59.4\% \\
6054 & Orthorhombic Pnma perovskite ABO\(_3\) & 364 & AMD-2966 & 52.7\%
& 30.9\% \\
1716 & Cubic perovskite ABO\(_3\) (SrTiO\(_3\)) & 21 & AMD-2358 & 100\%
& 11.3\% \\
2958 & AM\(_4\)X\(_8\) lacunar spinel (GaMo\(_4\)S\(_8\)) & 29 &
AMD-1950 & 51.7\% & 75.0\% \\
\end{longtable}

\textbf{The takeaway.} AMD is foundational for the database-hygiene
question it was built to answer (continuous, theorem-backed isometric
identity under perturbation; near-duplicate detection at million-crystal
scale\textsuperscript{19,20}) and is best read as the complementary
layer to ours rather than an alternative. Its representational scope ---
requiring definite atomic identity per site --- excludes exactly the
chemistry-rich families on which our renaissance-survey and
historical-densification analyses depend (cuprates, doped manganites,
Fe-pnictide 111/FeSe with mixed-chalcogenide sites). Where AMD does
compute, it merges sub-communities that the chemistry-aware embedding
distinguishes on sub-anion chemistry grounds, as the
ThCr\(_2\)Si\(_2\)-prototype 122-type pair (labels3 communities 296 and
1 both mapping to AMD-3420) demonstrates. The 213-d chemistry-aware
embedding the manuscript uses is the right representation for the
historical and discovery-dynamics questions on which the manuscript's
headline results rest.

\protect\phantomsection\label{refs}


\maketitle

\subsection{Extended Data figure
captions}\label{extended-data-figure-captions}

\begin{figure}
\centering
\pandocbounded{\includegraphics[keepaspectratio,alt={Extended Data Figure 1. End-to-end pipeline overview. Schematic of the analysis pipeline. Trunk: licensed ICSD CIFs (n = 167\{,\}500) are featurized into a per-site vector (7-d occupancy-weighted Element-property descriptor + 61-d matminer CrystalNNFingerprint("ops") = 68 dims per site), passed through three rounds of weighted-mean Weisfeiler--Lehman propagation on the CrystalNN bond graph, pooled to a 213-d structure-level vector (\textbackslash textit\{mean\} \textbackslash,\textbackslash Vert\textbackslash, \textbackslash textit\{max\} \textbackslash,\textbackslash Vert\textbackslash, \textbackslash textit\{var\} over sites + a 9-d global lattice block), standardized, and projected to 32 PCA components. The 32-d basis is frozen: identical for all downstream comparisons. The frozen basis feeds two analysis tracks. Left branch (green, ICSD self-comparison): a mutual k-nearest-neighbor graph with k=16, Gaussian-weighted edges (\textbackslash sigma = median k-NN Euclidean distance), and Louvain at resolution \textbackslash gamma = 1.0 produces the canonical Louvain partition; temporal replay of community membership through publication time produces the temporal cliff and renaissance-survey results in Figs. 1 and 2. Right branch (orange, external-source projection): five publicly-released computed-structure samples (GNoME, MatterGen-public, Materials Project theoretical-only, JARVIS-DFT off-hull, Alexandria-PBE off-hull) are featurized through the identical pipeline and projected into the frozen basis; per-community 95th-percentile within-community centroid-distance thresholds classify each projected entry as in-basin or frontier, producing the five-source calibration and synthesizability-prior quadrant in Figs. 3 and 4. The same geometry feeds both branches; this is what makes the held-out and external comparisons calibrated rather than independent partitionings.}]{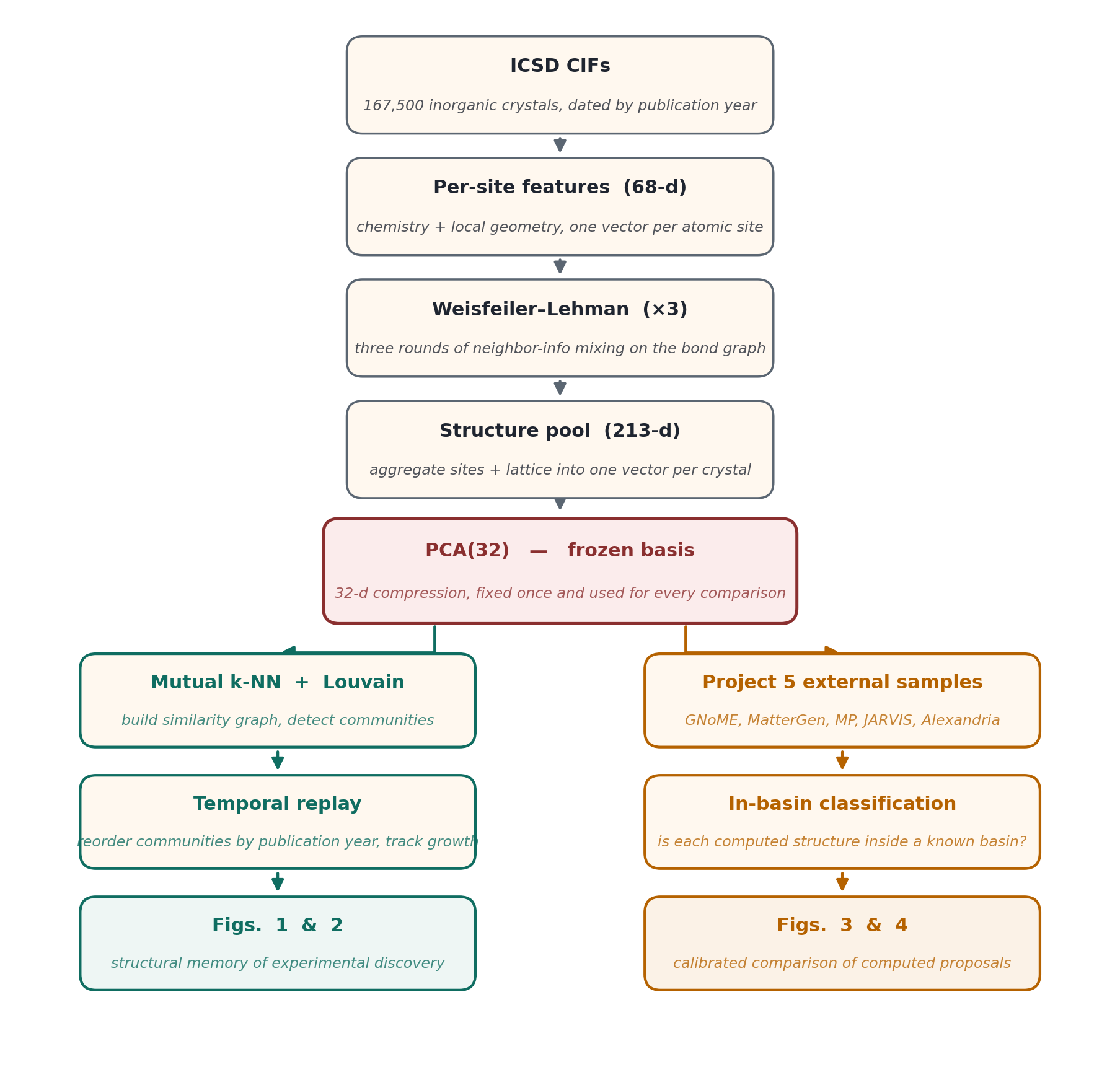}}
\caption{\textbf{Extended Data Figure 1. End-to-end pipeline overview.}
Schematic of the analysis pipeline. Trunk: licensed ICSD CIFs
(\(n = 167{,}500\)) are featurized into a per-site vector (\(7\)-d
occupancy-weighted Element-property descriptor + \(61\)-d matminer
\texttt{CrystalNNFingerprint("ops")} = \(68\) dims per site), passed
through three rounds of weighted-mean Weisfeiler--Lehman propagation on
the \texttt{CrystalNN} bond graph, pooled to a \(213\)-d structure-level
vector (\(\textit{mean} \,\Vert\, \textit{max} \,\Vert\, \textit{var}\)
over sites + a \(9\)-d global lattice block), standardized, and
projected to \(32\) PCA components. The \(32\)-d basis is \emph{frozen}:
identical for all downstream comparisons. The frozen basis feeds two
analysis tracks. \textbf{Left branch (green, ICSD self-comparison):} a
mutual \(k\)-nearest-neighbor graph with \(k=16\), Gaussian-weighted
edges (\(\sigma\) = median \(k\)-NN Euclidean distance), and Louvain at
resolution \(\gamma = 1.0\) produces the canonical Louvain partition;
temporal replay of community membership through publication time
produces the temporal cliff and renaissance-survey results in Figs. 1
and 2. \textbf{Right branch (orange, external-source projection):} five
publicly-released computed-structure samples (GNoME, MatterGen-public,
Materials Project theoretical-only, JARVIS-DFT off-hull, Alexandria-PBE
off-hull) are featurized through the identical pipeline and projected
into the frozen basis; per-community \(95\)th-percentile
within-community centroid-distance thresholds classify each projected
entry as in-basin or frontier, producing the five-source calibration and
synthesizability-prior quadrant in Figs. 3 and 4. The same geometry
feeds both branches; this is what makes the held-out and external
comparisons calibrated rather than independent
partitionings.}\label{fig:ed1}
\end{figure}

\begin{figure}
\centering
\pandocbounded{\includegraphics[keepaspectratio,alt={Extended Data Figure 2. Year-shuffle null model rules out the partition-artifact reading of the temporal cliff. Observed community-birth ratio (red) compared with the 5--95\% band of 200 random permutations of publication years over the same 163,339 community-assigned ICSD entries (gray), holding the final community labels fixed. The observed ratio is below the null in 1910s--1960s --- the literature densified existing structural neighborhoods faster than year-shuffling alone predicts --- and above the null in 1970s--2010s, by approximately an order of magnitude in the 1980s--2000s. Two distinct path-dependence regimes are present rather than a single mechanical front-loading artifact.}]{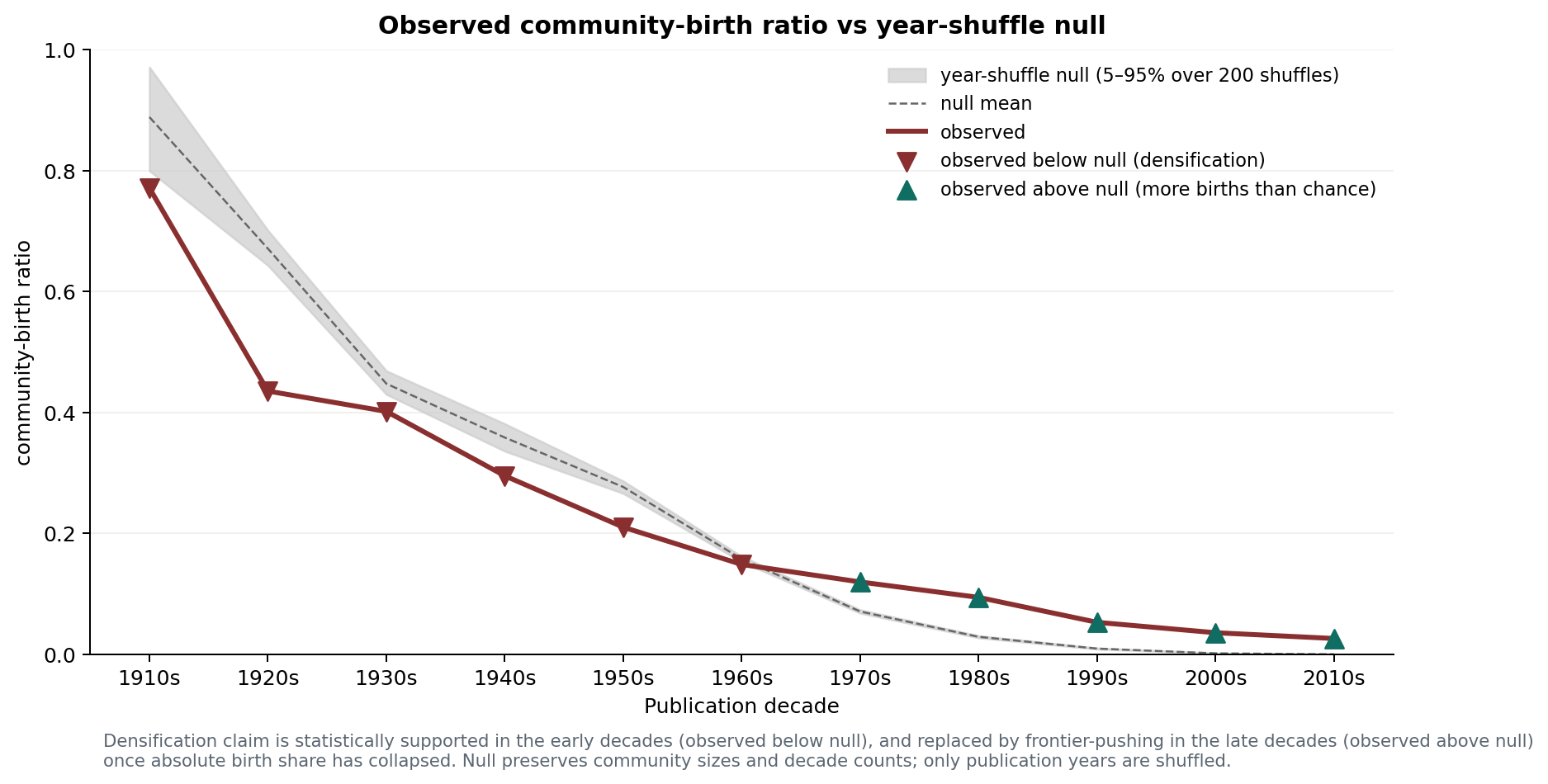}}
\caption{\textbf{Extended Data Figure 2. Year-shuffle null model rules
out the partition-artifact reading of the temporal cliff.} Observed
community-birth ratio (red) compared with the 5--95\% band of 200 random
permutations of publication years over the same 163,339
community-assigned ICSD entries (gray), holding the final community
labels fixed. The observed ratio is \emph{below} the null in
1910s--1960s --- the literature densified existing structural
neighborhoods faster than year-shuffling alone predicts --- and
\emph{above} the null in 1970s--2010s, by approximately an order of
magnitude in the 1980s--2000s. Two distinct path-dependence regimes are
present rather than a single mechanical front-loading
artifact.}\label{fig:ed2}
\end{figure}

\begin{figure}
\centering
\pandocbounded{\includegraphics[keepaspectratio,alt={Extended Data Figure 3. Targeted renaissance probes complementing the systematic survey. Three panels. (a) Fe-pnictide 1111 LaFeAsO family (community 5450, n=98): pre-2008 = 1.5 entries/yr, post-2008 = 5.0 entries/yr (3.3× fold-change after Kamihara \& Hosono 2008). The Fe-pnictide superconductor family also splits into 122-type and 111-type sister communities by structural prototype (community 4549 BaFe₂As₂ family with a 16.3× post-2008 fold-change of its own; and community 358 LiFeAs/FeSe family), an example of correct fine-grained discrimination. (b) Transition-metal dichalcogenide community 62 (n=95): pre-2010 = 0.4/yr, post-2010 = 0.5/yr (1.2×). Informative null result --- the post-graphene 2D-materials boom is property-driven on pre-existing bulk structures and does not produce new ICSD entries. (c) Deep-dive on rank-\#6 community 2349 (n=119): the Mo₂FeB₂-prototype 2:1:2 RE-In-TM ternary intermetallic family (La₂InCu₂, La₂InPd₂, etc.). Step at 1989 corresponds to the Pöttgen-era systematic crystal-chemistry survey (program-driven rather than publication-driven renaissance). Pre-event 0.1/yr, post-event 7.3/yr (73× fold-change).}]{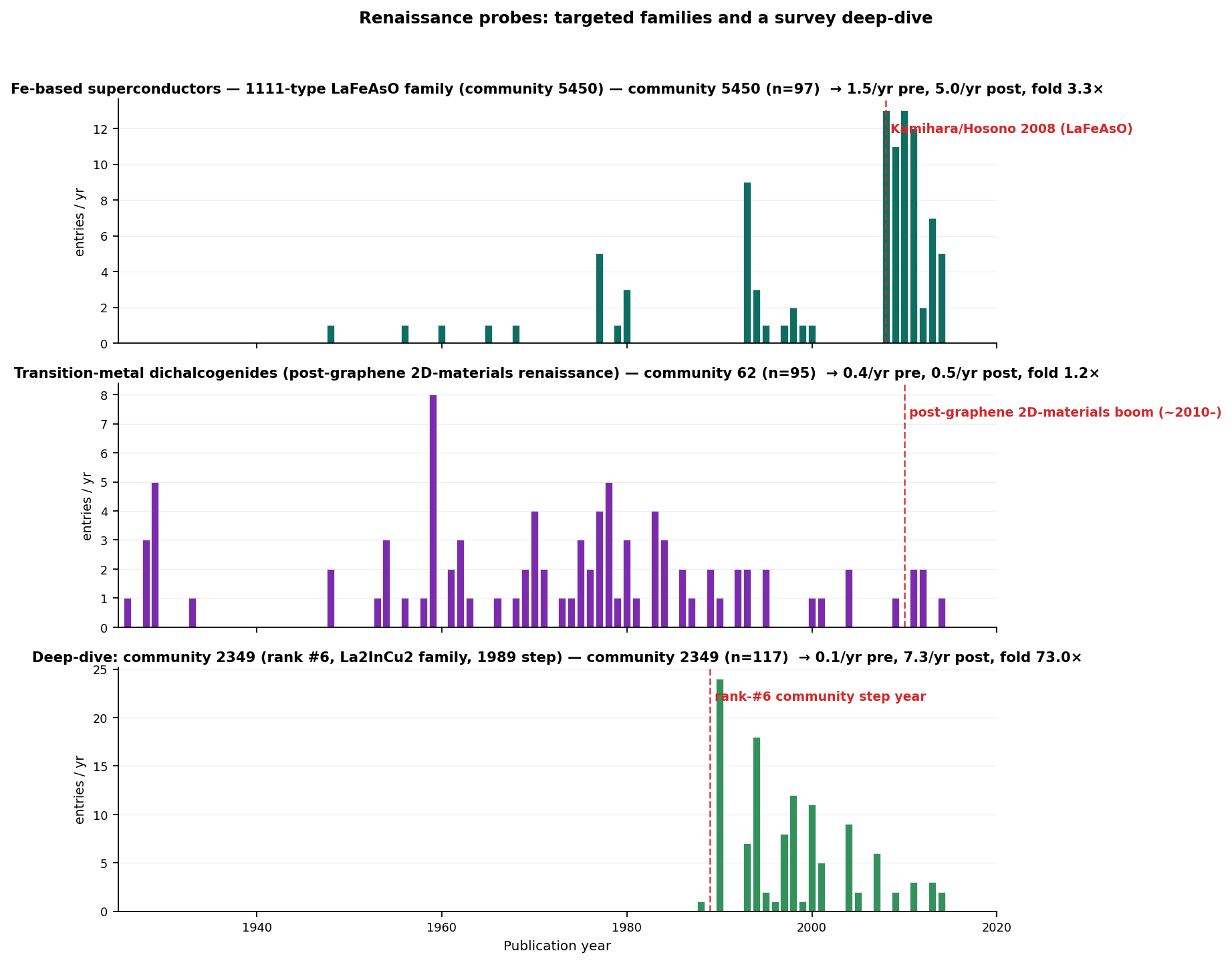}}
\caption{\textbf{Extended Data Figure 3. Targeted renaissance probes
complementing the systematic survey.} Three panels. (a) Fe-pnictide 1111
LaFeAsO family (community 5450, n=98): pre-2008 = 1.5 entries/yr,
post-2008 = 5.0 entries/yr (3.3× fold-change after Kamihara \& Hosono
2008). The Fe-pnictide superconductor family also splits into 122-type
and 111-type sister communities by structural prototype (community 4549
BaFe₂As₂ family with a 16.3× post-2008 fold-change of its own; and
community 358 LiFeAs/FeSe family), an example of correct fine-grained
discrimination. (b) Transition-metal dichalcogenide community 62 (n=95):
pre-2010 = 0.4/yr, post-2010 = 0.5/yr (1.2×). Informative null result
--- the post-graphene 2D-materials boom is property-driven on
pre-existing bulk structures and does not produce new ICSD entries. (c)
Deep-dive on rank-\#6 community 2349 (n=119): the Mo₂FeB₂-prototype
2:1:2 RE-In-TM ternary intermetallic family (La₂InCu₂, La₂InPd₂, etc.).
Step at 1989 corresponds to the Pöttgen-era systematic crystal-chemistry
survey (program-driven rather than publication-driven renaissance).
Pre-event 0.1/yr, post-event 7.3/yr (73× fold-change).}\label{fig:ed3}
\end{figure}

\begin{figure}
\centering
\pandocbounded{\includegraphics[keepaspectratio,alt={Extended Data Figure 4. Composition-matched in-basin rates across five external sources. Left panel: coarse anion / element-count / ratio strata. Right panel: pymatgen-anonymized stoichiometry. For each cutoff (1990, 2000, 2010) and each external source, in-basin rates are restricted to composition strata populated by both held-out ICSD and that external source. The exploration gap survives both stratifications for every source at every cutoff: anonymized 2010 gaps are +13 pp (MatterGen-public), +24 pp (MP-theoretical), +25 pp (GNoME), +32 pp (JARVIS-DFT), and +37 pp (Alexandria off-hull). Strata-count annotations show per-source intersection with held-out ICSD (G = GNoME, Mat = MatterGen, MP = Materials Project, Jar = JARVIS-DFT, Ale = Alexandria). The exploration gap is therefore not a stoichiometric drift artifact. Wilson 95\% confidence intervals throughout.}]{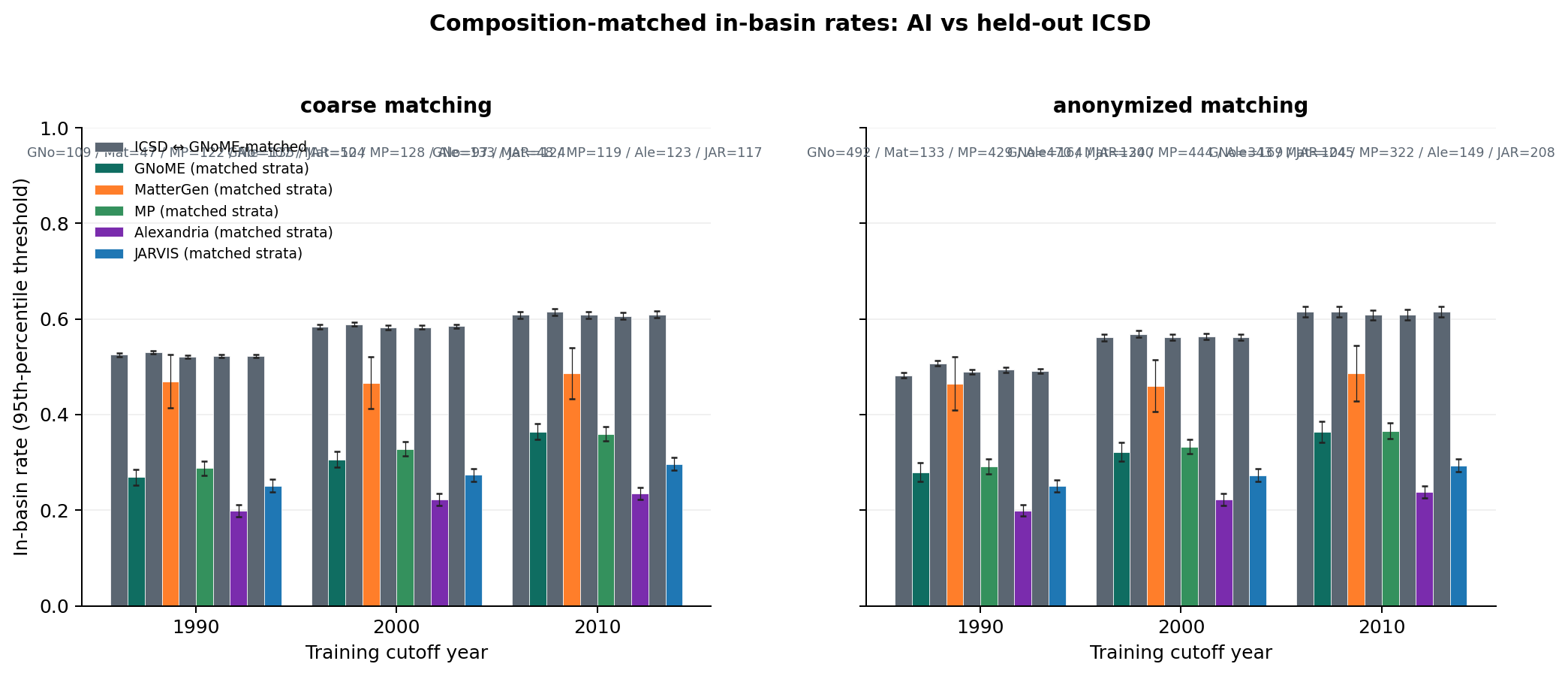}}
\caption{\textbf{Extended Data Figure 4. Composition-matched in-basin
rates across five external sources.} Left panel: coarse anion /
element-count / ratio strata. Right panel: pymatgen-anonymized
stoichiometry. For each cutoff (1990, 2000, 2010) and each external
source, in-basin rates are restricted to composition strata populated by
both held-out ICSD and that external source. The exploration gap
survives both stratifications for every source at every cutoff:
anonymized 2010 gaps are +13 pp (MatterGen-public), +24 pp
(MP-theoretical), +25 pp (GNoME), +32 pp (JARVIS-DFT), and +37 pp
(Alexandria off-hull). Strata-count annotations show per-source
intersection with held-out ICSD (G = GNoME, Mat = MatterGen, MP =
Materials Project, Jar = JARVIS-DFT, Ale = Alexandria). The exploration
gap is therefore not a stoichiometric drift artifact. Wilson 95\%
confidence intervals throughout.}\label{fig:ed4}
\end{figure}

\begin{figure}
\centering
\pandocbounded{\includegraphics[keepaspectratio,alt={Extended Data Figure 5. Structural-accessibility score across ICSD roles and public GNoME proposals. 𝒜ᵢ combines normalized centroid distance with the historical stabilizers of community size and community age. ICSD graph-core nodes have the lowest mean accessibility cost (-0.32), bridge nodes are intermediate (+0.72), and community-birth nodes are highest (+1.78). Public GNoME structures follow the same ordering: in-basin GNoME proposals (per-community 95th-percentile classification, consistent with Fig. 3c and Fig. 4) sit in the accessible non-core regime (mean +0.55), while frontier GNoME proposals shift toward the historical-birth regime (mean +1.20). The qualitative ordering ICSD core \textless{} ICSD periphery \textless{} ICSD bridge \textless{} GNoME in-basin \textless{} GNoME frontier ≈ ICSD birth is robust to the sensitivity-grid weighting of the score's three terms (Methods; held at 31 of the 36 (\textbackslash alpha, \textbackslash beta) cells of the weight grid). The corresponding analyses for MatterGen-public, MP-theoretical, JARVIS-DFT, and Alexandria off-hull are reported quantitatively in the five-source calibration in main-text Figure 3.}]{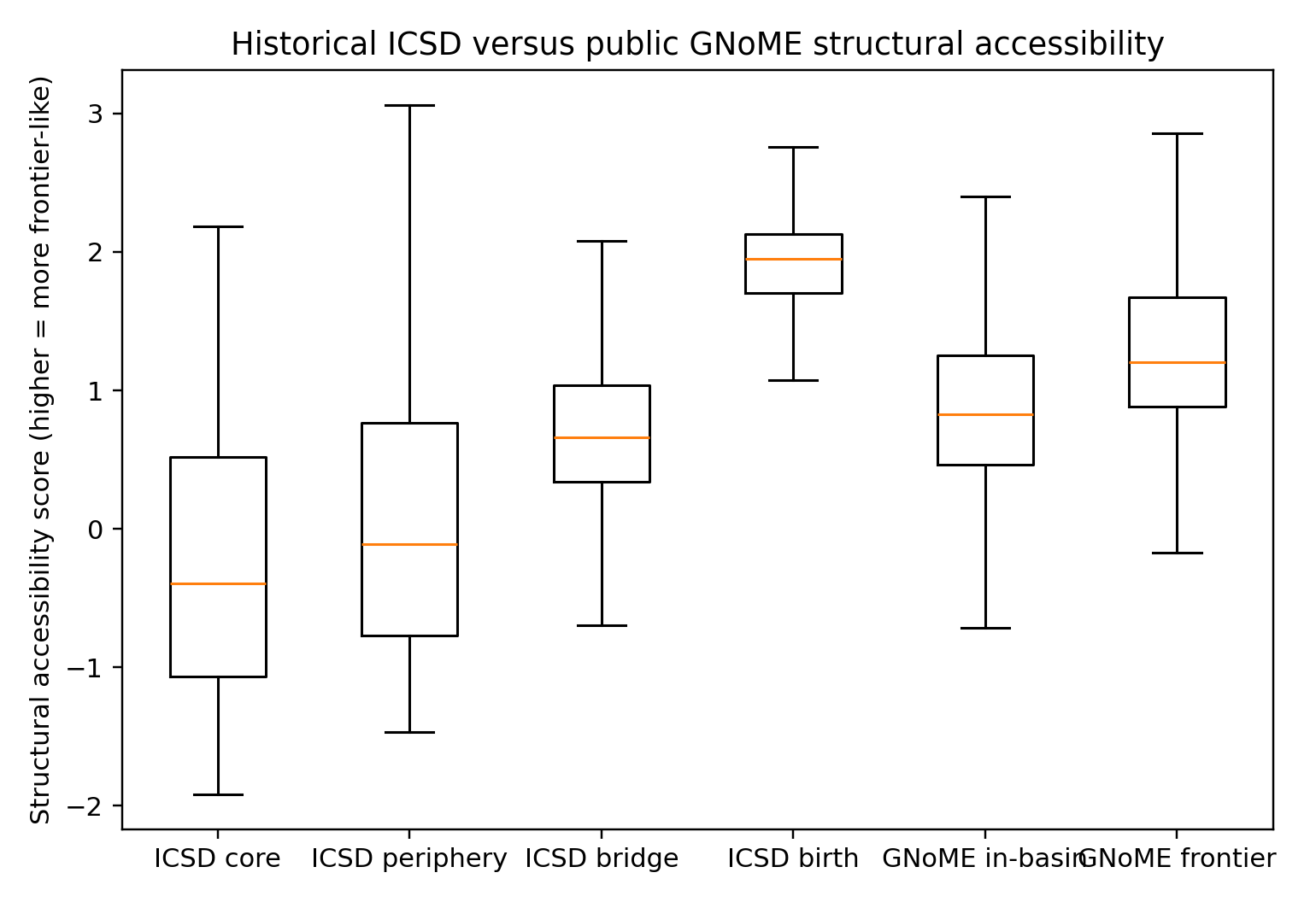}}
\caption{\textbf{Extended Data Figure 5. Structural-accessibility score
across ICSD roles and public GNoME proposals.} 𝒜ᵢ combines normalized
centroid distance with the historical stabilizers of community size and
community age. ICSD graph-core nodes have the lowest mean accessibility
cost (\(-0.32\)), bridge nodes are intermediate (\(+0.72\)), and
community-birth nodes are highest (\(+1.78\)). Public GNoME structures
follow the same ordering: in-basin GNoME proposals (per-community
95th-percentile classification, consistent with Fig. 3c and Fig. 4) sit
in the accessible non-core regime (mean \(+0.55\)), while frontier GNoME
proposals shift toward the historical-birth regime (mean \(+1.20\)). The
qualitative ordering ICSD core \textless{} ICSD periphery \textless{}
ICSD bridge \textless{} GNoME in-basin \textless{} GNoME frontier ≈ ICSD
birth is robust to the sensitivity-grid weighting of the score's three
terms (Methods; held at 31 of the 36 (\(\alpha, \beta\)) cells of the
weight grid). The corresponding analyses for MatterGen-public,
MP-theoretical, JARVIS-DFT, and Alexandria off-hull are reported
quantitatively in the five-source calibration in main-text Figure
3.}\label{fig:ed5}
\end{figure}

\subsection{Extended Data table
captions}\label{extended-data-table-captions}

\textbf{Extended Data Table 1. Datasets used in this study.} High-level
orientation: dataset, type, role in this paper, size used, and where
each is referenced. Per-source filter rules and projection success rates
for the five external samples are tabulated in Extended Data Table 3.

\begin{longtable}[]{@{}
  >{\raggedright\arraybackslash}p{(\linewidth - 8\tabcolsep) * \real{0.1875}}
  >{\raggedright\arraybackslash}p{(\linewidth - 8\tabcolsep) * \real{0.1875}}
  >{\raggedright\arraybackslash}p{(\linewidth - 8\tabcolsep) * \real{0.1875}}
  >{\raggedleft\arraybackslash}p{(\linewidth - 8\tabcolsep) * \real{0.2500}}
  >{\raggedright\arraybackslash}p{(\linewidth - 8\tabcolsep) * \real{0.1875}}@{}}
\toprule\noalign{}
\begin{minipage}[b]{\linewidth}\raggedright
Dataset
\end{minipage} & \begin{minipage}[b]{\linewidth}\raggedright
Type
\end{minipage} & \begin{minipage}[b]{\linewidth}\raggedright
Role in this paper
\end{minipage} & \begin{minipage}[b]{\linewidth}\raggedleft
Size used
\end{minipage} & \begin{minipage}[b]{\linewidth}\raggedright
Referenced in
\end{minipage} \\
\midrule\noalign{}
\endhead
\bottomrule\noalign{}
\endlastfoot
ICSD (FIZ Karlsruhe)\textsuperscript{1} & Experimental crystal-structure
database & Baseline of experimentally-realized inorganic structures;
defines the frozen 32-d structural geometry & 167,500 of 181,362
requested (92.4 \%) & Methods (\emph{Crystal representation}); Figs. 1,
2 \\
GNoME\textsuperscript{2} & Generative AI, graph network (DeepMind) &
External test sample projected into the frozen ICSD basis & 5,000
(public CIF release) & Figs. 3, 4 \\
MatterGen-public\textsuperscript{3} & Generative AI, diffusion model
(Microsoft Research) & External test sample & 386 (public CIF release) &
Figs. 3, 4 \\
Materials Project theoretical-only\textsuperscript{4} & DFT screening,
prototype substitution & External test sample, no ICSD provenance &
5,000 random of 20,479 candidates & Figs. 3, 4 \\
JARVIS-DFT off-hull\textsuperscript{5} & DFT screening & External test
sample, no ICSD provenance & 5,000 random of 22,022 candidates & Figs.
3, 4 \\
Alexandria-PBE off-hull\textsuperscript{6} & DFT screening & External
test sample & 5,000 random of 154,942 candidates & Figs. 3, 4 \\
TRI thermodynamic-stability network\textsuperscript{7} &
Thermodynamic-stability network (Toyota Research Institute) &
Formula-overlap reference for the stepping-stone analysis & 18,821
shared formulas (16,582 classifiable) & Fig. 1 (the 82.9 \%
stepping-stone claim); Supporting Information §S2, §S5 \\
\end{longtable}

\textbf{Extended Data Table 2. Top-20 communities by birth-year
step-change score.} For each of the 378 communities of size ≥ 50
(excluding HDBSCAN noise), we score the strongest pre/post fold-change
over symmetric ±10-year windows scanning candidate event years
1970--2010 (Methods). The top-20 by score are tabulated with their
best-fit step year, top-five member reduced formulas, and best-effort
identification of the corresponding scientific event. The survey
positively identifies nine textbook field-defining renaissances among
the top-20 step-change communities: Sm₂Fe₁₇N permanent magnets (rank 1,
1985), CMR manganites (rank 2, 1993), SOFC perovskite cathodes (rank 5,
1994), MAX phases (rank 9, 2004), NaCoO₂ thermoelectrics (rank 12,
1999), ordered double perovskites for spintronics (ranks 11 and 14,
1999/1997), Li-ion battery cathodes (rank 15, 1992), dilute magnetic
semiconductors (rank 8, 2005), and post-1986 layered Ruddlesden-Popper
cuprates (rank 16, 1987). The remaining ranks are program-driven
systematic surveys and applied-chemistry surges at equivalent
statistical strength --- not missed renaissances --- including the
rank-\#6 Pöttgen-era 2:1:2 RE-In-TM intermetallic family (1989, 73×
fold-change) and post-Steglich 1:1:2 heavy-fermion campaigns (\#4); the
full breakdown is in Supporting Information §S8.4.

\textbf{Extended Data Table 3. Per-source filter specifications and
projection success rates.}

\begin{longtable}[]{@{}
  >{\raggedright\arraybackslash}p{(\linewidth - 14\tabcolsep) * \real{0.1034}}
  >{\raggedright\arraybackslash}p{(\linewidth - 14\tabcolsep) * \real{0.1034}}
  >{\raggedright\arraybackslash}p{(\linewidth - 14\tabcolsep) * \real{0.1034}}
  >{\raggedleft\arraybackslash}p{(\linewidth - 14\tabcolsep) * \real{0.1379}}
  >{\raggedleft\arraybackslash}p{(\linewidth - 14\tabcolsep) * \real{0.1379}}
  >{\raggedleft\arraybackslash}p{(\linewidth - 14\tabcolsep) * \real{0.1379}}
  >{\raggedleft\arraybackslash}p{(\linewidth - 14\tabcolsep) * \real{0.1379}}
  >{\raggedleft\arraybackslash}p{(\linewidth - 14\tabcolsep) * \real{0.1379}}@{}}
\toprule\noalign{}
\begin{minipage}[b]{\linewidth}\raggedright
Source
\end{minipage} & \begin{minipage}[b]{\linewidth}\raggedright
Database / version
\end{minipage} & \begin{minipage}[b]{\linewidth}\raggedright
Filter rule
\end{minipage} & \begin{minipage}[b]{\linewidth}\raggedleft
Seed
\end{minipage} & \begin{minipage}[b]{\linewidth}\raggedleft
Max sites
\end{minipage} & \begin{minipage}[b]{\linewidth}\raggedleft
Candidates after filter
\end{minipage} & \begin{minipage}[b]{\linewidth}\raggedleft
Sampled
\end{minipage} & \begin{minipage}[b]{\linewidth}\raggedleft
Successfully projected
\end{minipage} \\
\midrule\noalign{}
\endhead
\bottomrule\noalign{}
\endlastfoot
GNoME\textsuperscript{2} & DeepMind \texttt{materials\_discovery} repo,
public 5,000-CIF release & none (use full release) & 42 & 256 & 5,000 &
5,000 & 5,000 \\
MatterGen-public\textsuperscript{3} & Public CIF release accompanying
the MatterGen publication & none (use full release) & 42 & 256 & 386 &
386 & 386 \\
Materials Project (theoretical-only)\textsuperscript{4} & Modern API via
\texttt{mp-api}, queried 2026-04-27 & \texttt{theoretical\ =\ True};
\texttt{e\_above\_hull} \(\leq\) 0.2 eV/atom; empty
\texttt{database\_IDs.icsd} & 42 & 256 & 20,479 & 5,000 & 4,999 \\
JARVIS-DFT\textsuperscript{5} & Figshare release
\texttt{dft\_3d-12-12-2022.json} (75,993 entries) loaded via
\texttt{jarvis-tools}, downloaded 2026-04-27 & empty \texttt{icsd}
field; \texttt{ehull} \(\in [0.05, 0.5]\) eV/atom & 42 & 256 & 22,022 &
5,000 & 4,964 \\
Alexandria-PBE\textsuperscript{6} & 2025.07.02 PBE bulk JSON release,
three source files (indices 0, 19, 38) downloaded 2026-04-27 &
\texttt{e\_above\_hull} \(\in [0.05, 0.5]\) eV/atom & 42 & 256 & 154,942
& 5,000 & 5,000 \\
\end{longtable}

For each source, ``Successfully projected'' counts entries that were
featurized through the matminer-based three-round message-passing
pipeline (Methods), assigned to the nearest community centroid in the
frozen 32-D PCA space, and assignable to a parseable composition
stratum. The small drops between Sampled and Successfully projected for
MP-theoretical (1 entry) and JARVIS-DFT (36 entries) reflect entries
with NaN matminer features (typically noble-gas-containing structures
whose Pauling electronegativity is undefined). Composition-matched and
unmatched in-basin rates per cutoff for each source are reported in
main-text Figure 3 and Supporting Information §S7.

\protect\phantomsection\label{refs}